\documentclass[12pt]{article}

\usepackage{epsf}
\usepackage[dvips]{graphicx}

\begin{document}

\begin{center}
{\bf\large New parameterization of
effective nucleon-nucleon $t$-matrix interaction for scattering
at intermediate energies.}
\end{center}

\vspace{1cm}

\begin{center}
{\large N.B.Ladygina }
\end{center}

\vspace{0.5cm}

\begin{center}
{\it E-mail:nladygina@jinr.ru}
\end{center}

\vspace{1cm}

\begin{abstract}
The model suggested by Love and Franey for description of
the nucleon-nucleon interaction
was used as the base. The new fitting of the model parameters
was done in the energy range from 100 MeV up to 1100 MeV.
It is based on the modern partial-wave-analysis solution for NN-amplitudes.
The three observables: differential cross section, vector analyzing
power, and spin correlation coefficient -- were obtained at every
energy. The results are compared with existing the experimental data.
\end{abstract}

\newpage

In order to solve the problems of the nucleon-nucleus scattering,
it is very important to know the coupling between two nucleons.
A quantitative knowledge of the nucleon-nucleon coupling from first
principles over a wide energy range must await a more complete
theory of strong interaction.

In this connection it is useful to have a model which describe the
effective nucleon-nucleon interaction. Such phenomenological model was
suggested by W.G.Love and M.A.Franey in ref.\cite{LF}. According to
this paper $t$-matrix can be expressed in the following form
\begin{eqnarray}
\label{lfampl}
t^{sym}(E^*, \Theta^*)&=&A^\prime (E^*, \Theta^*) P_S+B^\prime(E^*, \Theta^*) P_T
+C^\prime(E^*, \Theta^*)((\vec\sigma_1 +\vec\sigma_2 )\cdot\hat n)+
\nonumber\\
&+&E^\prime(E^*, \Theta^*)S_{12}(\hat q) +F^\prime(E^*, \Theta^*)S_{12}(\hat Q),
\end{eqnarray}
where $E^*$ is the center-of-mass energy; $\Theta^*$ is the 
scattering angle in the c.m. system. The unit vectors
$\hat q, \hat Q , \hat n$ form a right-handed coordinate system
with
\begin{eqnarray}
\hat q=\frac{\vec k -\vec k^\prime}{|\vec k -\vec k^\prime|},~~~~
\hat Q=\frac{\vec k +\vec k^\prime}{|\vec k +\vec k^\prime|},~~~~
\hat n=\hat q\times \hat Q,
\end{eqnarray}
where $\vec k$ and $\vec k^\prime$ are the initial and final momenta
in the c.m. system.

The usual tensor operator $S_{12}(\hat k)=3(\vec\sigma_1 \hat k)(\vec\sigma_2 \hat k)-
(\vec\sigma_1\vec\sigma_2)$,
the singlet spin-projection operator
$P_S=\frac{1}{4}(1-(\vec\sigma_1\vec\sigma_2))$, and
the triplet spin-projection operator
$P_T=\frac{1}{4}(3+(\vec\sigma_1\vec\sigma_2))$ 
were used in Eq.(\ref{lfampl}).
It should be noted, that t-matrix, here, is the antisymmetrized one,
i.e. $t^{sym}=[1-P_{12}]t$, where $P_{12}$ is the permutation operator of the 
particles of "1" and "2".

This $t$-matrix is connected with $NN$-amplitude $M$ by the relationship
\begin{eqnarray}
t_{NN}(E^*, \theta^*)=-\frac{4\pi}{E^*}M(E^*, \theta^*).
\end{eqnarray}

The amplitudes $A^\prime,~B^\prime,~C^\prime,~E^\prime,~F^\prime$
are parameterized according to the symmetric properties.
In case, when isospin NN-pare equals to zero, $I=0$, these
amplitudes have the following form

\begin{eqnarray}
A_0^\prime&=&\frac{4\pi R_i^3}{(2\pi)^3}V_i^{SO}\left(\frac{1}{1+q^2R_i^2}-
\frac{1}{1+Q^2R_i^2}\right)
\nonumber\\
B_0^\prime&=&\frac{4\pi R_i^3}{(2\pi)^3}V_i^{TE}\left(\frac{1}{1+q^2R_i^2}+
\frac{1}{1+Q^2R_i^2}\right)
\nonumber\\
C_0^\prime&=&\frac{2\pi i R_i^5}{(2\pi)^3}V_i^{LSE}qQ
\left(\frac{1}{(1+q^2R_i^2)^2}-
\frac{1}{(1+Q^2R_i^2)^2}\right)
\\
E_0^\prime&=&-\frac{32\pi i R_i^7}{(2\pi)^3}V_i^{TNE}
\frac{q^2}{(1+q^2R_i^2)^3}
\nonumber\\
F_0^\prime&=&-\frac{32\pi i R_i^7}{(2\pi)^3}V_i^{TNE}
\frac{Q^2}{(1+Q^2R_i^2)^3}.
\nonumber
\end{eqnarray}

\newpage
For $I=1$ these amplitudes are
\begin{eqnarray}
A_1^\prime&=&\frac{4\pi R_i^3}{(2\pi)^3}V_i^{SE}\left(\frac{1}{1+q^2R_i^2}+
\frac{1}{1+Q^2R_i^2}\right)
\nonumber\\
B_1^\prime&=&\frac{4\pi R_i^3}{(2\pi)^3}V_i^{TO}\left(\frac{1}{1+q^2R_i^2}-
\frac{1}{1+Q^2R_i^2}\right)
\nonumber\\
C_1^\prime&=&\frac{2\pi i R_i^5}{(2\pi)^3}V_i^{LSO}qQ
\left(\frac{1}{(1+q^2R_i^2)^2}+
\frac{1}{(1+Q^2R_i^2)^2}\right)
\\
E_1^\prime&=&-\frac{32\pi i R_i^7}{(2\pi)^3}V_i^{TNO}
\frac{q^2}{(1+q^2R_i^2)^3}
\nonumber\\
F_1^\prime&=&\frac{32\pi i R_i^7}{(2\pi)^3}V_i^{TNO}
\frac{Q^2}{(1+Q^2R_i^2)^3}.
\nonumber
\end{eqnarray}

In this  paper the new set of the coefficients $R_i$
and $V_i$ were obtained by fitting SP07 \cite{said}
$NN$ amplitudes. The energy region from 100 MeV up to 1100 MeV
was considered. In order to get the amplitudes subset for isospin
equal to zero, the "NP" and "NP1" subsets were used. The new sets 
of parameters are  presented in the tables below. The three
observables: differential cross section, vector analyzing power
and spin correlation coefficient -- were obtained for every energy
and were compared with the  experimental data \cite{said}. 
 On the figures below the fitting results are presented by the
 red lines. The black curves correspond to the results obtained
 with the amplitudes from SAID  \cite{said}.

The NN $t$-matrix can be also expressed in the standard form
\begin{eqnarray}
\label{ampl}
t(E^*, \Theta^*)&=&A(E^*, \Theta^*)+
B(E^*, \Theta^*)(\vec\sigma_1 \hat n)(\vec\sigma_2 \hat n)+
C(E^*, \Theta^*)((\vec\sigma_1 +\vec\sigma_2 )\cdot\hat n)+
\nonumber\\
&+&E(E^*, \Theta^*)(\vec\sigma_1 \hat q)(\vec\sigma_2 \hat q)+
F(E^*, \Theta^*)(\vec\sigma_1 \hat Q)(\vec\sigma_2 \hat Q).
\end{eqnarray}

The relationships between amplitudes 
$A^\prime, ~B^\prime,~C^\prime, ~E^\prime,~F^\prime$ and
$A, ~B,~C, ~E,~F$ can be got with comparison of the two definitions
of $t$-matrix, Eq.(\ref{lfampl}) and Eq.(\ref{ampl}).
\begin{eqnarray}
A&=&\frac{1}{4}(A^\prime+3B^\prime)
\nonumber\\
B&=&\frac{1}{4}(B^\prime -A^\prime)-(E^\prime +F^\prime)
\nonumber\\
C&=&C^\prime
\\
E&=&\frac{1}{4}(B^\prime -A^\prime)+2E^\prime-F^\prime
\nonumber\\
F&=&\frac{1}{4}(B^\prime -A^\prime)-E^\prime+2F^\prime.
\nonumber
\end{eqnarray}

All notations used in this paper are the same as in ref.\cite{LF}.

The work has been supported in part by the Russian Foundation for Basic Research
under grant 07-02-00102a.

\newpage
$T_{lab}=100 MeV$, Re part

\begin{tabular}{|l|l|l|l|l|}
\hline
R & VSE & VTO & VLSO & VTNO\\ \hline
0.11 & 8352.18 & 4669.53 & 0 & -600000\\
0.15 & -680.002 & -2523.94 & 0 & 23340.9\\
0.25 & 383.456 & 142.706 & 0 & -200\\
0.4 & -753.348 & -895.424 & 0 & 288.248\\
0.55 & 220.39 & 360.761 & 0 & -47.7974\\
0.7 & -50.1382 & -91.3391 & 0 & 5.97377\\
1.4 & -0.966063 & 0.55162 & 0 & -0.00201629\\
\hline
\end{tabular}

\vspace{2cm}
$T_{lab}=100 MeV$, Im part

\begin{tabular}{|l|l|l|l|l|}
\hline
R & VSE & VTO & VLSO & VTNO\\ \hline
0.11 & -6519.78 & -6037.04 & 17149.7 & -1.17288e+06\\
0.15 & 7112.78 & 2053.56 & -39.1544 & 857602\\
0.25 & -7831.77 & -5679.67 & 228.483 & -55084.2\\
0.4 & 1901.84 & 962.641 & 1076.17 & 2887.14\\
0.55 & -1.10827 & 21.4285 & -79.9082 & -310.513\\
0.7 & 62.9981 & 33.4869 & 2.20799 & 10.0006\\
1.4 & 9.77397 & -3.48397 & -0.173829 & -0.0710843\\
\hline
\end{tabular}

\vspace{2cm}
$T_{lab}=100 MeV$, Re part

\begin{tabular}{|l|l|l|l|l|}
\hline
R & VSO & VTE & VLSE & VTNE\\ \hline
0.11 & 29001.4 & -3280.68 & 0 & 1.27875e+06\\
0.15 & -78611.4 & 43315.1 & 0 & -20000\\
0.25 & 2329.43 & -4559.44 & 0 & 920\\
0.4 & -9198.27 & 704.897 & 0 & -800\\
0.55 & 4694.37 & -1299.19 & 0 & 145.923\\
0.7 & -936.359 & 64.2434 & 0 & -38.7549\\
1.4 & 14.9161 & -6.81892 & 0 & -0.026915\\
\hline
\end{tabular}

\vspace{2cm}
$T_{lab}=100 MeV$, Im part

\begin{tabular}{|l|l|l|l|l|}
\hline
R & VSO & VTE & VLSE & VTNE\\ \hline
0.11 & -2103.44 & 56000 & -40003.8 & -400000\\
0.15 & 220004 & -2500 & 4999.68 & 400\\
0.25 & -15781.1 & 450 & -50.032 & -30000\\
0.4 & 800.06 & -3476.29 & 1370.94 & 4251.08\\
0.55 & -2027.65 & 1400 & -80.0005 & -514.627\\
0.7 & 680 & -220 & 11.1557 & 92.3779\\
1.4 & -40.7887 & 2.67444 & -0.65136 & 0.140121\\
\hline

\end{tabular}

\newpage
$T_{lab}=100 MeV$, pp

\begin{figure}[hbtp]
 \centering
    \resizebox{9.8cm}{!}{\includegraphics{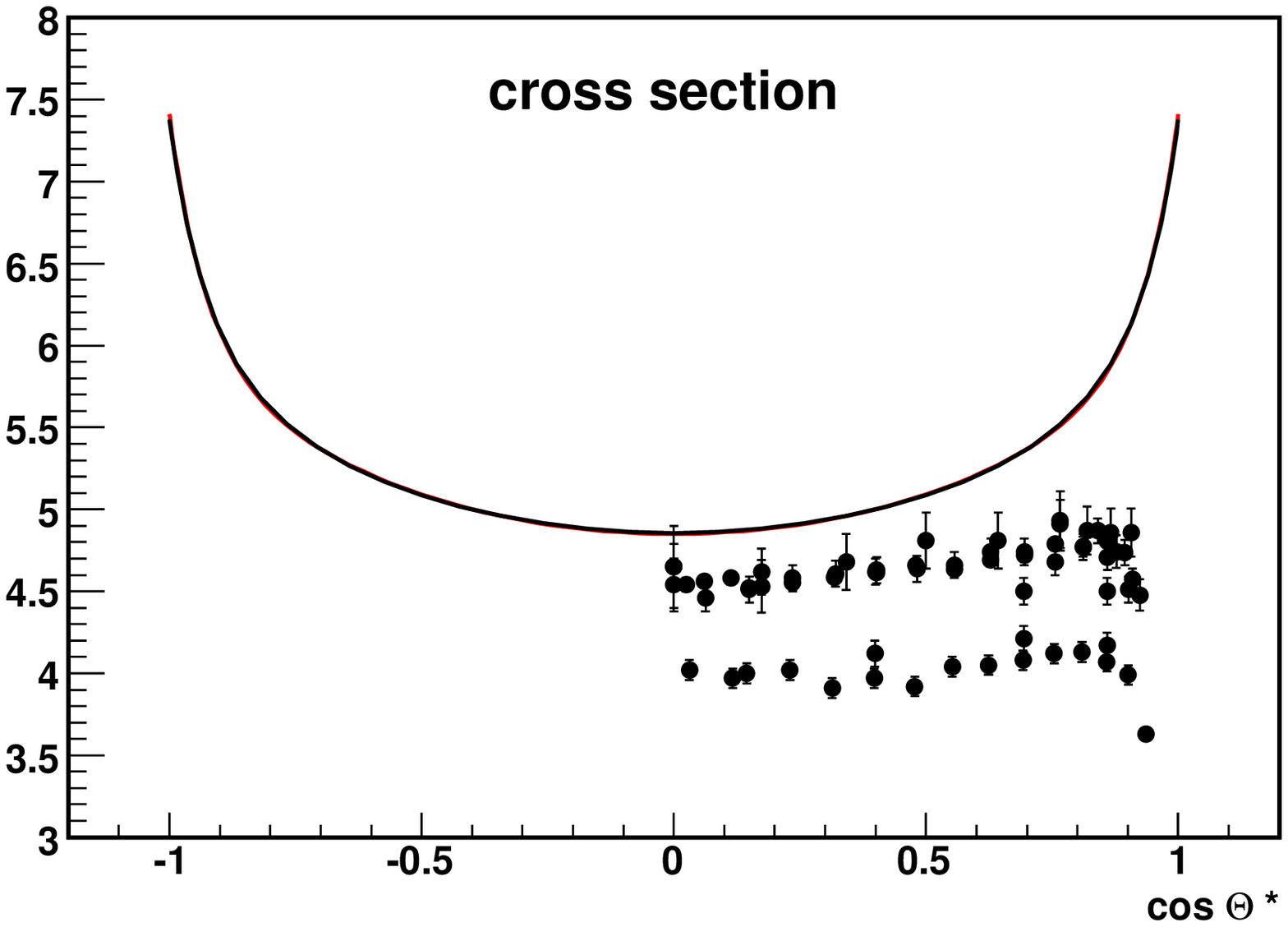}}    
\end{figure}


\begin{figure}[hbtp]
 \centering
    \resizebox{9.8cm}{!}{\includegraphics{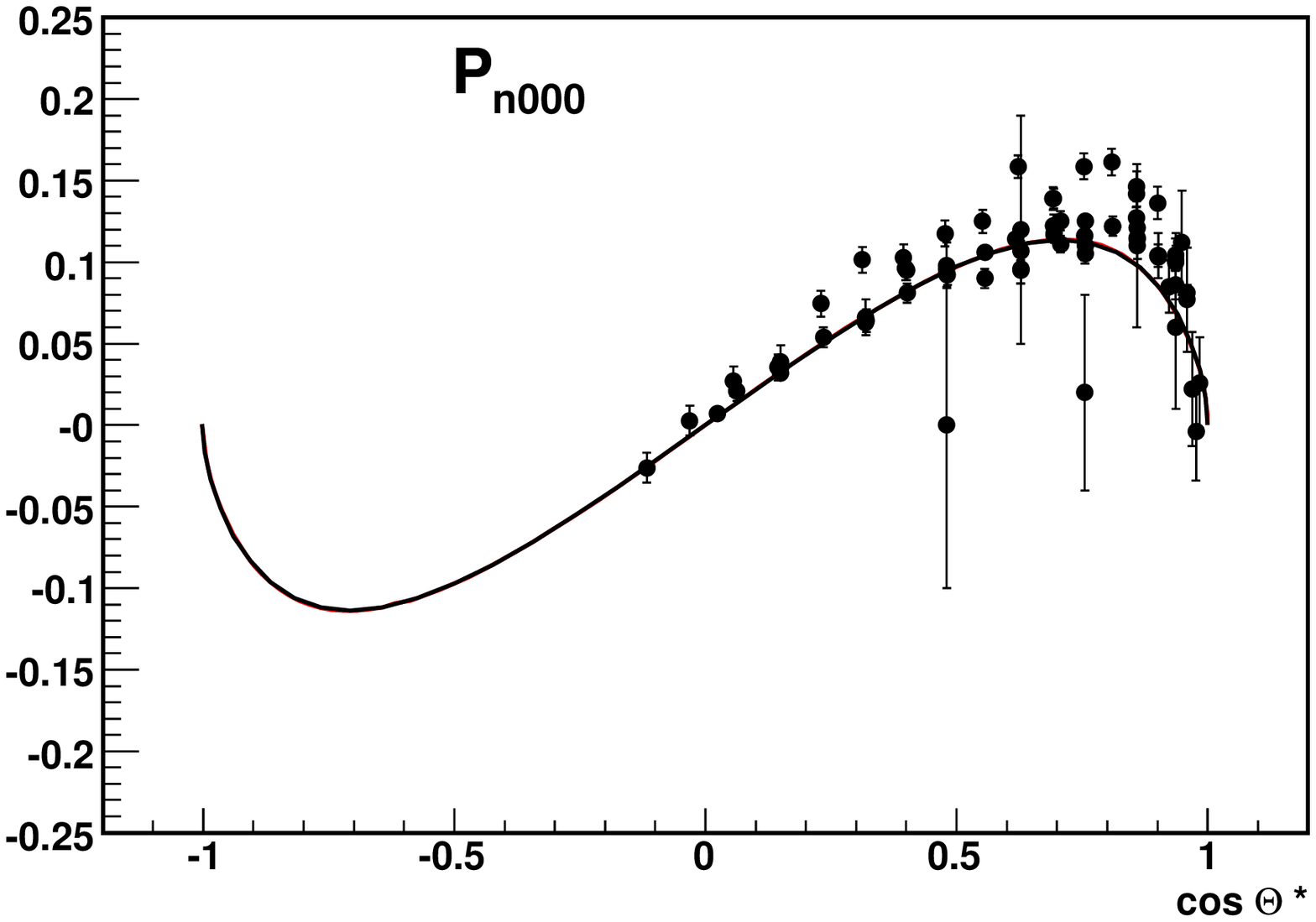}}
\end{figure}


\begin{figure}[hbtp]
 \centering
    \resizebox{9.8cm}{!}{\includegraphics{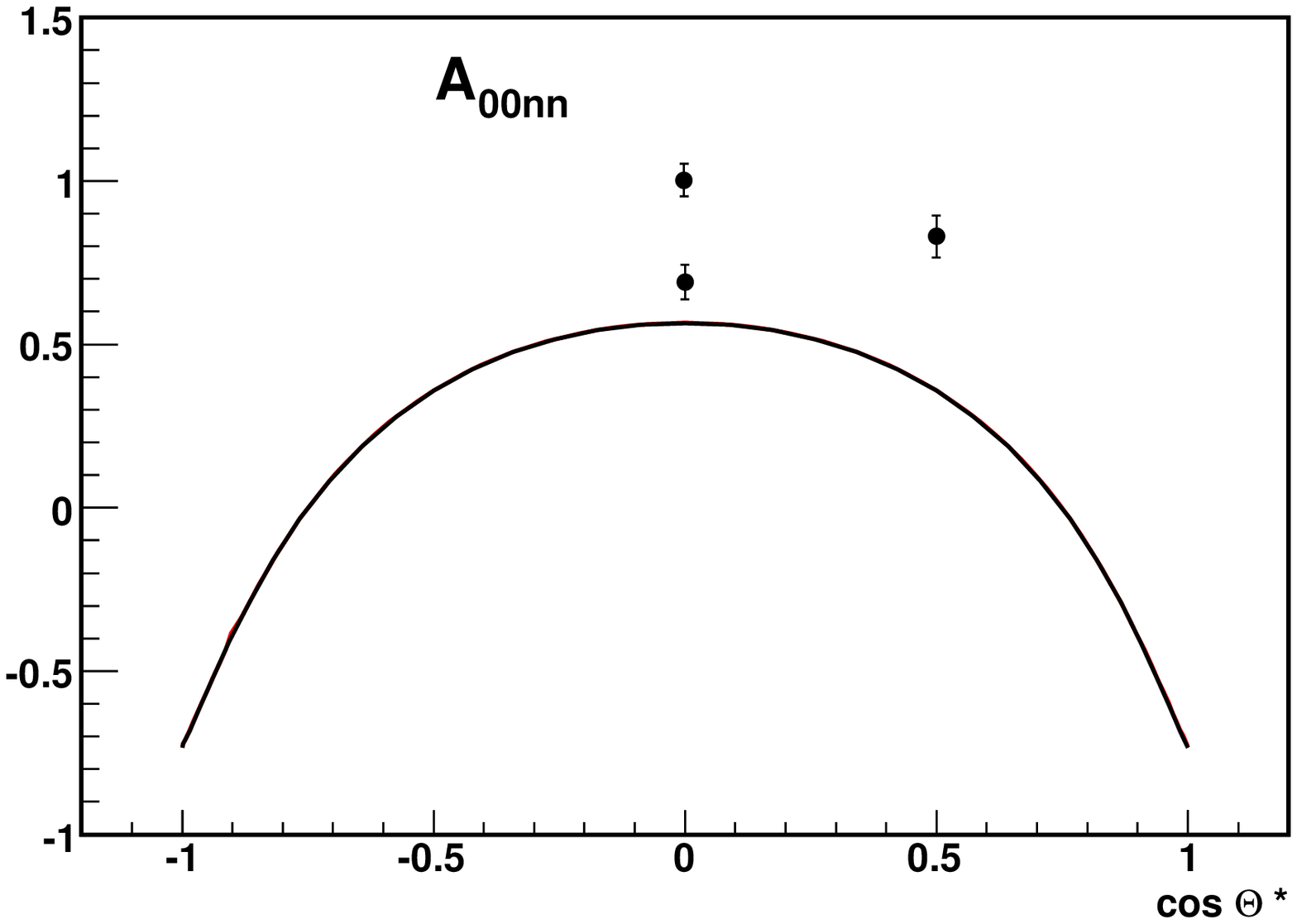}}
\end{figure}

\newpage
$T_{lab}=100 MeV$, np

\begin{figure}[hbtp]
 \centering
    \resizebox{9.8cm}{!}{\includegraphics{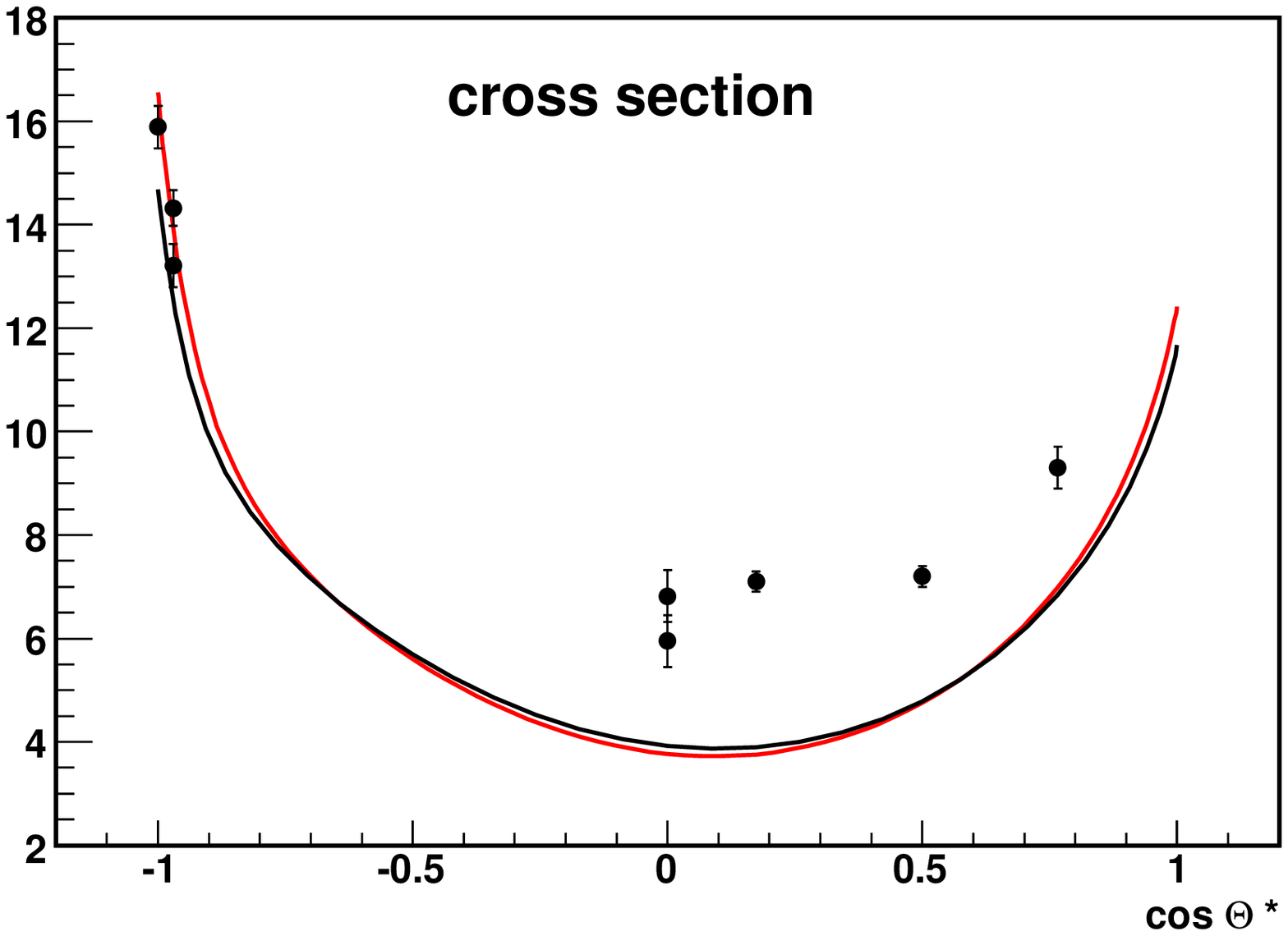}}
\end{figure}


\begin{figure}[hbtp]
 \centering
    \resizebox{9.8cm}{!}{\includegraphics{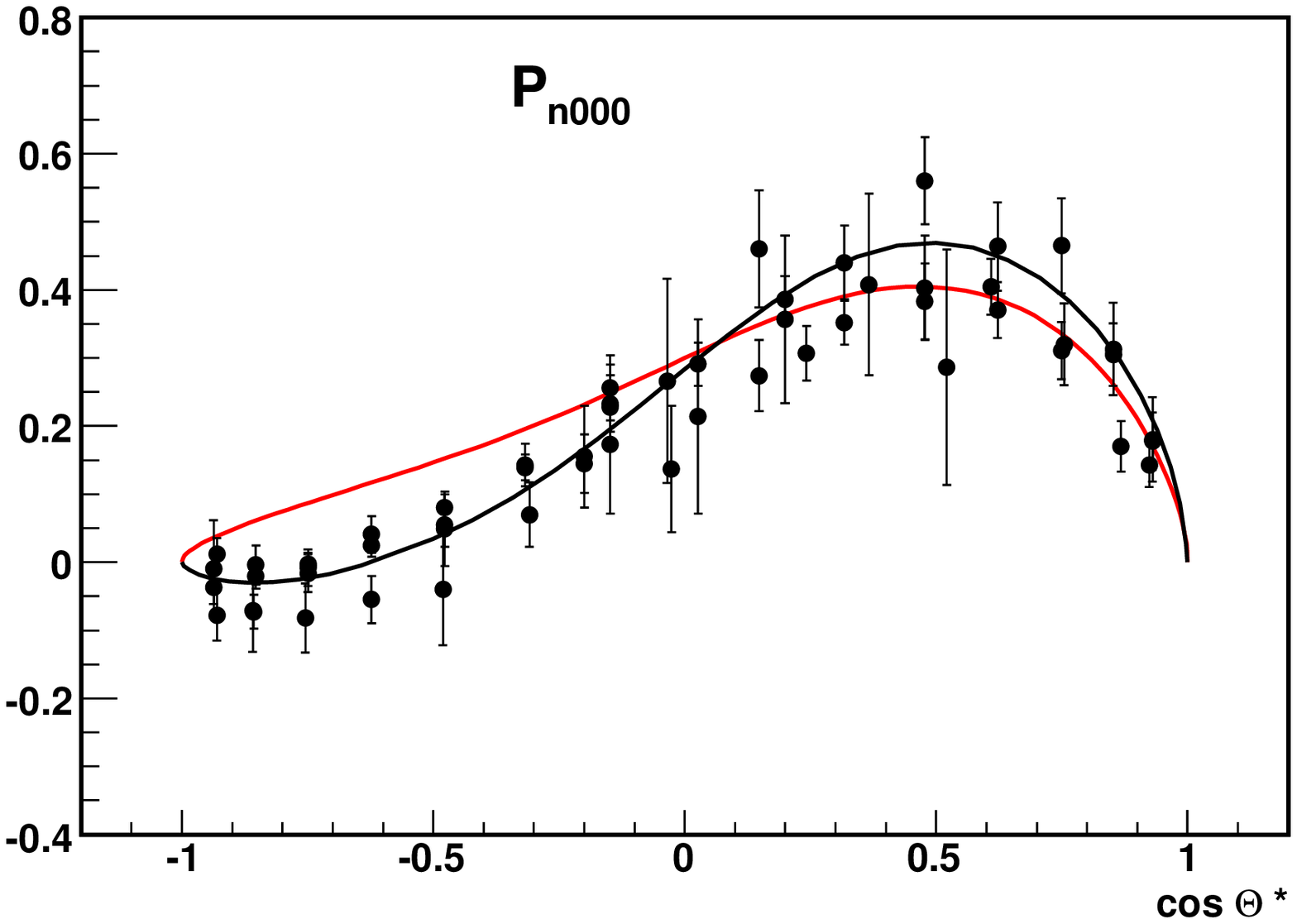}}
\end{figure}


\begin{figure}[hbtp]
 \centering
    \resizebox{9.8cm}{!}{\includegraphics{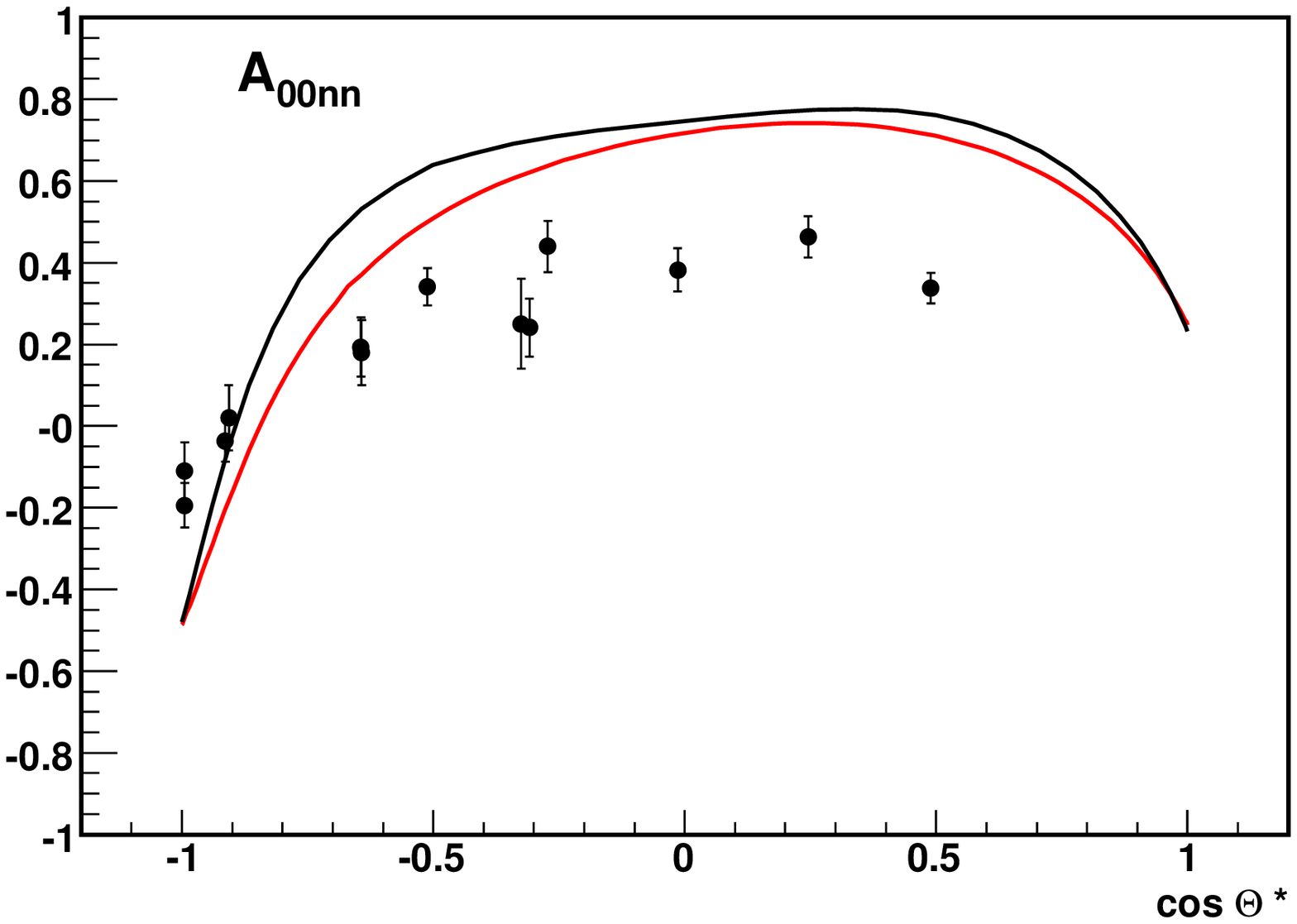}}
\end{figure}

\newpage

$T_{lab}=200 MeV$, Re part

\begin{tabular}{|l|l|l|l|l|}
\hline
R & VSE & VTO & VLSO & VTNO\\ \hline
0.11 & 21605.3 & 5564.15 & 0 & -1e+06\\
0.15 & -1247.82 & -5599.74 & 0 & 124038\\
0.25 & 333.265 & 149.951 & 0 & -800\\
0.4 & -1299.38 & -908.962 & 0 & 153.371\\
0.55 & 434.175 & 359.347 & 0 & -35.7073\\
0.7 & -91.1403 & -82.007 & 0 & 5.46177\\
1.4 & 0.00100003 & 0.498954 & 0 & -0.00872548\\
\hline
\end{tabular}

\vspace{2cm}
$T_{lab}=200 MeV$, Im part

\begin{tabular}{|l|l|l|l|l|}
\hline
R & VSE & VTO & VLSO & VTNO\\ \hline
0.11 & -8624.53 & -8271.96 & 18202.5 & -816552\\
0.15 & 7230.76 & 5172.77 & -87.9493 & 300032\\
0.25 & -8081.44 & -6332.04 & 230.637 & -20160.7\\
0.4 & 2166.57 & 1008.42 & 1149.09 & 1677.25\\
0.55 & -164.434 & 24.2187 & -151.771 & -290.396\\
0.7 & 131.628 & 20.3244 & 7.41174 & 19.102\\
1.4 & 7.79064 & -3.39373 & 0.0362834 & -0.111059\\
\hline
\end{tabular}

\vspace{2cm}
$T_{lab}=200 MeV$, Re part

\begin{tabular}{|l|l|l|l|l|}
\hline
R & VSO & VTE & VLSE & VTNE\\ \hline
0.11 & 25999.9 & -5359.08 & 0 & 850000\\
0.15 & -41962.5 & 27454.7 & 0 & -40000\\
0.25 & 2185.95 & -3999.81 & 0 & 972.163\\
0.4 & -5768.09 & 700.001 & 0 & -800\\
0.55 & 3625.01 & -799.977 & 0 & 117.309\\
0.7 & -950 & 37.101 & 0 & -26.6337\\
1.4 & 23.5844 & -7.1173 & 0 & -0.0774762\\
\hline
\end{tabular}

\vspace{2cm}
$T_{lab}=200 MeV$, Im part

\begin{tabular}{|l|l|l|l|l|}
\hline
R & VSO & VTE & VLSE & VTNE\\ \hline
0.11 & -4000.02 & 35929.1 & -60000.4 & -200000\\
0.15 & 249994 & -4938.93 & 2999.99 & 76436.7\\
0.25 & -20027.1 & 280.001 & -100.002 & -16000\\
0.4 & 1226.74 & -2999.9 & 1099.95 & 2810.16\\
0.55 & -2389.81 & 1370.36 & -159.609 & -351.581\\
0.7 & 704.295 & -150 & 22.2239 & 68\\
1.4 & -33.4615 & 2.34164 & -0.450722 & 0.320879\\
\hline
\end{tabular}

\newpage
$T_{lab}=200 MeV$, pp

\begin{figure}[hbtp]
 \centering
    \resizebox{9.8cm}{!}{\includegraphics{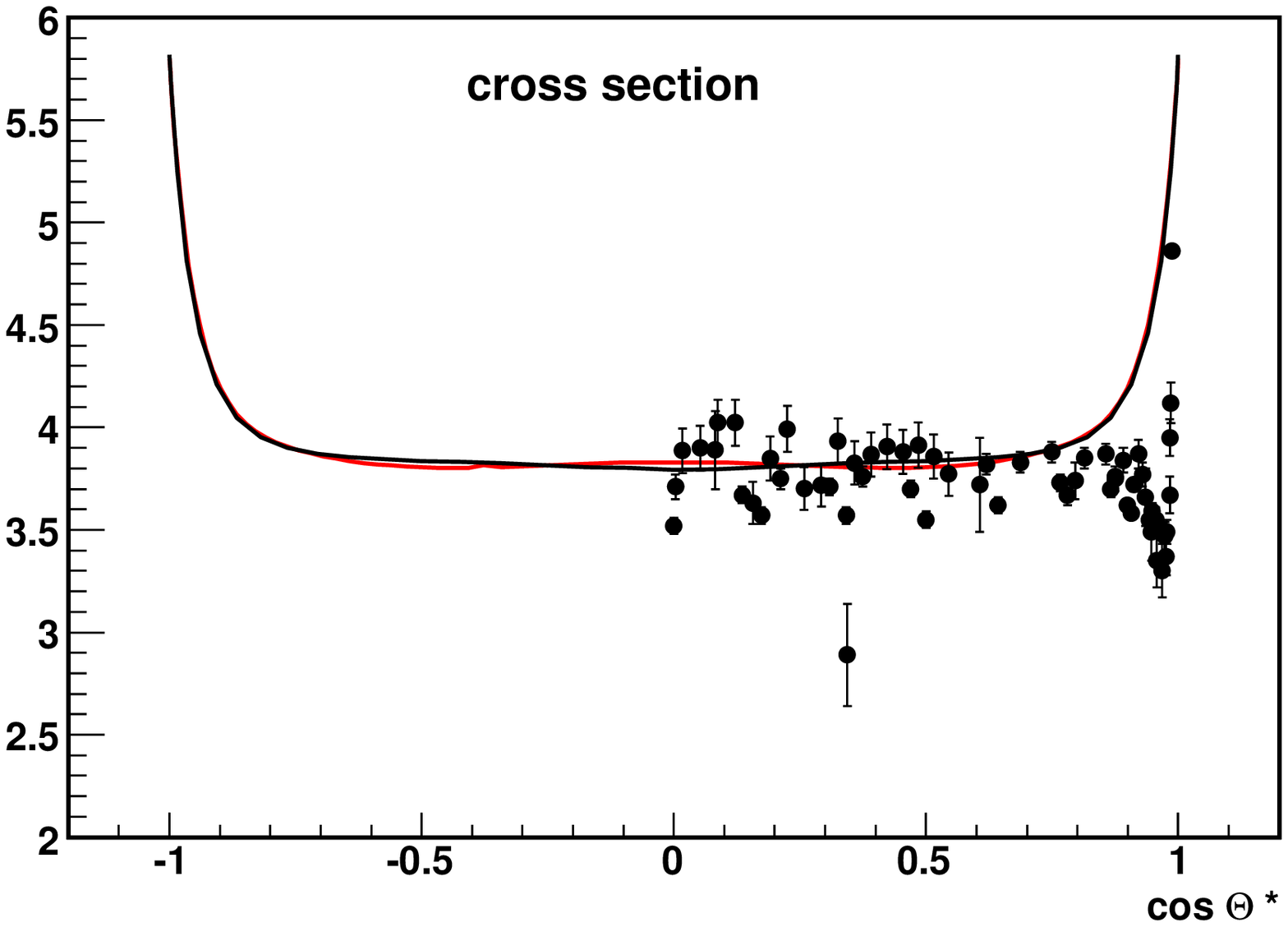}}
\end{figure}


\begin{figure}[hbtp]
 \centering
    \resizebox{9.8cm}{!}{\includegraphics{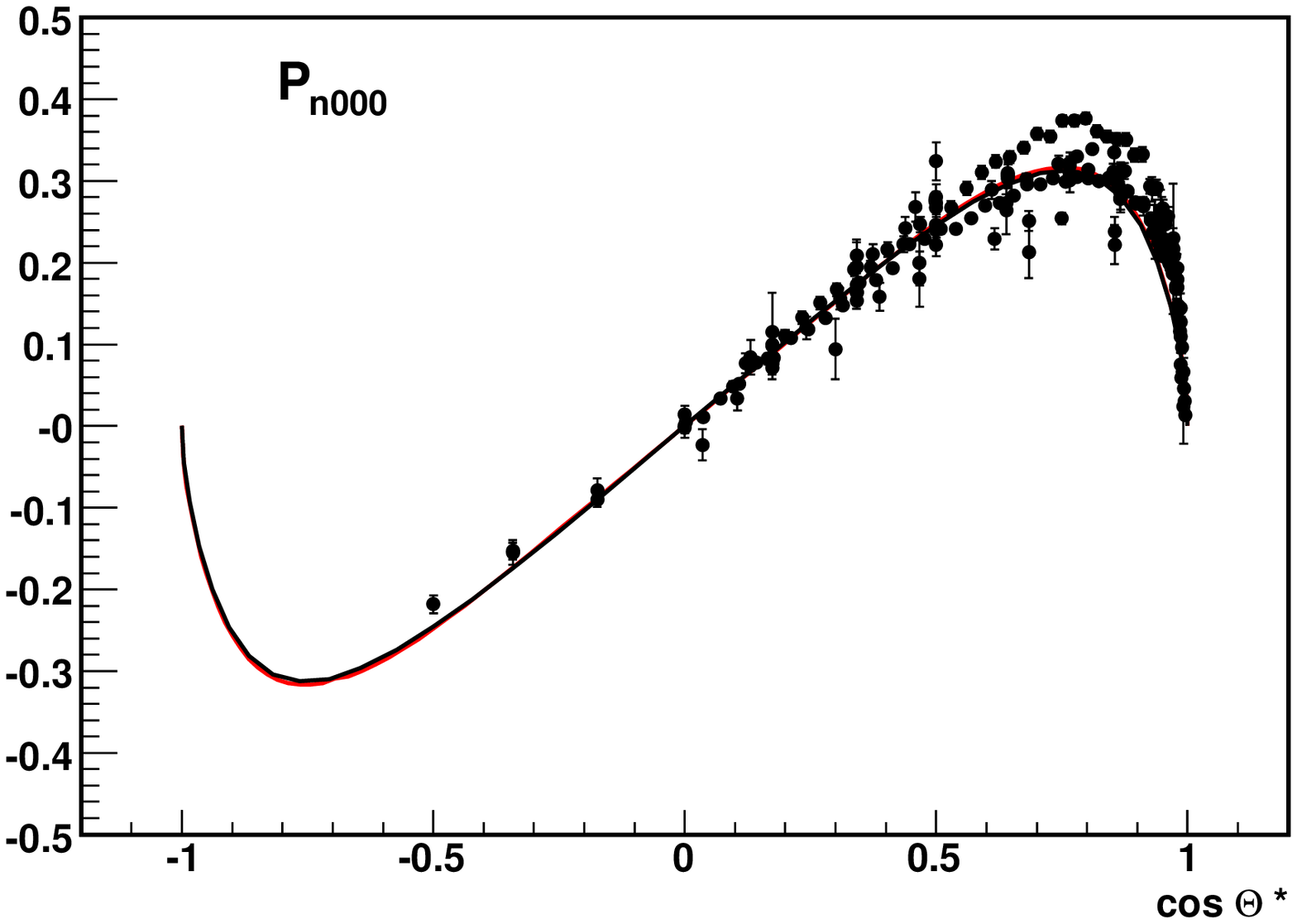}}
\end{figure}


\begin{figure}[hbtp]
 \centering
    \resizebox{9.8cm}{!}{\includegraphics{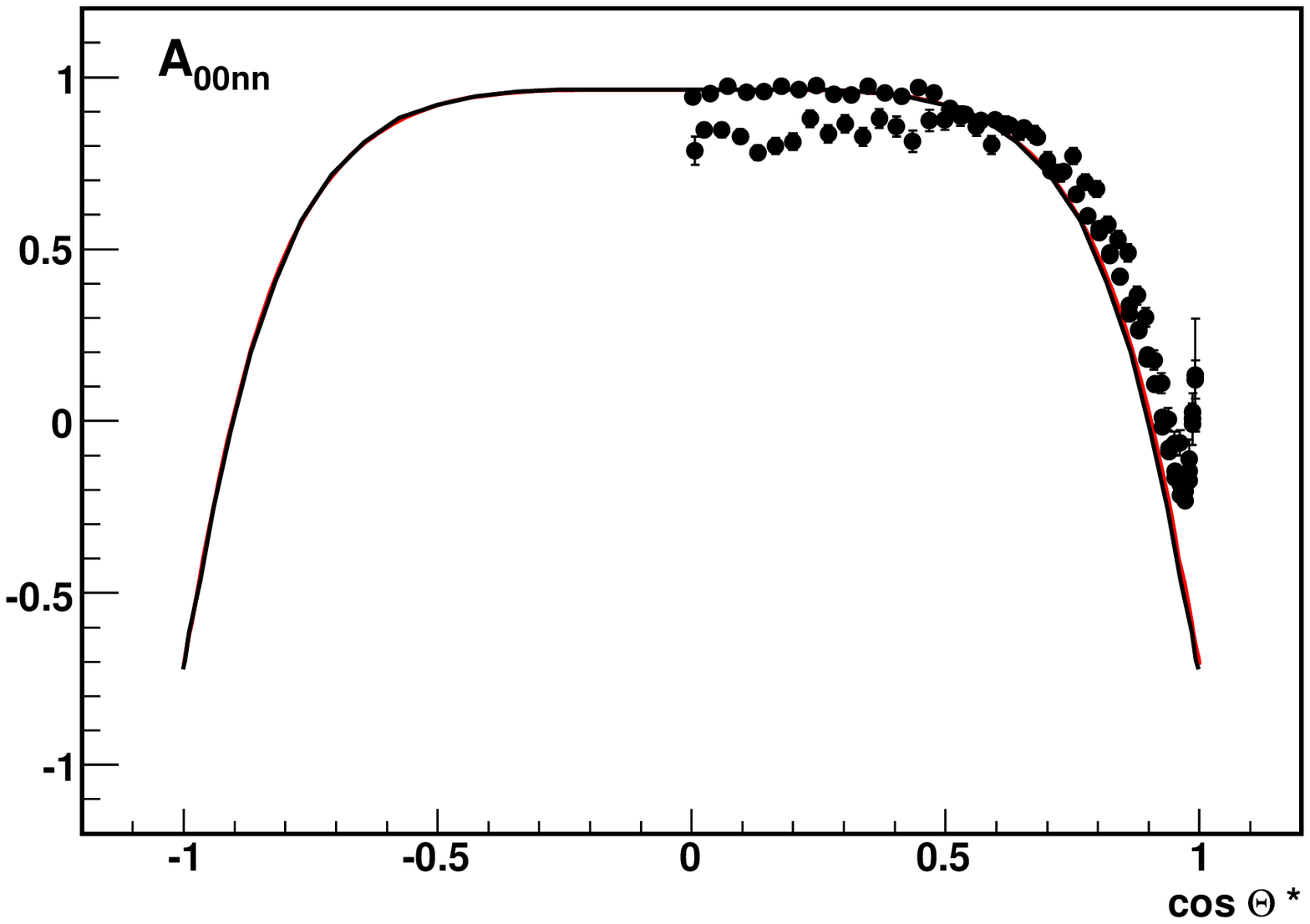}}
\end{figure}

\newpage
$T_{lab}=200 MeV$, np

\begin{figure}[hbtp]
 \centering
    \resizebox{9.8cm}{!}{\includegraphics{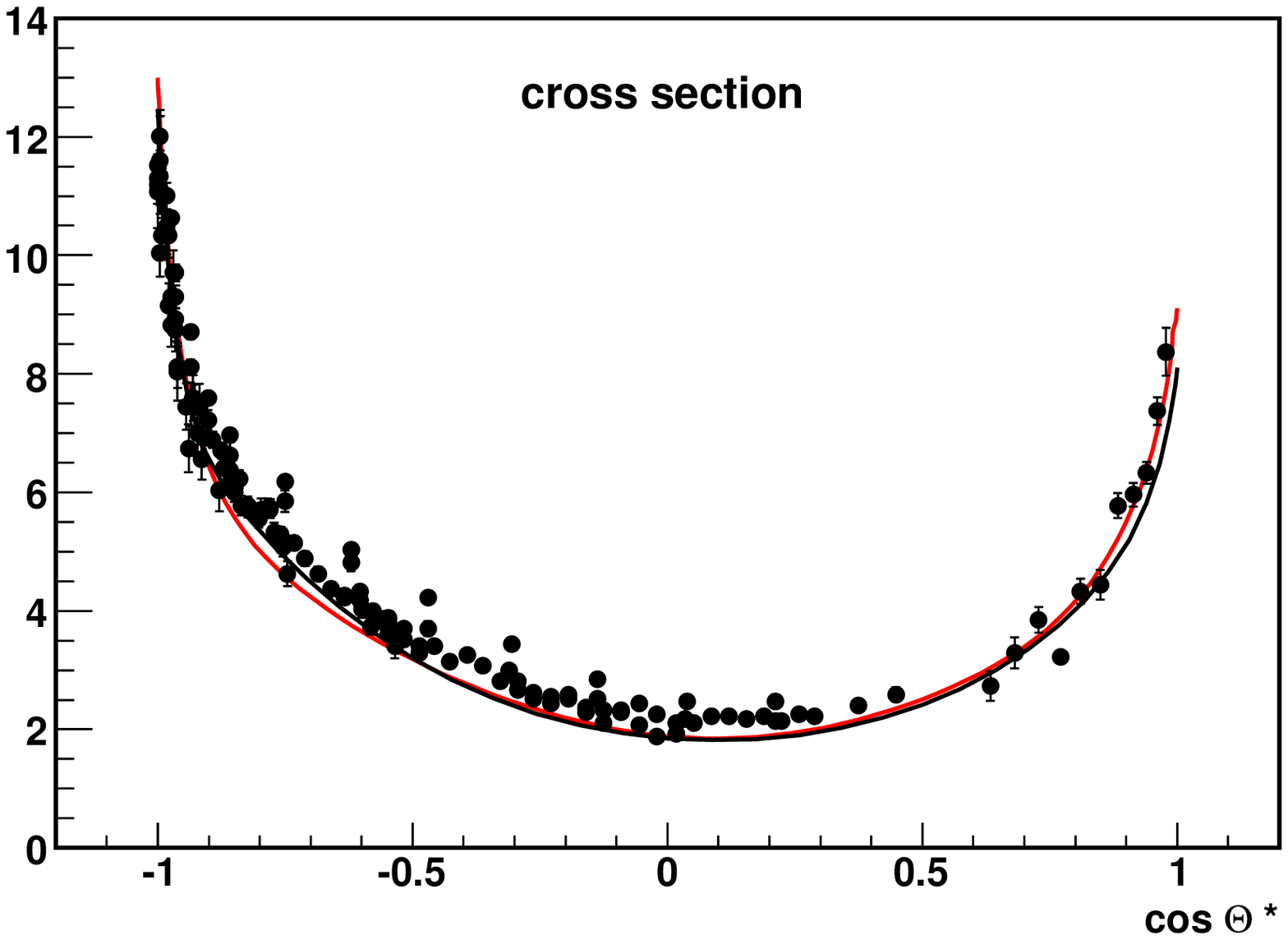}}
\end{figure}


\begin{figure}[hbtp]
 \centering
    \resizebox{9.8cm}{!}{\includegraphics{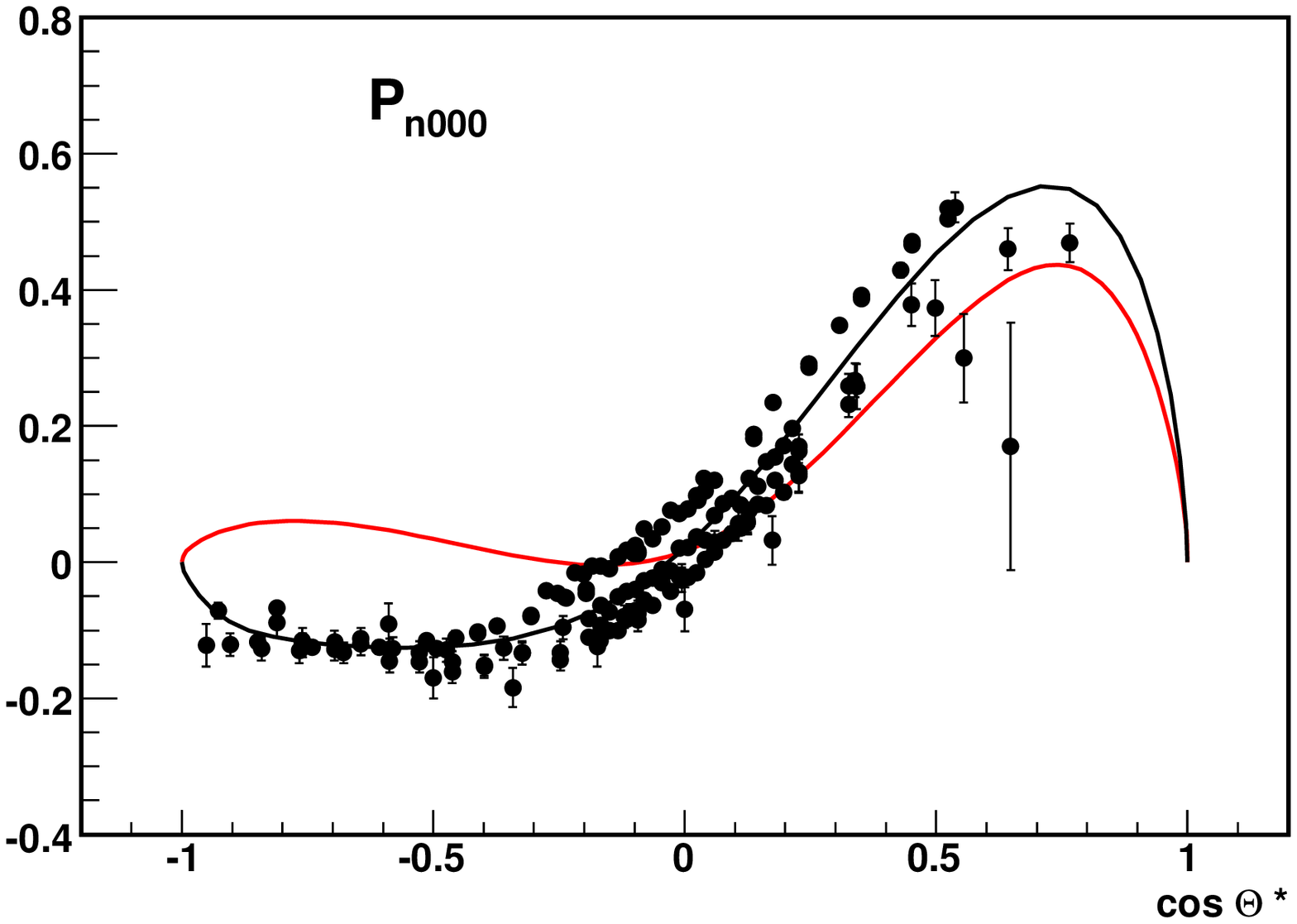}}
\end{figure}


\begin{figure}[hbtp]
 \centering
    \resizebox{9.8cm}{!}{\includegraphics{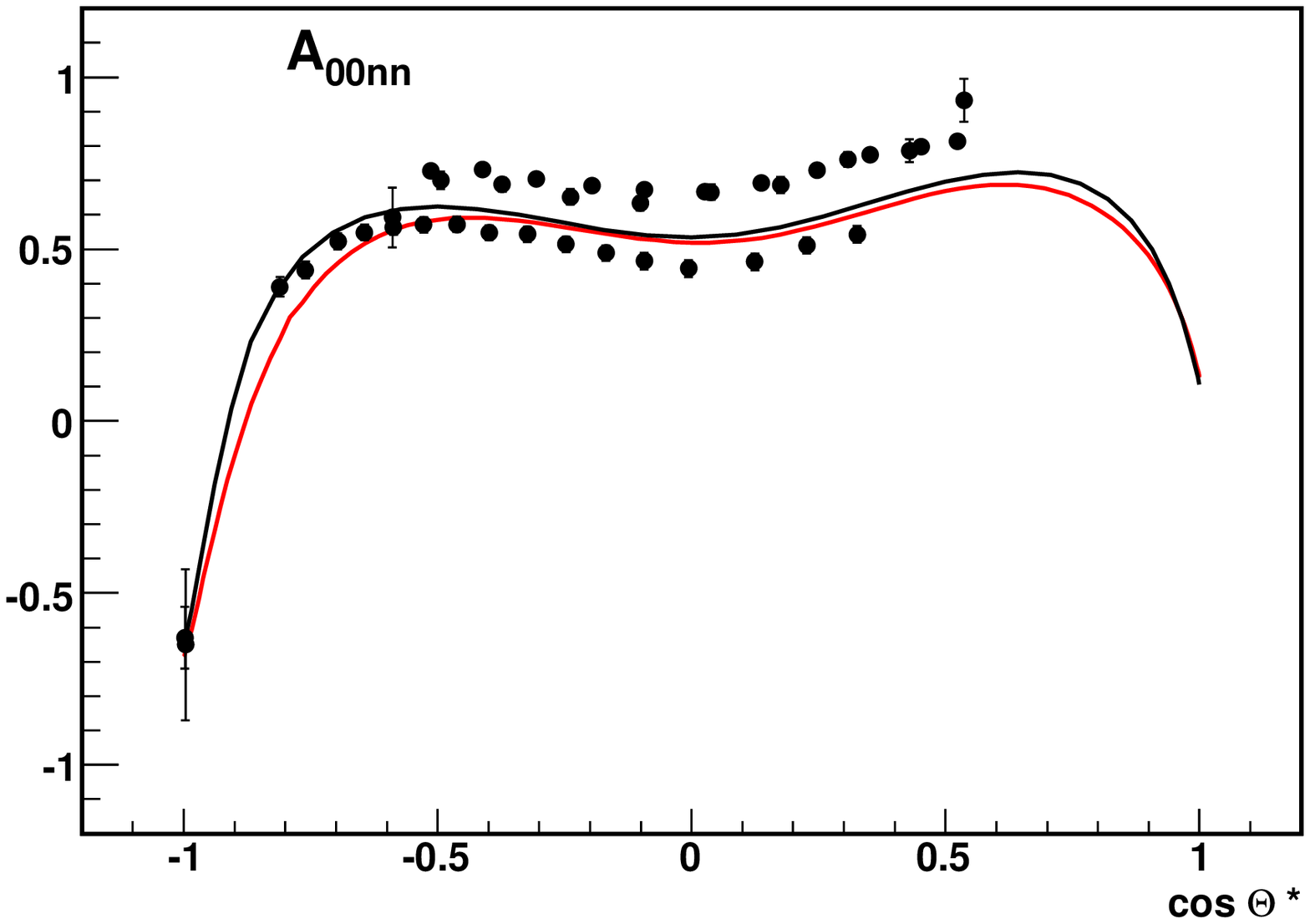}}
\end{figure}

\newpage

$T_{lab}=300 MeV$, Re part

\begin{tabular}{|l|l|l|l|l|}
\hline
R & VSE & VTO & VLSO & VTNO\\ \hline
0.11 & 26672.7 & 6019.47 & 0 & -1.18252e+06\\
0.15 & -440.936 & -7940.31 & 0 & 209184\\
0.25 & 338.321 & 462.706 & 0 & -4417.48\\
0.4 & -1955.99 & -910.192 & 0 & 330.769\\
0.55 & 662.155 & 357.713 & 0 & -64.9937\\
0.7 & -124.504 & -81.137 & 0 & 10.3199\\
1.4 & -0.211034 & 0.654291 & 0 & -0.0221523\\
\hline
\end{tabular}

\vspace{2cm}
$T_{lab}=300 MeV$, Im part

\begin{tabular}{|l|l|l|l|l|}
\hline
R & VSE & VTO & VLSO & VTNO\\ \hline
0.11 & -8700.55 & -8520.21 & 30105 & -315784\\
0.15 & 8994.24 & 6830.08 & -559.126 & 95000\\
0.25 & -8691.5 & -6492.94 & 259.858 & -7914.95\\
0.4 & 2398.75 & 1018.76 & 963.971 & 972.465\\
0.55 & -248.855 & 49.6393 & -151.998 & -229.982\\
0.7 & 132.005 & -2.2 & 12.2475 & 17.9979\\
1.4 & 7.21693 & -3.03145 & 0.0281493 & -0.13225\\
\hline
\end{tabular}

\vspace{2cm}
$T_{lab}=300 MeV$, Re part

\begin{tabular}{|l|l|l|l|l|}
\hline
R & VSO & VTE & VLSE & VTNE\\ \hline
0.11 & 15872.8 & -9959.96 & 0 & 700000\\
0.15 & -41011.9 & 16785.5 & 0 & -74674.8\\
0.25 & 1519.37 & -2500.03 & 0 & 1000\\
0.4 & -3005.64 & 699.998 & 0 & -581.599\\
0.55 & 2834.9 & -498.014 & 0 & 95.6785\\
0.7 & -976.34 & 20.0001 & 0 & -25.4931\\
1.4 & 27.6585 & -9.56093 & 0 & -0.0611397\\
\hline
\end{tabular}

\vspace{2cm}
$T_{lab}=300 MeV$, Im part

\begin{tabular}{|l|l|l|l|l|}
\hline
R & VSO & VTE & VLSE & VTNE\\ \hline
0.11 & -6059.03 & 25992.6 & -83997.3 & -100000\\
0.15 & 199774 & -5155.48 & 2001.06 & 115271\\
0.25 & -21601.1 & 400.025 & -200.298 & -14500\\
0.4 & 1505.84 & -2299.71 & 1177.83 & 2146.74\\
0.55 & -2451.54 & 1098.88 & -248.234 & -220\\
0.7 & 829.524 & -155.576 & 39.9428 & 53.6106\\
1.4 & -36.9757 & 2.83388 & -0.474437 & 0.276692\\
\hline
\end{tabular}

\newpage
$T_{lab}=300 MeV$, pp

\begin{figure}[hbtp]
 \centering
    \resizebox{9.8cm}{!}{\includegraphics{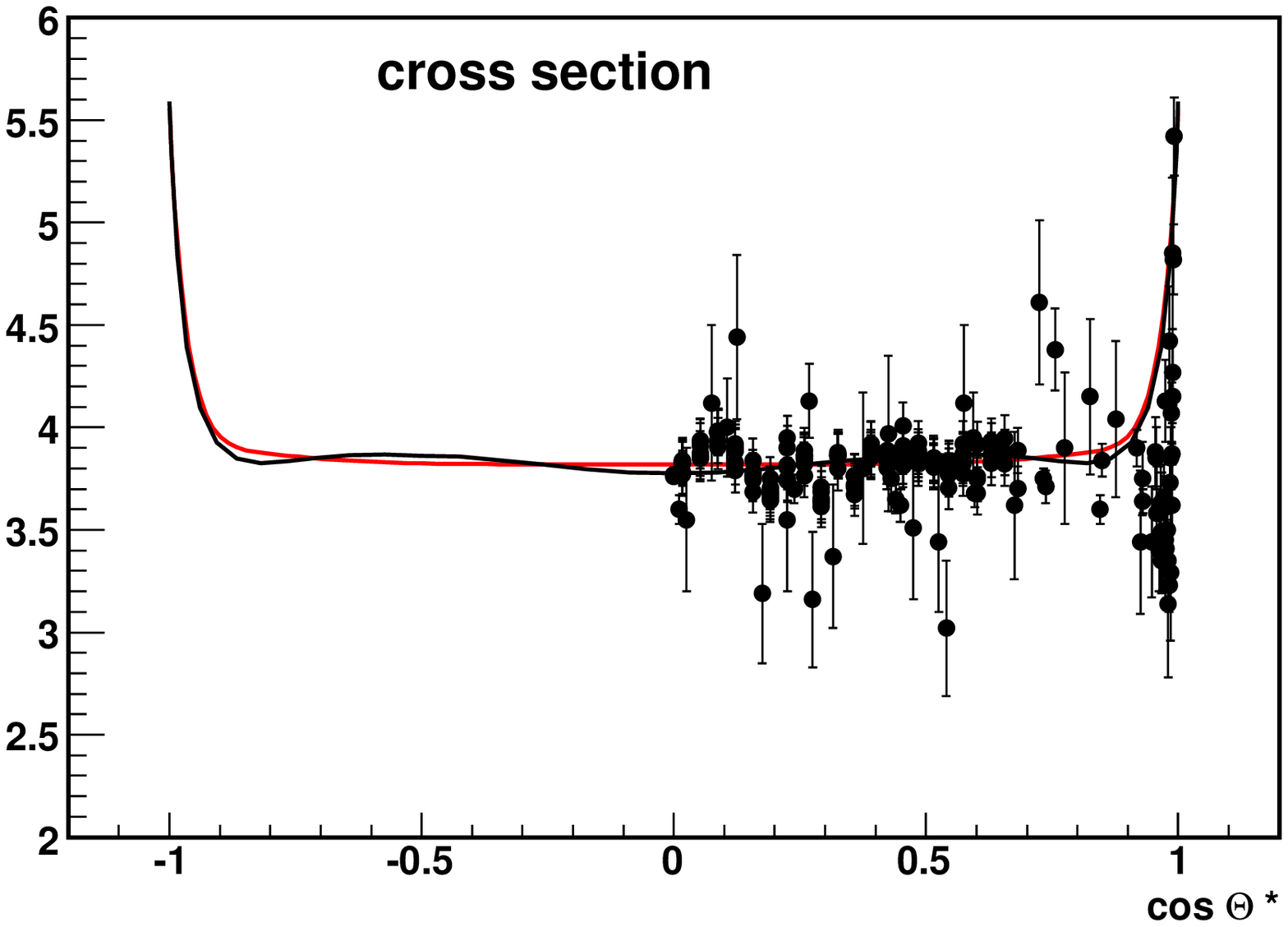}}
\end{figure}


\begin{figure}[hbtp]
 \centering
    \resizebox{9.8cm}{!}{\includegraphics{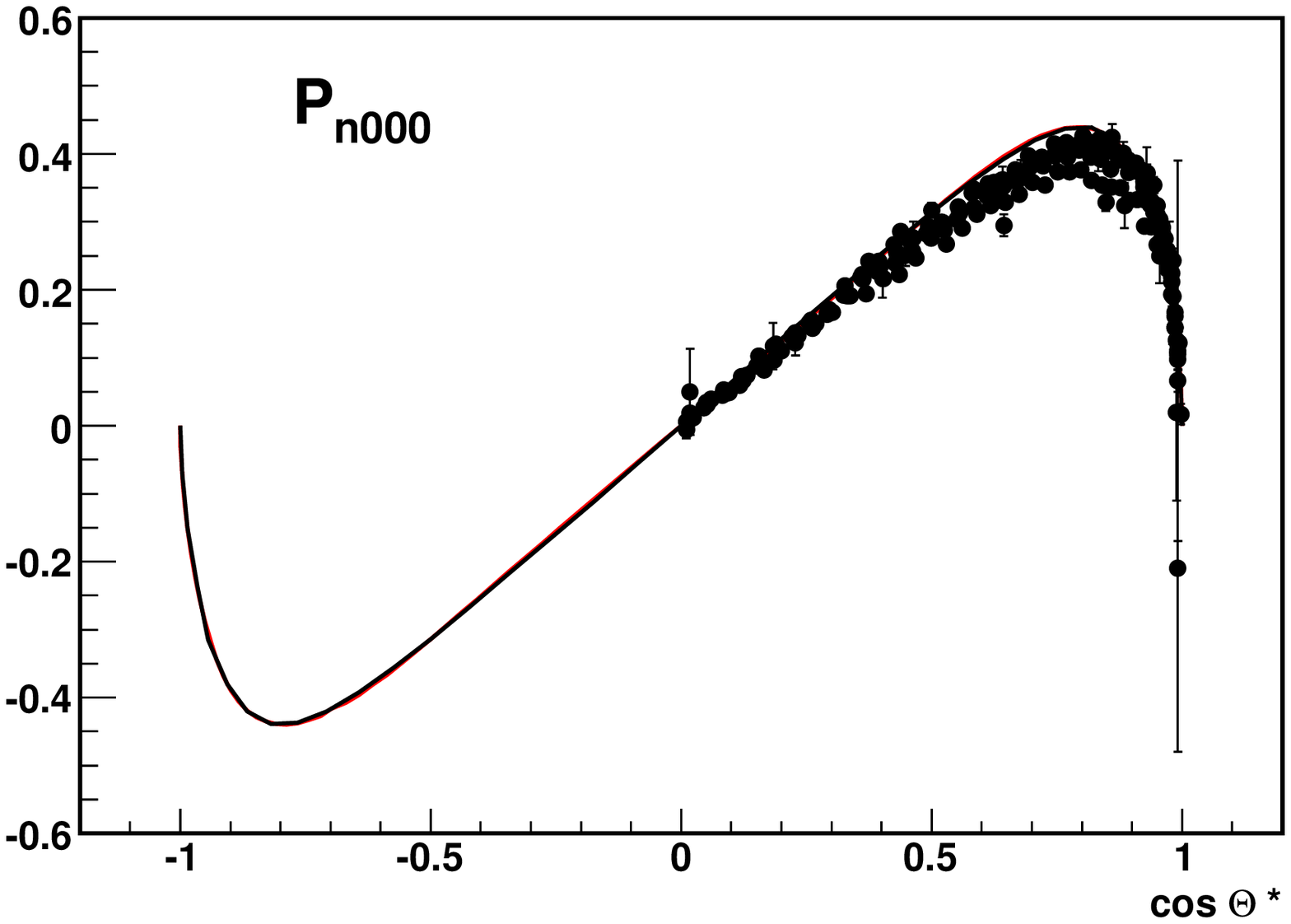}}
\end{figure}


\begin{figure}[hbtp]
 \centering
    \resizebox{9.8cm}{!}{\includegraphics{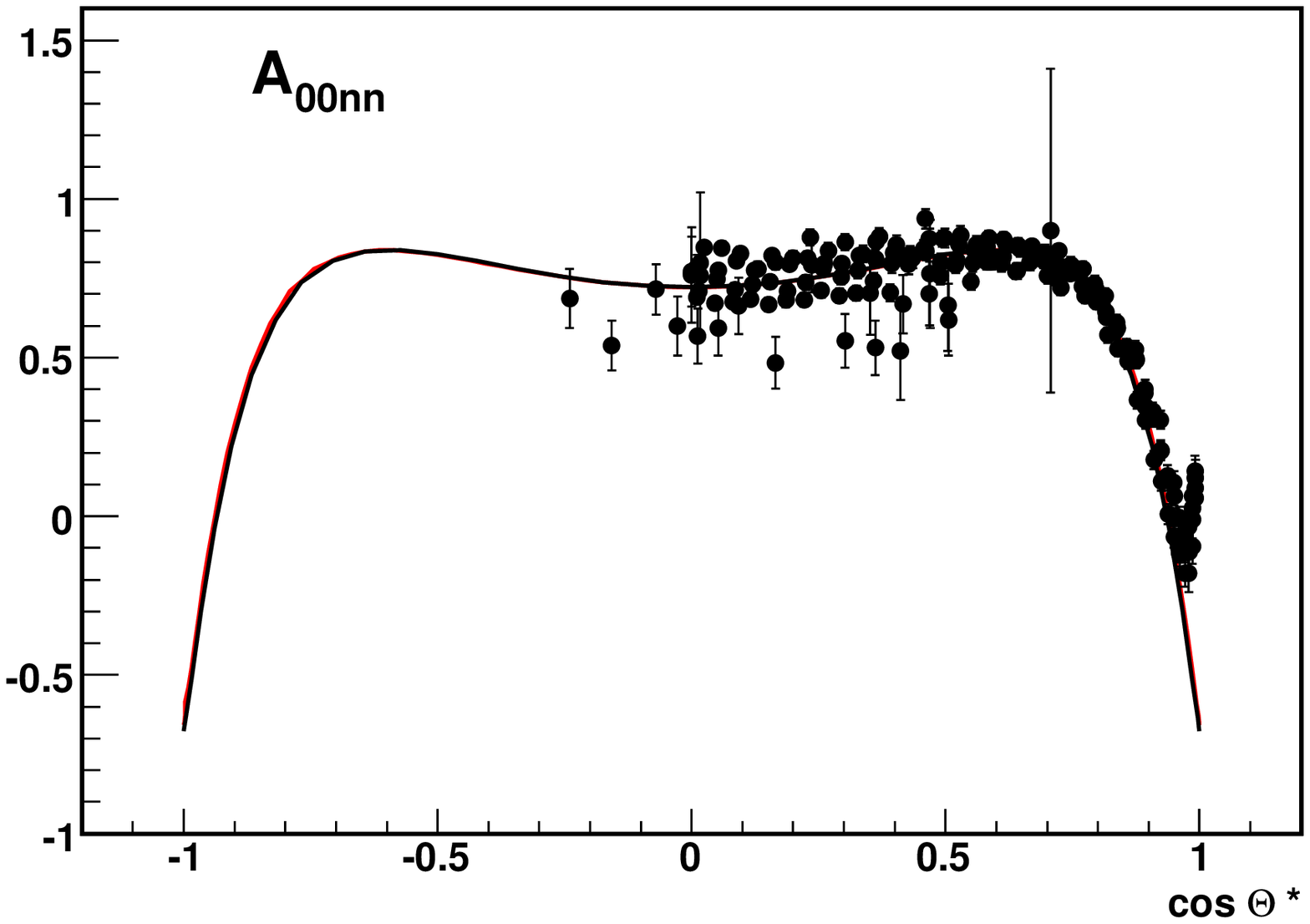}}
\end{figure}

\newpage
$T_{lab}=300 MeV$, np

\begin{figure}[hbtp]
 \centering
    \resizebox{9.8cm}{!}{\includegraphics{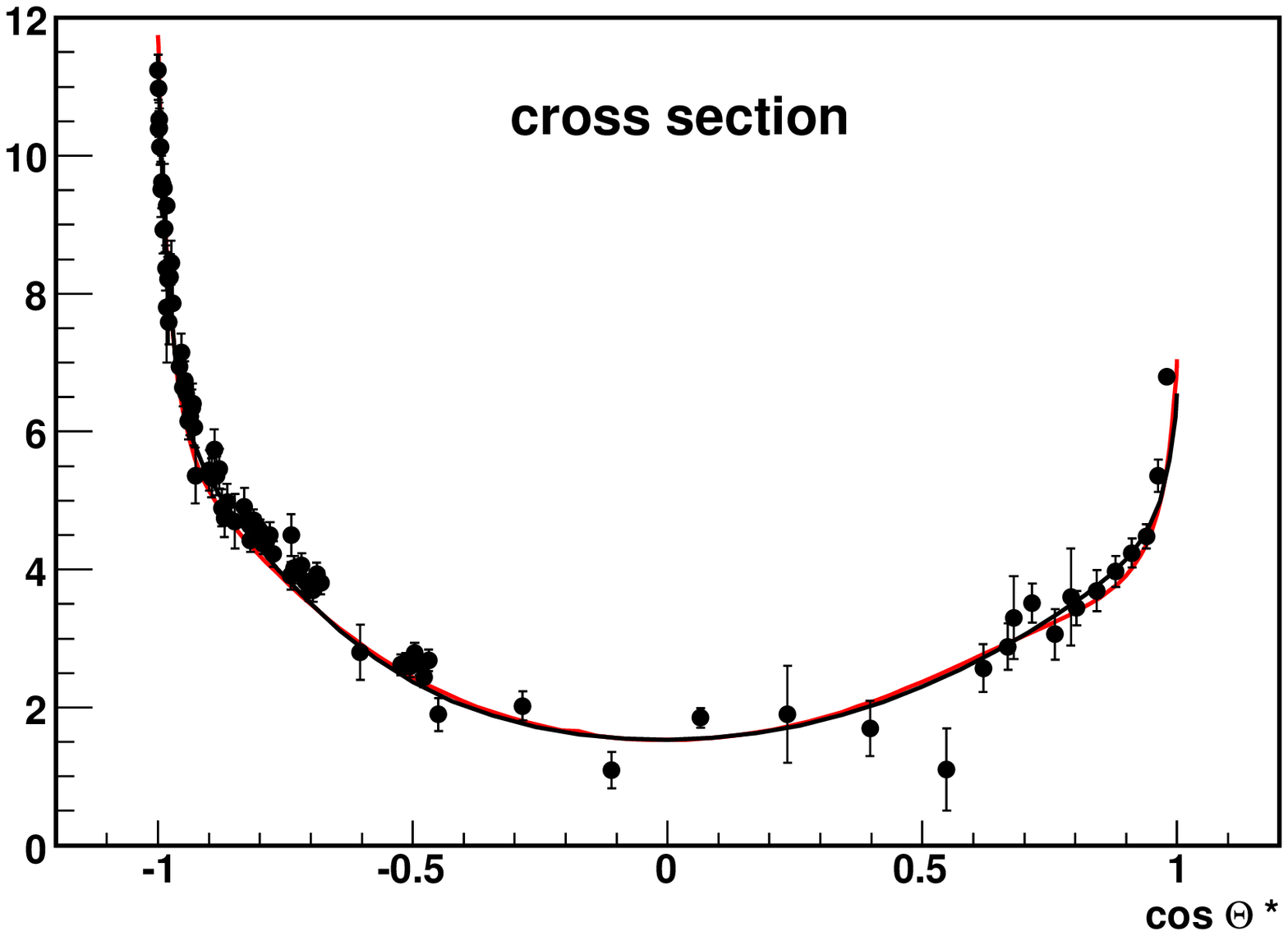}}
\end{figure}


\begin{figure}[hbtp]
 \centering
    \resizebox{9.8cm}{!}{\includegraphics{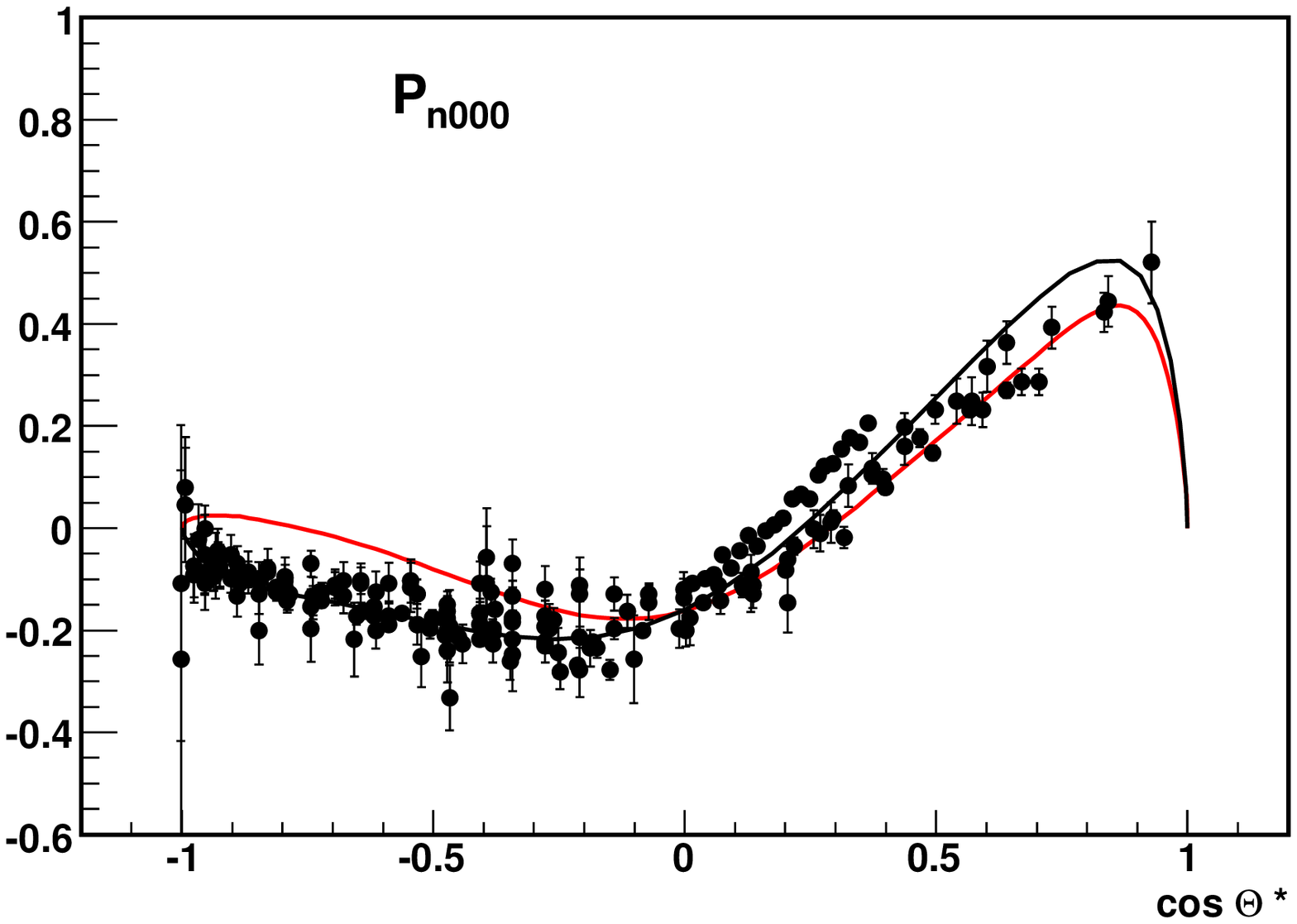}}
\end{figure}


\begin{figure}[hbtp]
 \centering
    \resizebox{9.8cm}{!}{\includegraphics{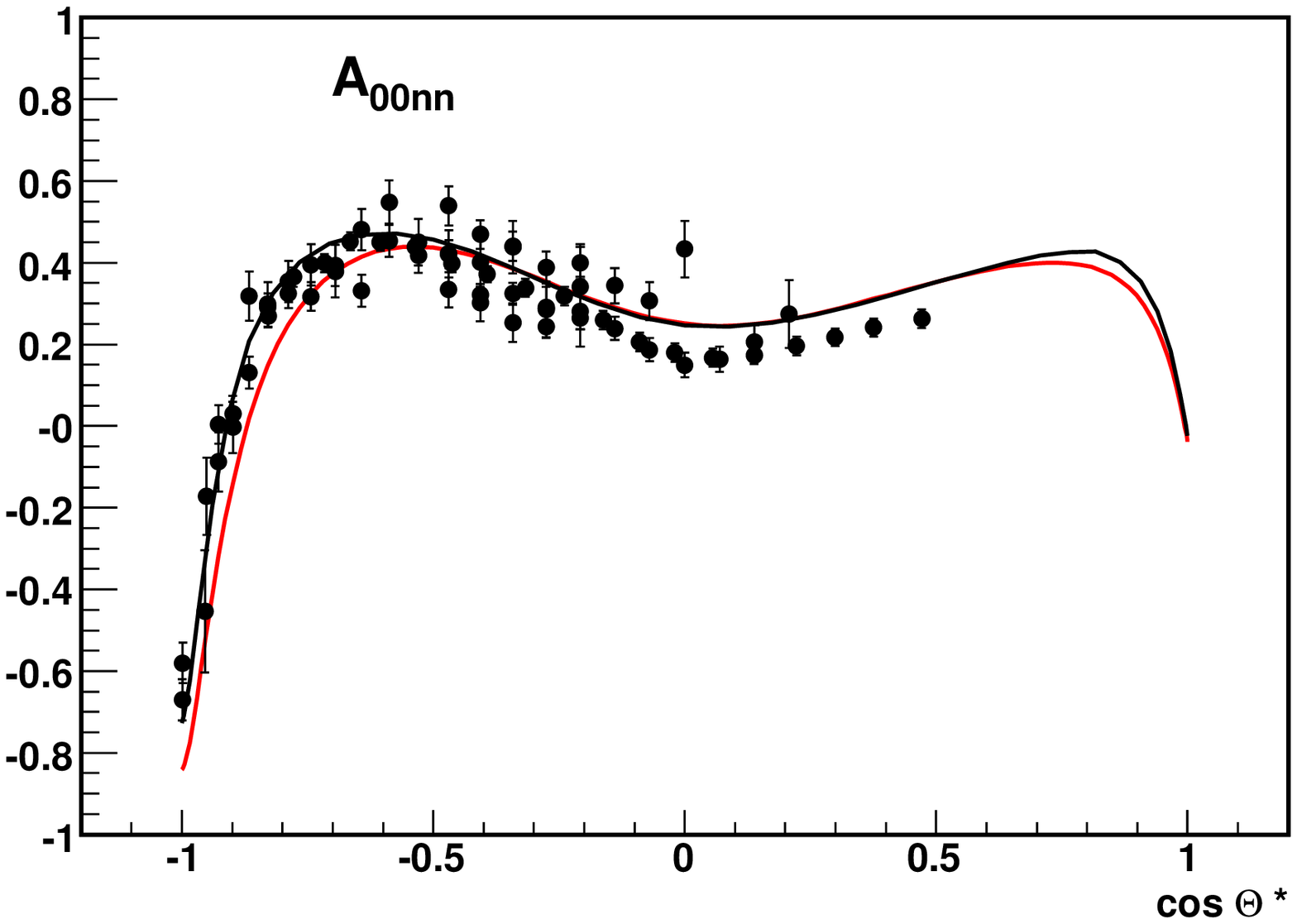}}
\end{figure}

\newpage

$T_{lab}=400 MeV$, Re part

\begin{tabular}{|l|l|l|l|l|}
\hline
R & VSE & VTO & VLSO & VTNO\\ \hline
0.11 & 35128.7 & 8969.36 & 0 & -1.4e+06\\
0.15 & -573.388 & -7987.51 & 0 & 305229\\
0.25 & 343.641 & 823.923 & 0 & -9317.37\\
0.4 & -2767.96 & -915.473 & 0 & 529.21\\
0.55 & 929.159 & 300.001 & 0 & -65.1609\\
0.7 & -164.348 & -65.5461 & 0 & 7.57547\\
1.4 & -0.657018 & 0.697277 & 0 & 0.00274612\\
\hline
\end{tabular}

\vspace{2cm}
$T_{lab}=400 MeV$, Im part

\begin{tabular}{|l|l|l|l|l|}
\hline
R & VSE & VTO & VLSO & VTNO\\ \hline
0.11 & -14196.8 & -9755.82 & 34347 & -174487\\
0.15 & 30704.2 & 7612.04 & -1027.46 & 30005.5\\
0.25 & -16339.6 & -6599.66 & 259.974 & -3000.55\\
0.4 & 3549.85 & 1022.54 & 839.925 & 599.838\\
0.55 & -428.424 & 49.9533 & -152 & -178.924\\
0.7 & 144.327 & -4.00007 & 16.5247 & 14.6032\\
1.4 & 6.439 & -3.02676 & -0.00171169 & -0.138766\\
\hline
\end{tabular}

\vspace{2cm}
$T_{lab}=400 MeV$, Re part

\begin{tabular}{|l|l|l|l|l|}
\hline
R & VSO & VTE & VLSE & VTNE\\ \hline
0.11 & 10003.5 & -6500.01 & 0 & 700000\\
0.15 & -37996.9 & 19984.3 & 0 & -105666\\
0.25 & 3011.6 & -4313.26 & 0 & 2500\\
0.4 & -1899.16 & 509.9 & 0 & -605.942\\
0.55 & 1997.57 & -400.106 & 0 & 101.344\\
0.7 & -786.388 & 66.1521 & 0 & -26.8545\\
1.4 & 27.8255 & -8.85684 & 0 & -0.0519602\\
\hline
\end{tabular}

\vspace{2cm}
$T_{lab}=400 MeV$, Im part

\begin{tabular}{|l|l|l|l|l|}
\hline
R & VSO & VTE & VLSE & VTNE\\ \hline
0.11 & -9000.74 & 26096.1 & -86161.3 & -140000\\
0.15 & 150000 & -6880.2 & 3602.31 & 147643\\
0.25 & -20695.4 & 800.026 & -500.093 & -14459.2\\
0.4 & 2202.03 & -2394.61 & 1199.98 & 1878.77\\
0.55 & -2500.01 & 1161.22 & -267.554 & -180\\
0.7 & 850.16 & -199.975 & 40.3628 & 52.5685\\
1.4 & -38.5451 & 2.80001 & -0.392178 & 0.178394\\
\hline
\end{tabular}

\newpage
$T_{lab}=400 MeV$, pp

\begin{figure}[hbtp]
 \centering
    \resizebox{9.8cm}{!}{\includegraphics{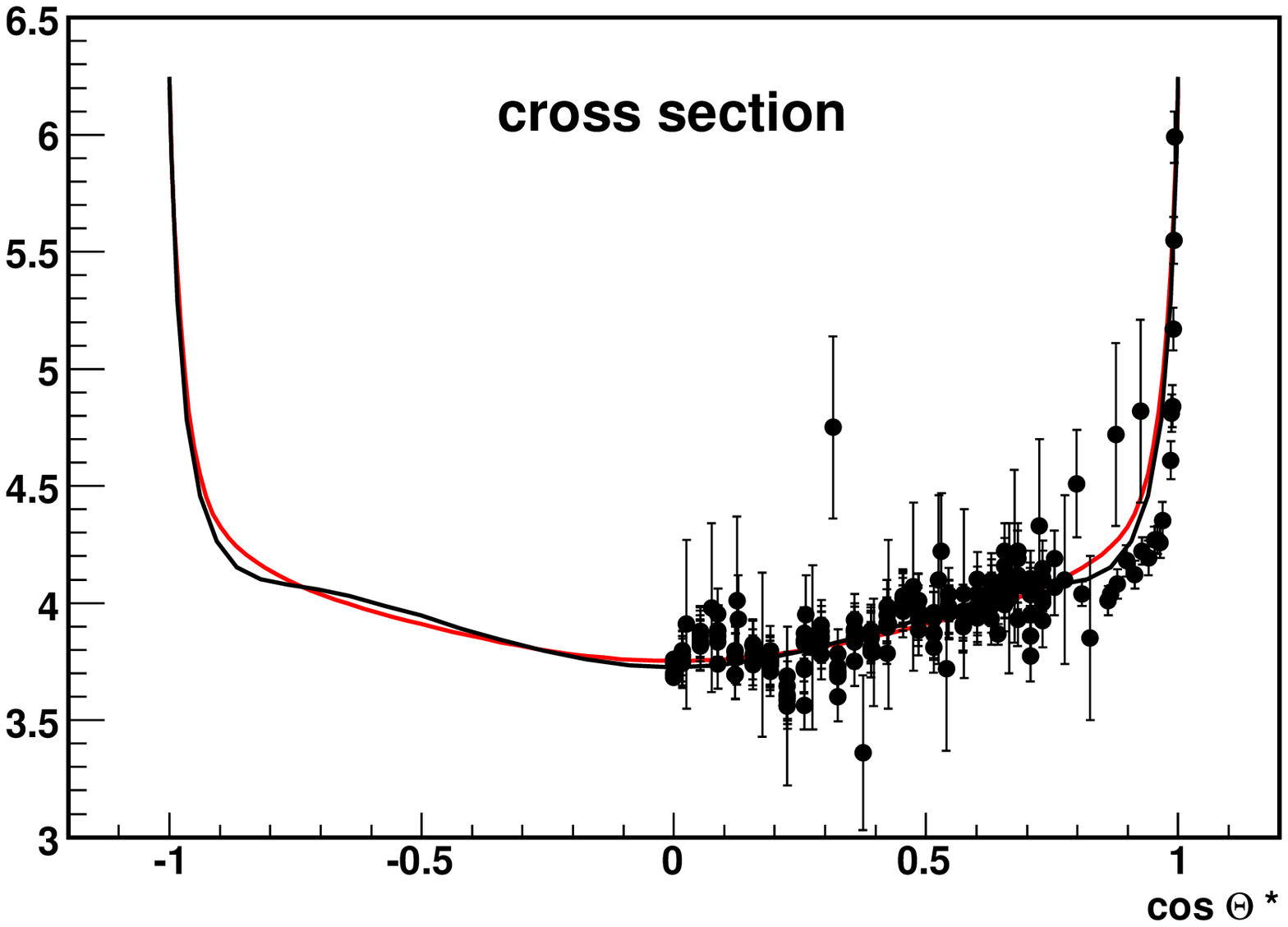}}
\end{figure}


\begin{figure}[hbtp]
 \centering
    \resizebox{9.8cm}{!}{\includegraphics{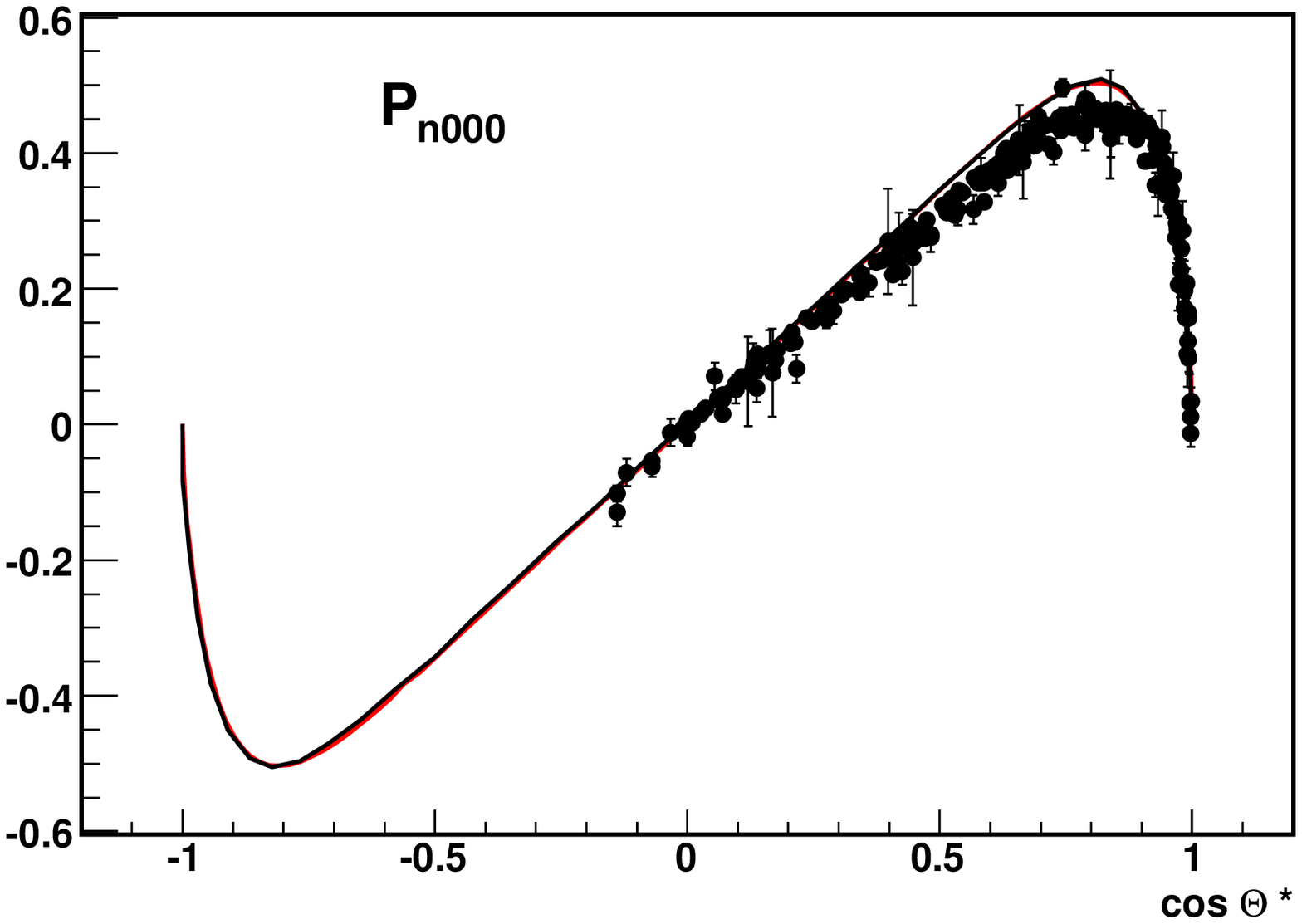}}
\end{figure}


\begin{figure}[hbtp]
 \centering
    \resizebox{9.8cm}{!}{\includegraphics{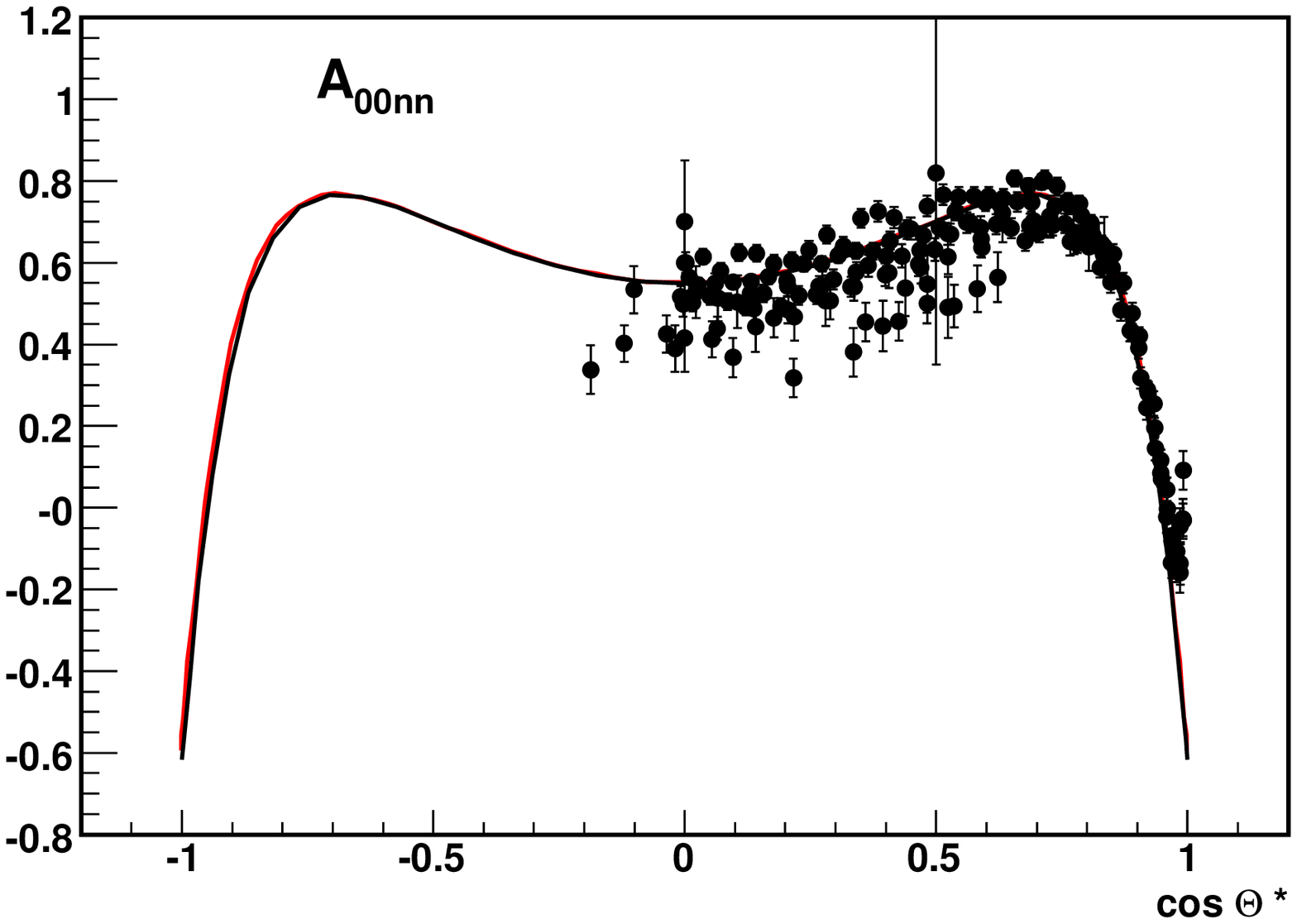}}
\end{figure}

\newpage
$T_{lab}=400 MeV$, np

\begin{figure}[hbtp]
 \centering
    \resizebox{9.8cm}{!}{\includegraphics{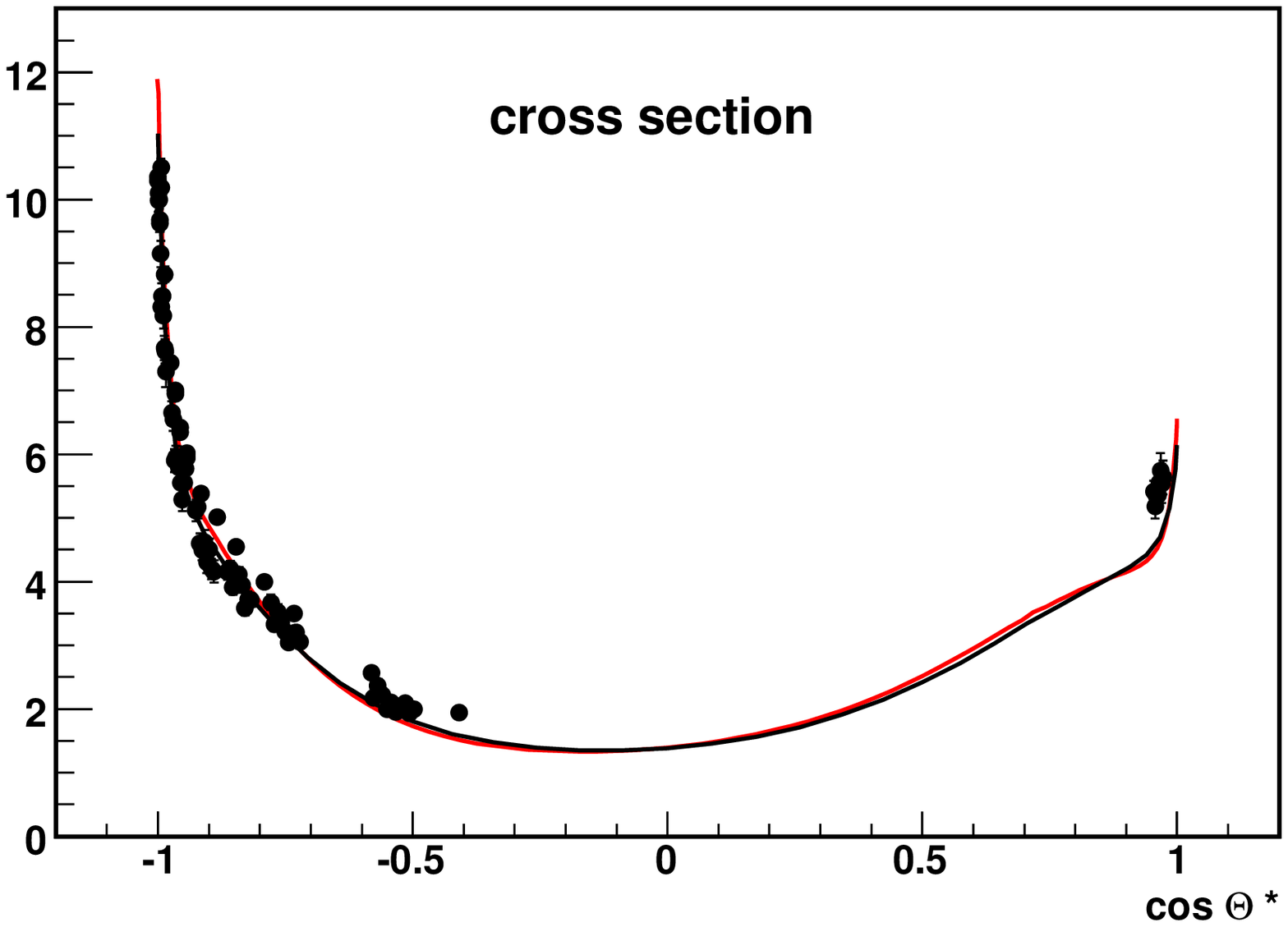}}
\end{figure}


\begin{figure}[hbtp]
 \centering
    \resizebox{9.8cm}{!}{\includegraphics{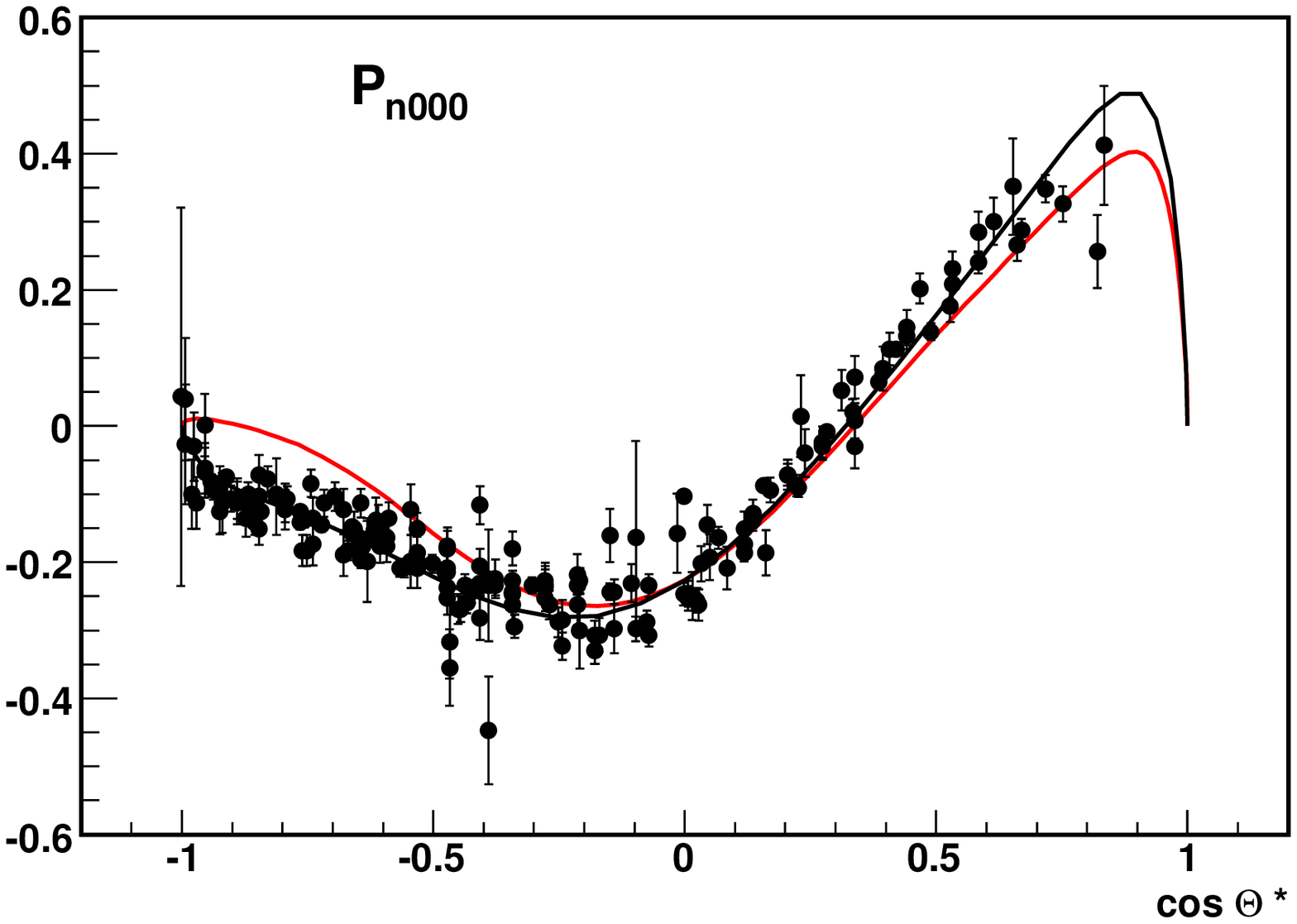}}
\end{figure}


\begin{figure}[hbtp]
 \centering
    \resizebox{9.8cm}{!}{\includegraphics{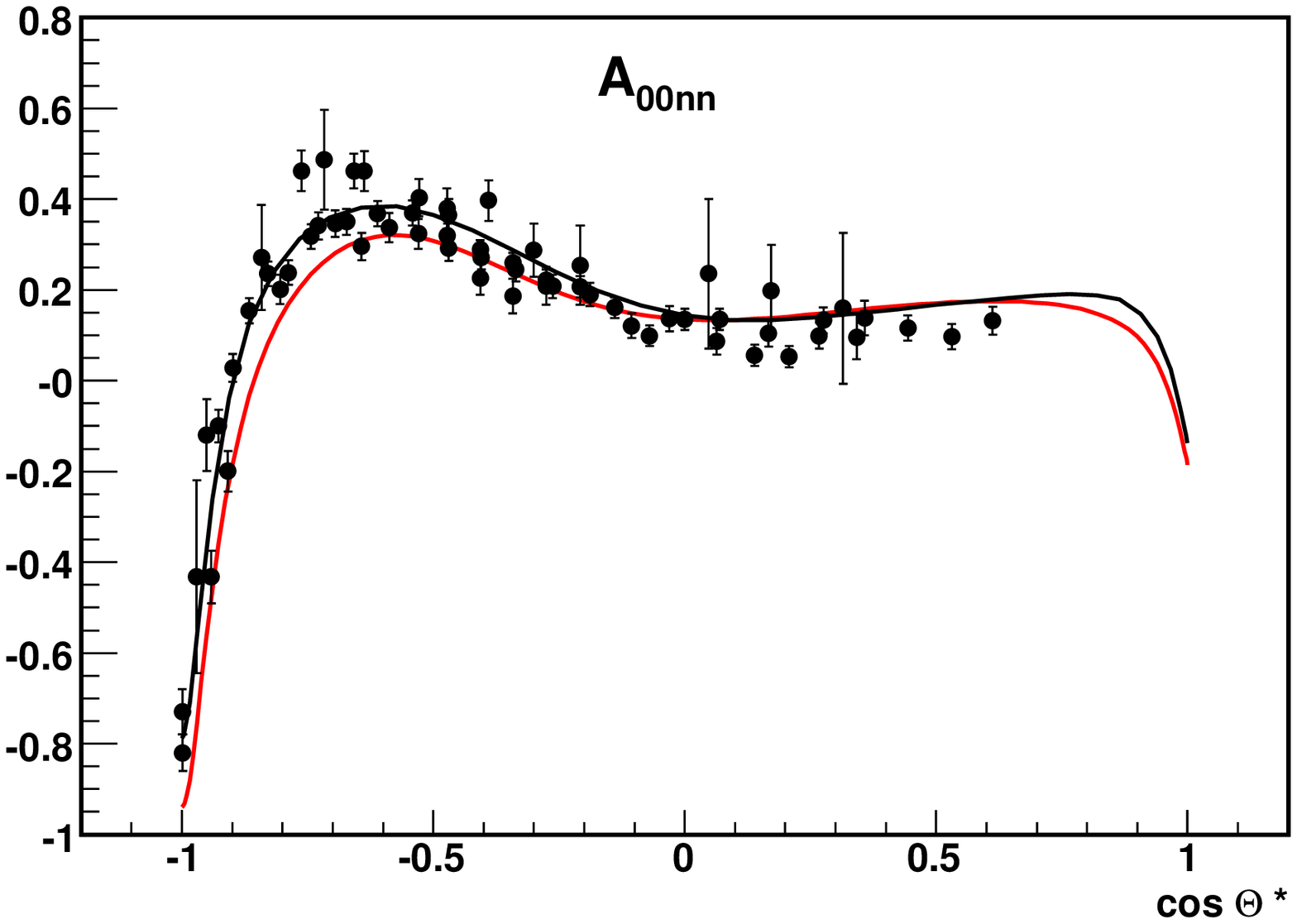}}
\end{figure}

\newpage

$T_{lab}=500 MeV$, Re part

\begin{tabular}{|l|l|l|l|l|}
\hline
R & VSE & VTO & VLSO & VTNO\\ \hline
0.11 & 37862.8 & 14504.7 & 0 & -1.74716e+06\\
0.15 & -4954.58 & -837.824 & 0 & 455147\\
0.25 & 3767.37 & 1500.01 & 0 & -19986.4\\
0.4 & -4346.01 & -1839.63 & 0 & 1541.32\\
0.55 & 1293.77 & 637.077 & 0 & -251.049\\
0.7 & -208.814 & -134.014 & 0 & 31.7401\\
1.4 & -0.995248 & 1.26681 & 0 & -0.0407424\\
\hline
\end{tabular}

\vspace{2cm}
$T_{lab}=500 MeV$, Im part

\begin{tabular}{|l|l|l|l|l|}
\hline
R & VSE & VTO & VLSO & VTNO\\ \hline
0.11 & -14492.5 & -20517.6 & 60377 & -194388\\
0.15 & 89149.1 & 9483.94 & -8021.8 & 16807.9\\
0.25 & -43841.8 & -6649.3 & 260.464 & -800\\
0.4 & 9572.99 & 1074.55 & 758.391 & 424.257\\
0.55 & -1771.73 & 49.9999 & -152.473 & -150.147\\
0.7 & 294.519 & -6.0001 & 19.8721 & 12.6668\\
1.4 & 6.39985 & -3.09034 & -0.0364889 & -0.142317\\
\hline
\end{tabular}

\vspace{2cm}
$T_{lab}=500 MeV$, Re part

\begin{tabular}{|l|l|l|l|l|}
\hline
R & VSO & VTE & VLSE & VTNE\\ \hline
0.11 & 33722.8 & -12212.7 & 0 & 750000\\
0.15 & -43521.8 & 23397.1 & 0 & -161042\\
0.25 & 4500.25 & -5499.99 & 0 & 7500\\
0.4 & -1350.43 & 1000.49 & 0 & -1068.63\\
0.55 & 1098.49 & -664.344 & 0 & 179.948\\
0.7 & -449.229 & 150 & 0 & -36.1737\\
1.4 & 23.9514 & -6.09722 & 0 & -0.0409734\\
\hline
\end{tabular}

\vspace{2cm}
$T_{lab}=500 MeV$, Im part

\begin{tabular}{|l|l|l|l|l|}
\hline
R & VSO & VTE & VLSE & VTNE\\ \hline
0.11 & -15019.4 & 29775.1 & -93983.5 & -220000\\
0.15 & 99907.1 & -10333 & 8001.18 & 152897\\
0.25 & -16195.6 & 1528.95 & -728.691 & -14042.5\\
0.4 & 3005.14 & -2544.99 & 1246.66 & 1927.52\\
0.55 & -2900.47 & 1099.96 & -321.017 & -200\\
0.7 & 946.45 & -170.001 & 52.1427 & 51.7188\\
1.4 & -40.3178 & 2.75396 & -0.413609 & 0.191684\\
\hline
\end{tabular}

\newpage
$T_{lab}=500 MeV$, pp

\begin{figure}[hbtp]
 \centering
    \resizebox{9.8cm}{!}{\includegraphics{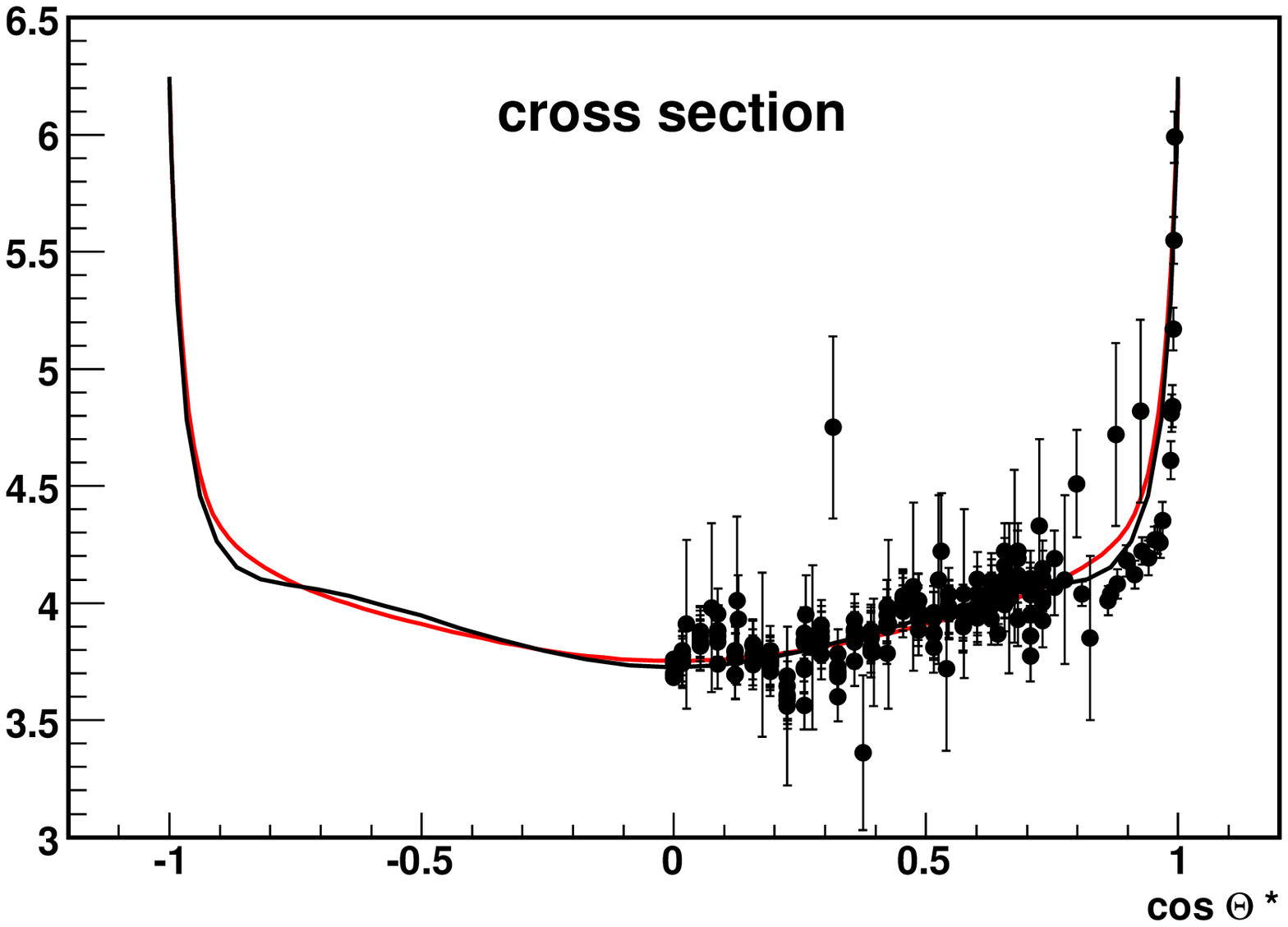}}
\end{figure}


\begin{figure}[hbtp]
 \centering
    \resizebox{9.8cm}{!}{\includegraphics{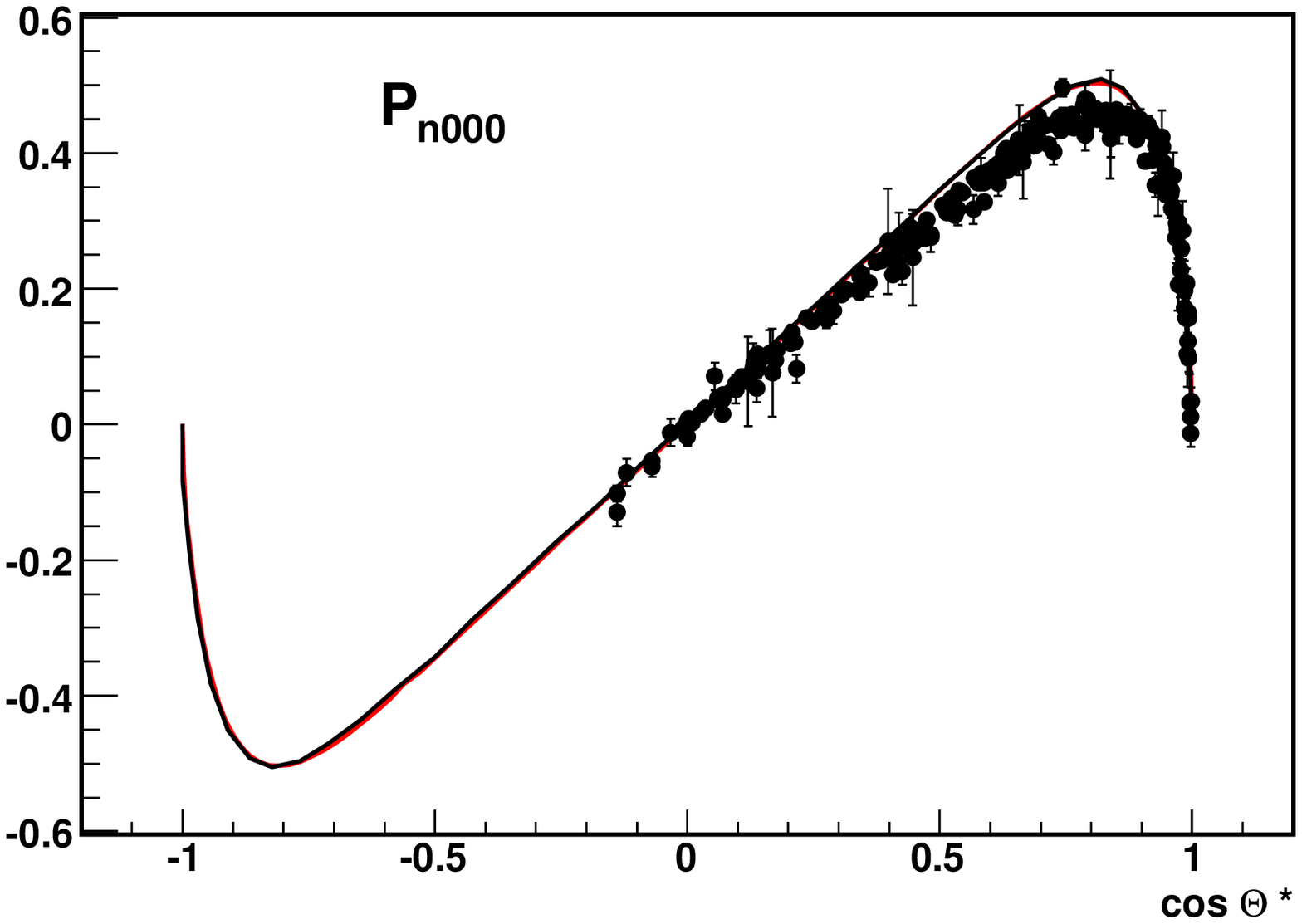}}
\end{figure}


\begin{figure}[hbtp]
 \centering
    \resizebox{9.8cm}{!}{\includegraphics{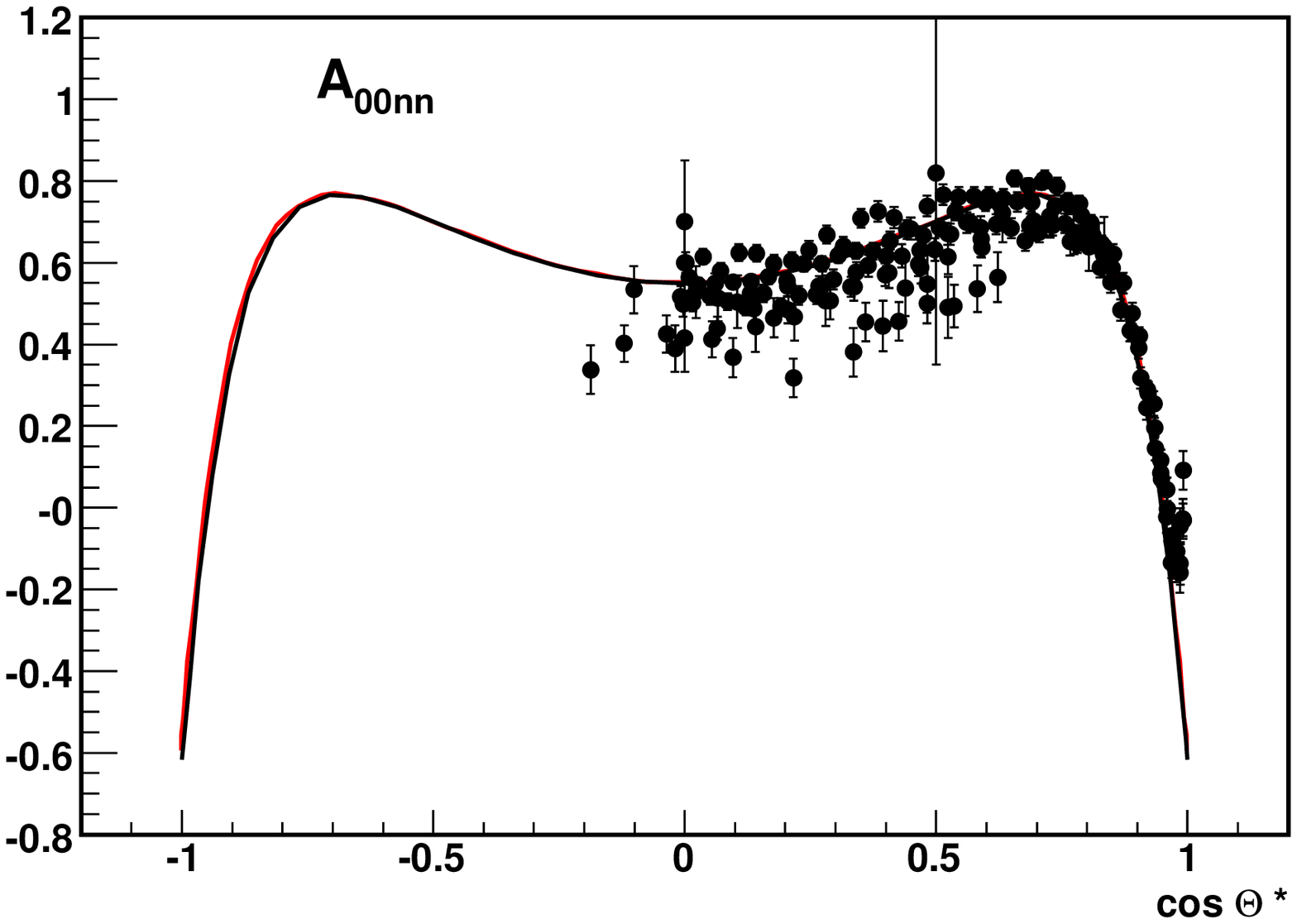}}
\end{figure}

\newpage
$T_{lab}=500 MeV$, np

\begin{figure}[hbtp]
 \centering
    \resizebox{9.8cm}{!}{\includegraphics{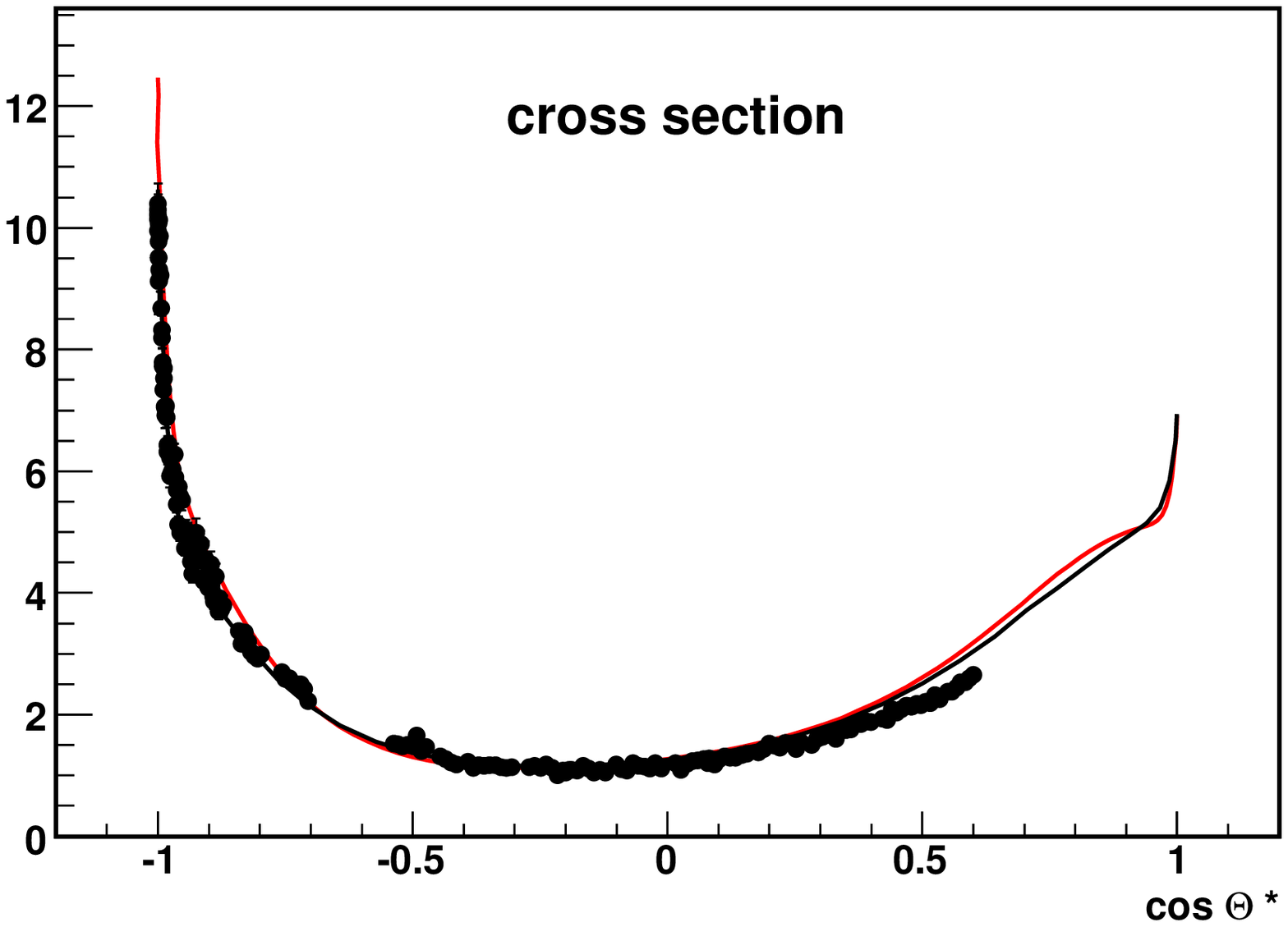}}
\end{figure}


\begin{figure}[hbtp]
 \centering
    \resizebox{9.8cm}{!}{\includegraphics{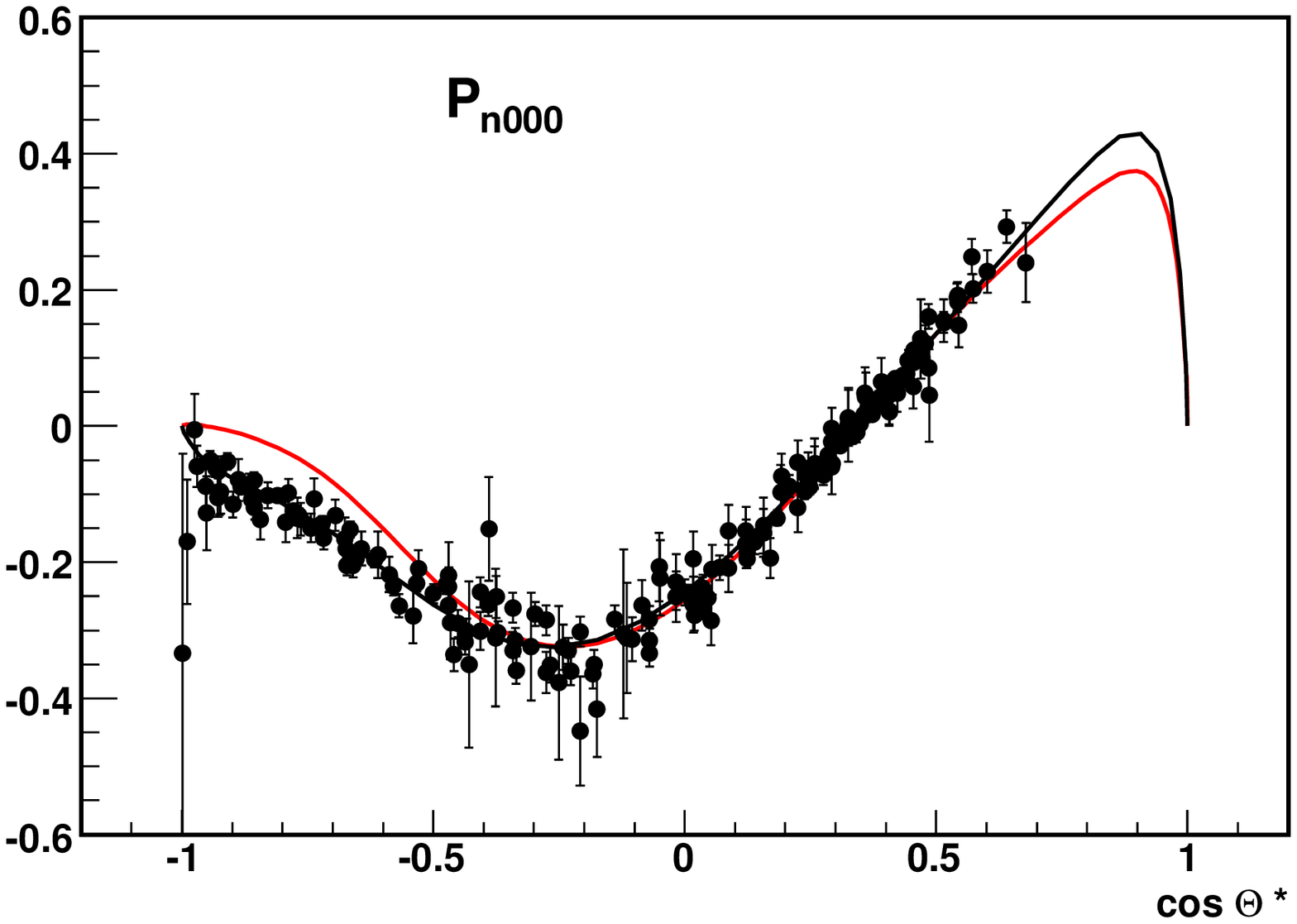}}
\end{figure}


\begin{figure}[hbtp]
 \centering
    \resizebox{9.8cm}{!}{\includegraphics{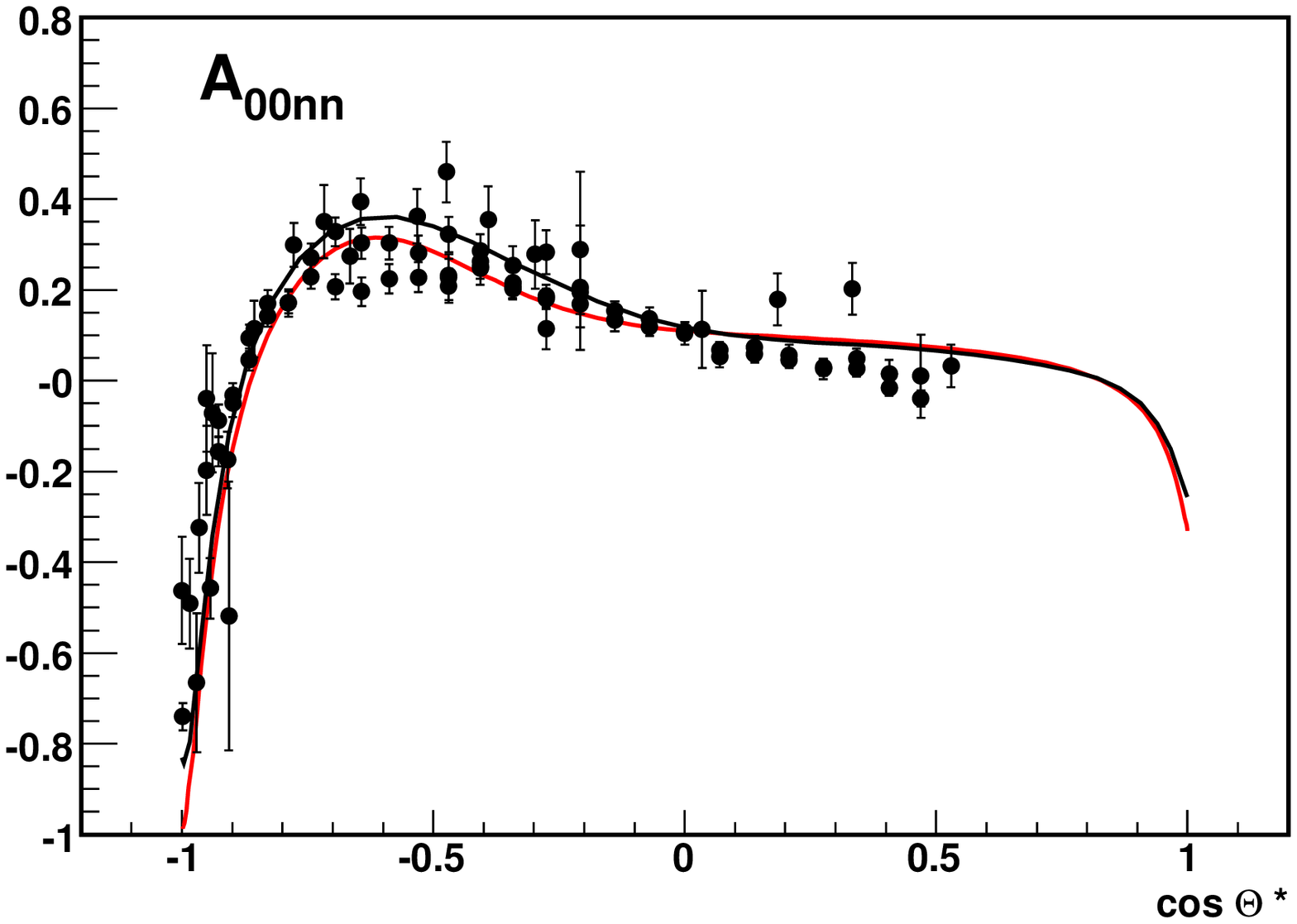}}
\end{figure}

\newpage

$T_{lab}=600 MeV$, Re part

\begin{tabular}{|l|l|l|l|l|}
\hline
R & VSE & VTO & VLSO & VTNO\\ \hline
0.11 & 39223.1 & 39999.2 & 0 & -1.9818e+06\\
0.15 & -16823.7 & -859.535 & 0 & 594371\\
0.25 & 8156.47 & 3249.9 & 0 & -32689.4\\
0.4 & -4906.45 & -3297.6 & 0 & 2833.79\\
0.55 & 1302.73 & 1137.8 & 0 & -501.582\\
0.7 & -214.687 & -233.693 & 0 & 65.2895\\
1.4 & -1.00004 & 2.0221 & 0 & -0.108178\\
\hline
\end{tabular}

\vspace{2cm}
$T_{lab}=600 MeV$, Im part

\begin{tabular}{|l|l|l|l|l|}
\hline
R & VSE & VTO & VLSO & VTNO\\ \hline
0.11 & -22964.1 & -26608.4 & 71451 & -489266\\
0.15 & 116414 & 25000.2 & -15177.1 & 85000\\
0.25 & -53112.9 & -9611.59 & 861.466 & -1897.11\\
0.4 & 10672.9 & 1353.9 & 609.355 & 318.405\\
0.55 & -1961.67 & 50.3023 & -127.863 & -97.8803\\
0.7 & 300.803 & -17.8269 & 17.5571 & 4.03453\\
1.4 & 6.39 & -2.90834 & -0.0387547 & -0.113632\\
\hline
\end{tabular}

\vspace{2cm}
$T_{lab}=600 MeV$, Re part

\begin{tabular}{|l|l|l|l|l|}
\hline
R & VSO & VTE & VLSE & VTNE\\ \hline
0.11 & 34999.8 & -23999.8 & 0 & 762619\\
0.15 & -46778.3 & 25015 & 0 & -197296\\
0.25 & 8064.22 & -5701.42 & 0 & 11698\\
0.4 & -2764.45 & 1834.76 & 0 & -1422.61\\
0.55 & 1191.14 & -965.507 & 0 & 220\\
0.7 & -256.618 & 180.001 & 0 & -37.7607\\
1.4 & 10.4361 & -3.30028 & 0 & -0.0672715\\
\hline
\end{tabular}

\vspace{2cm}
$T_{lab}=600 MeV$, Im part

\begin{tabular}{|l|l|l|l|l|}
\hline
R & VSO & VTE & VLSE & VTNE\\ \hline
0.11 & -30002.5 & 35000.3 & -95999.2 & -350000\\
0.15 & 84998.8 & -15973.9 & 11000 & 191095\\
0.25 & -15228.7 & 3114.64 & -971.086 & -16483.9\\
0.4 & 3300.84 & -3000 & 1295.6 & 2200\\
0.55 & -2866.78 & 1170.99 & -379.887 & -260.55\\
0.7 & 907.596 & -170.001 & 67.0043 & 57.0783\\
1.4 & -39.6339 & 2.5781 & -0.479642 & 0.180795\\
\hline
\end{tabular}

\newpage
$T_{lab}=600 MeV$, pp

\begin{figure}[hbtp]
 \centering
    \resizebox{9.8cm}{!}{\includegraphics{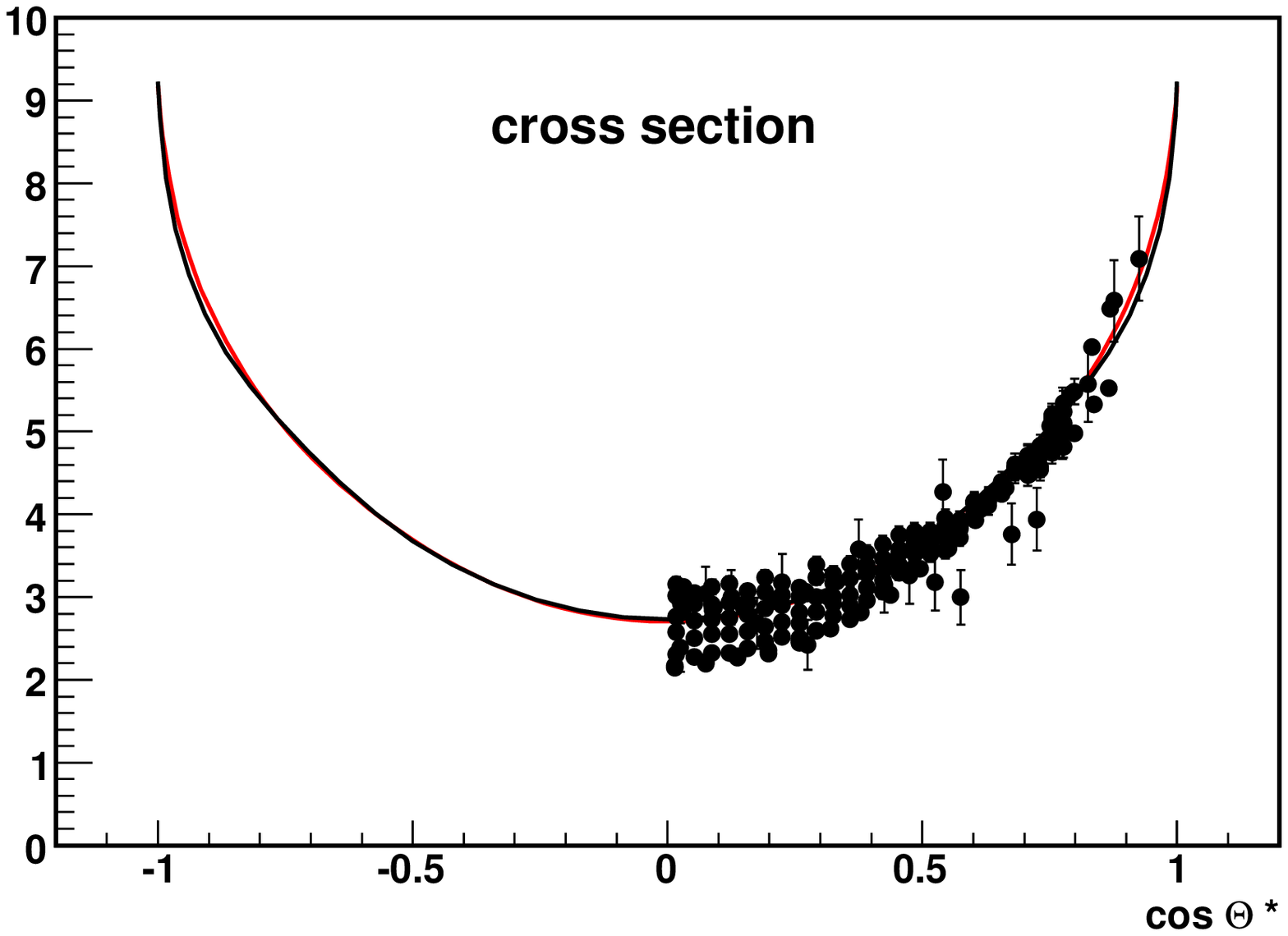}}
\end{figure}


\begin{figure}[hbtp]
 \centering
    \resizebox{9.8cm}{!}{\includegraphics{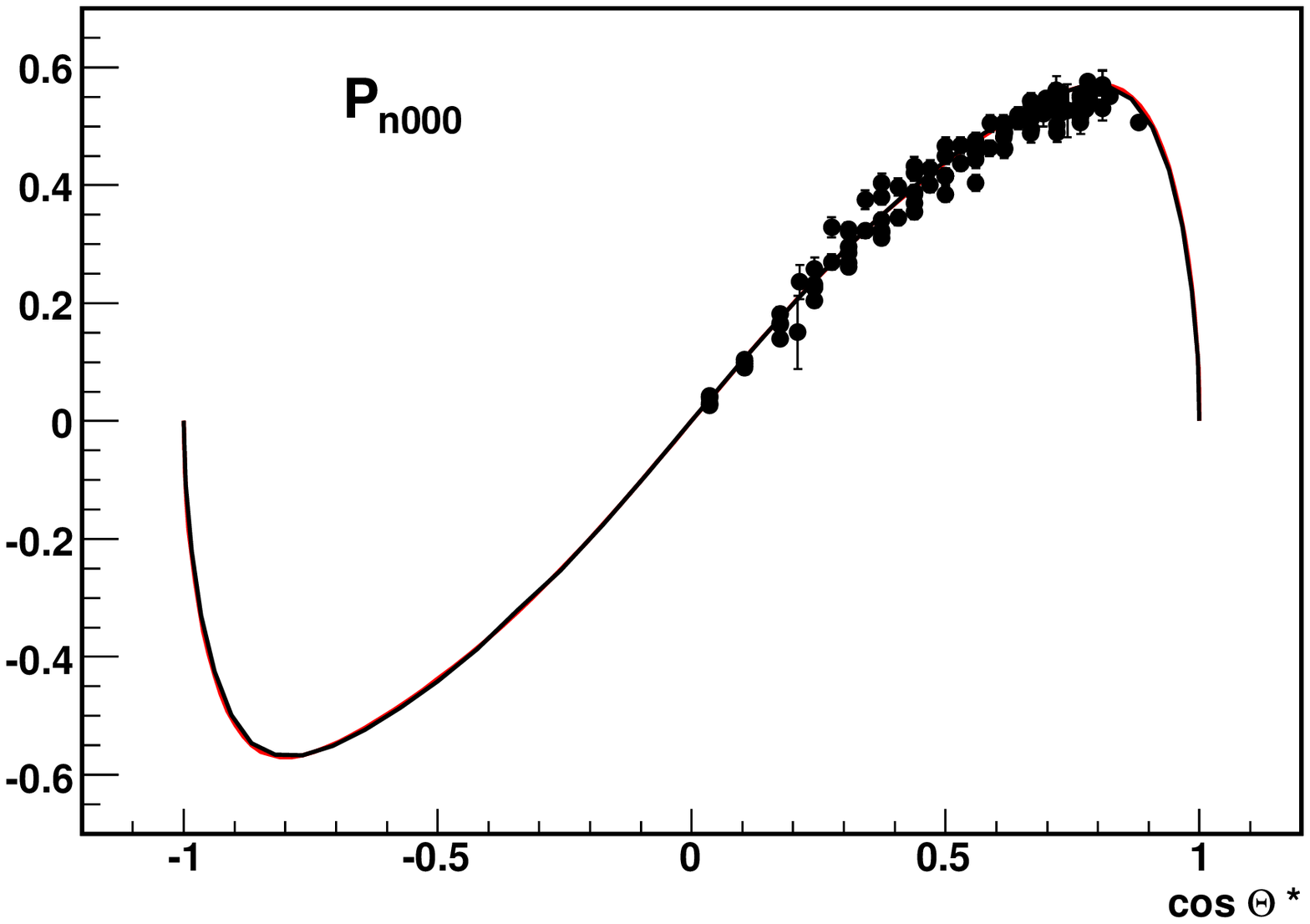}}
\end{figure}


\begin{figure}[hbtp]
 \centering
    \resizebox{9.8cm}{!}{\includegraphics{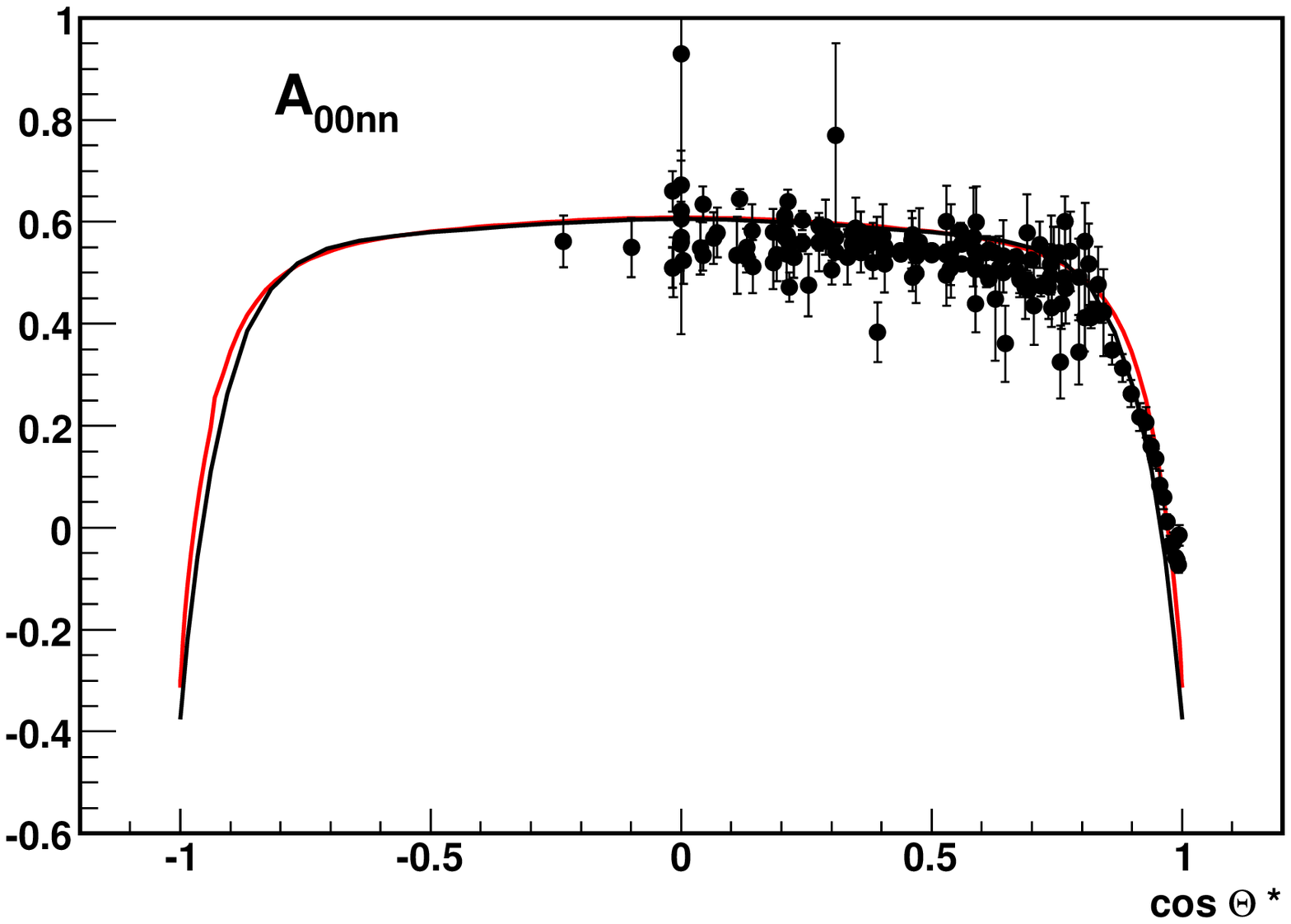}}
\end{figure}

\newpage
$T_{lab}=600 MeV$, np

\begin{figure}[hbtp]
 \centering
    \resizebox{9.8cm}{!}{\includegraphics{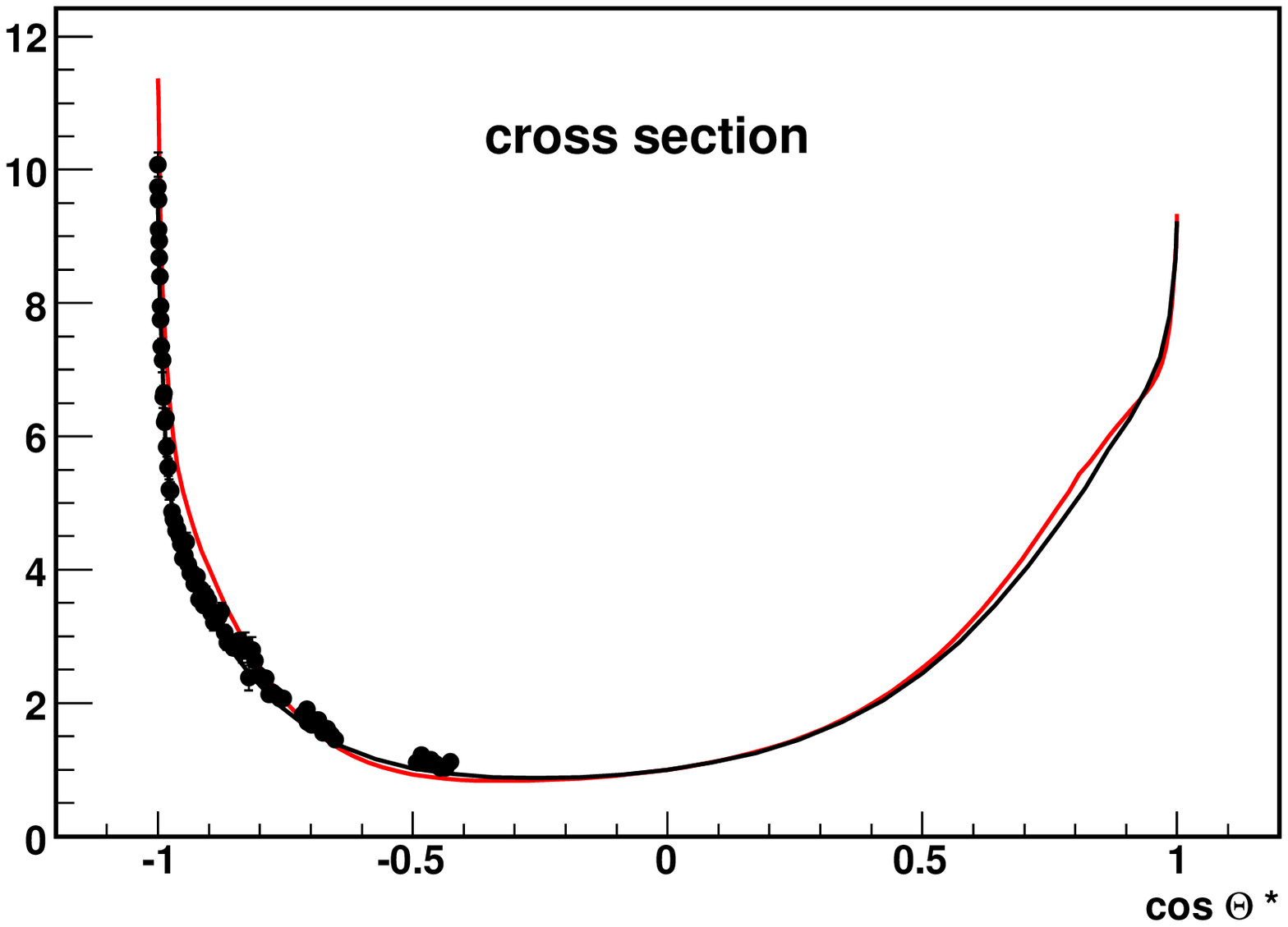}}
\end{figure}


\begin{figure}[hbtp]
 \centering
    \resizebox{9.8cm}{!}{\includegraphics{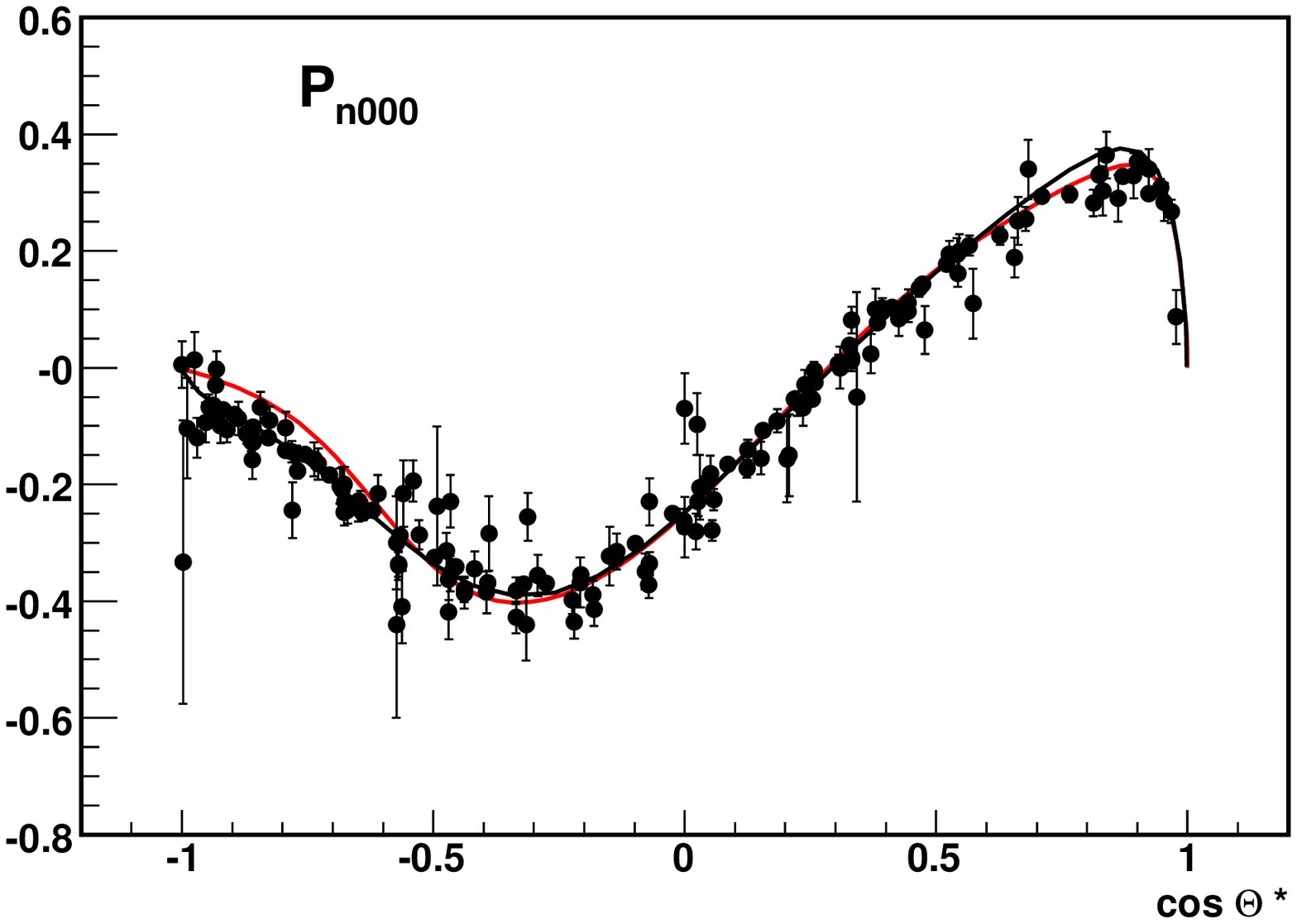}}
\end{figure}


\begin{figure}[hbtp]
 \centering
    \resizebox{9.8cm}{!}{\includegraphics{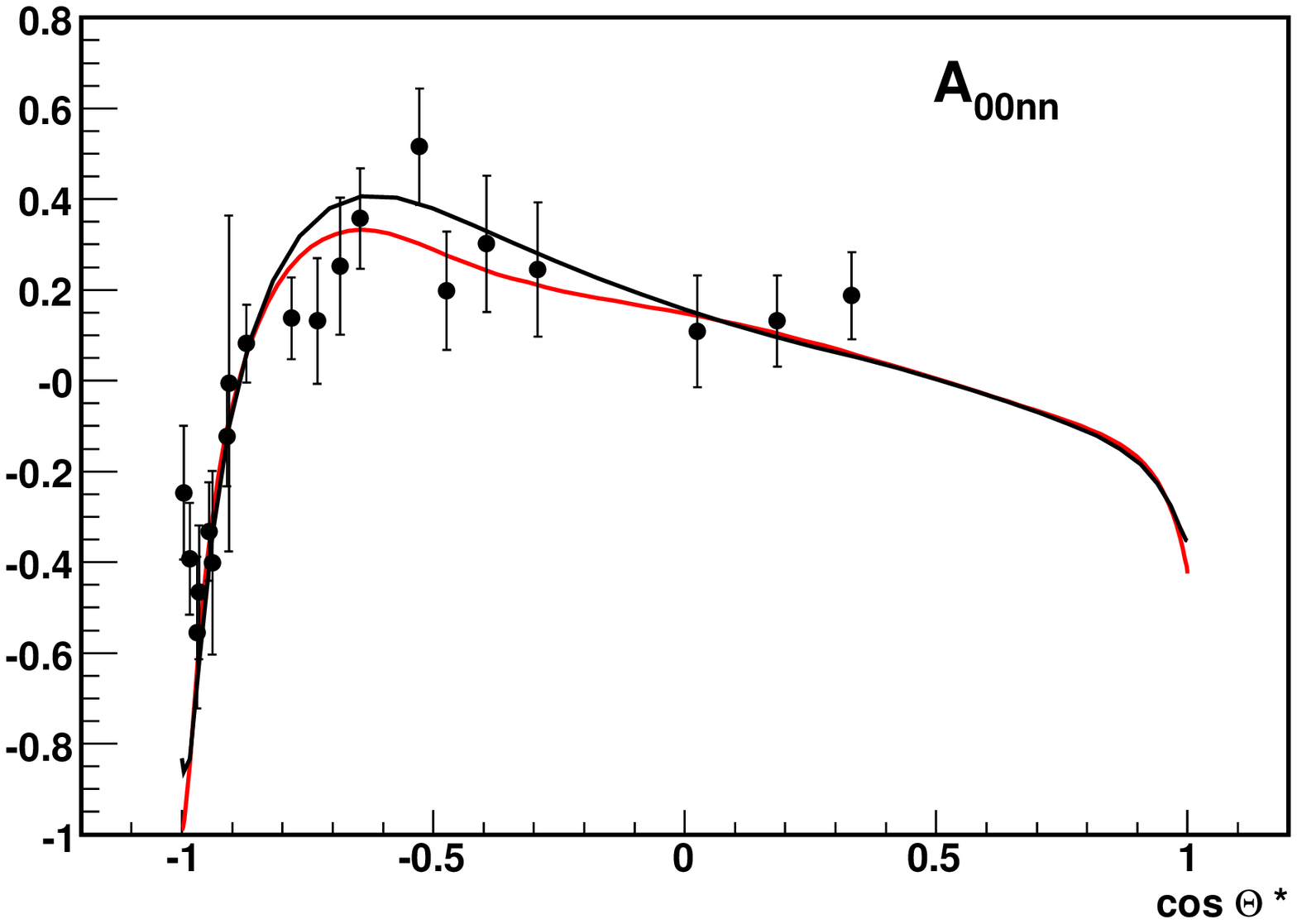}}
\end{figure}

\newpage

$T_{lab}=700 MeV$, Re part

\begin{tabular}{|l|l|l|l|l|}
\hline
R & VSE & VTO & VLSO & VTNO\\ \hline
0.11 & 45263.7 & 34976.6 & 0 & -1.80674e+06\\
0.15 & -61485 & -1082.61 & 0 & 616534\\
0.25 & 26601 & 8006.56 & 0 & -40869.1\\
0.4 & -7902.72 & -5585.28 & 0 & 3907.26\\
0.55 & 1662.15 & 1885.61 & 0 & -737.9\\
0.7 & -216.65 & -379.801 & 0 & 99.4035\\
1.4 & -1.94428 & 3.12775 & 0 & -0.183929\\
\hline
\end{tabular}

\vspace{2cm}
$T_{lab}=700 MeV$, Im part

\begin{tabular}{|l|l|l|l|l|}
\hline
R & VSE & VTO & VLSO & VTNO\\ \hline
0.11 & -34920 & -26764.2 & 46699.4 & -656304\\
0.15 & 120864 & 75446.9 & -18638.6 & 160970\\
0.25 & -53431.5 & -19034.7 & 2628.57 & -6500\\
0.4 & 11362.7 & 1408.82 & 304.796 & 458.73\\
0.55 & -2532.68 & 348.91 & -66.2242 & -89\\
0.7 & 445.215 & -111.874 & 9.37959 & 1.91424\\
1.4 & 4.84994 & -1.92047 & -0.0220682 & -0.112063\\
\hline
\end{tabular}

\vspace{2cm}
$T_{lab}=700 MeV$, Re part

\begin{tabular}{|l|l|l|l|l|}
\hline
R & VSO & VTE & VLSE & VTNE\\ \hline
0.11 & 38678.4 & -33538.1 & 0 & 819999\\
0.15 & -60050 & 26277.9 & 0 & -241958\\
0.25 & 17152.3 & -5845.67 & 0 & 17400.3\\
0.4 & -5998.57 & 3101.48 & 0 & -2300.32\\
0.55 & 1622.22 & -1813.63 & 0 & 432.924\\
0.7 & -84.9992 & 399.808 & 0 & -70.4326\\
1.4 & 1.64853 & -2.06173 & 0 & 0.0163029\\
\hline
\end{tabular}

\vspace{2cm}
$T_{lab}=700 MeV$, Im part

\begin{tabular}{|l|l|l|l|l|}
\hline
R & VSO & VTE & VLSE & VTNE\\ \hline
0.11 & -53124.9 & 49726.5 & -98084.6 & -680000\\
0.15 & 64437.3 & -31519.1 & 28523.1 & 270000\\
0.25 & -11324.3 & 6008.46 & -2698.22 & -18737.3\\
0.4 & 3287 & -3045.78 & 1332.24 & 2252.89\\
0.55 & -2808.41 & 1306.58 & -387.504 & -286.462\\
0.7 & 857.454 & -319.919 & 69.3883 & 61.7145\\
1.4 & -37.3773 & 9.29109 & -0.485126 & 0.131033\\
\hline
\end{tabular}

\newpage
$T_{lab}=700 MeV$, pp

\begin{figure}[hbtp]
 \centering
    \resizebox{9.8cm}{!}{\includegraphics{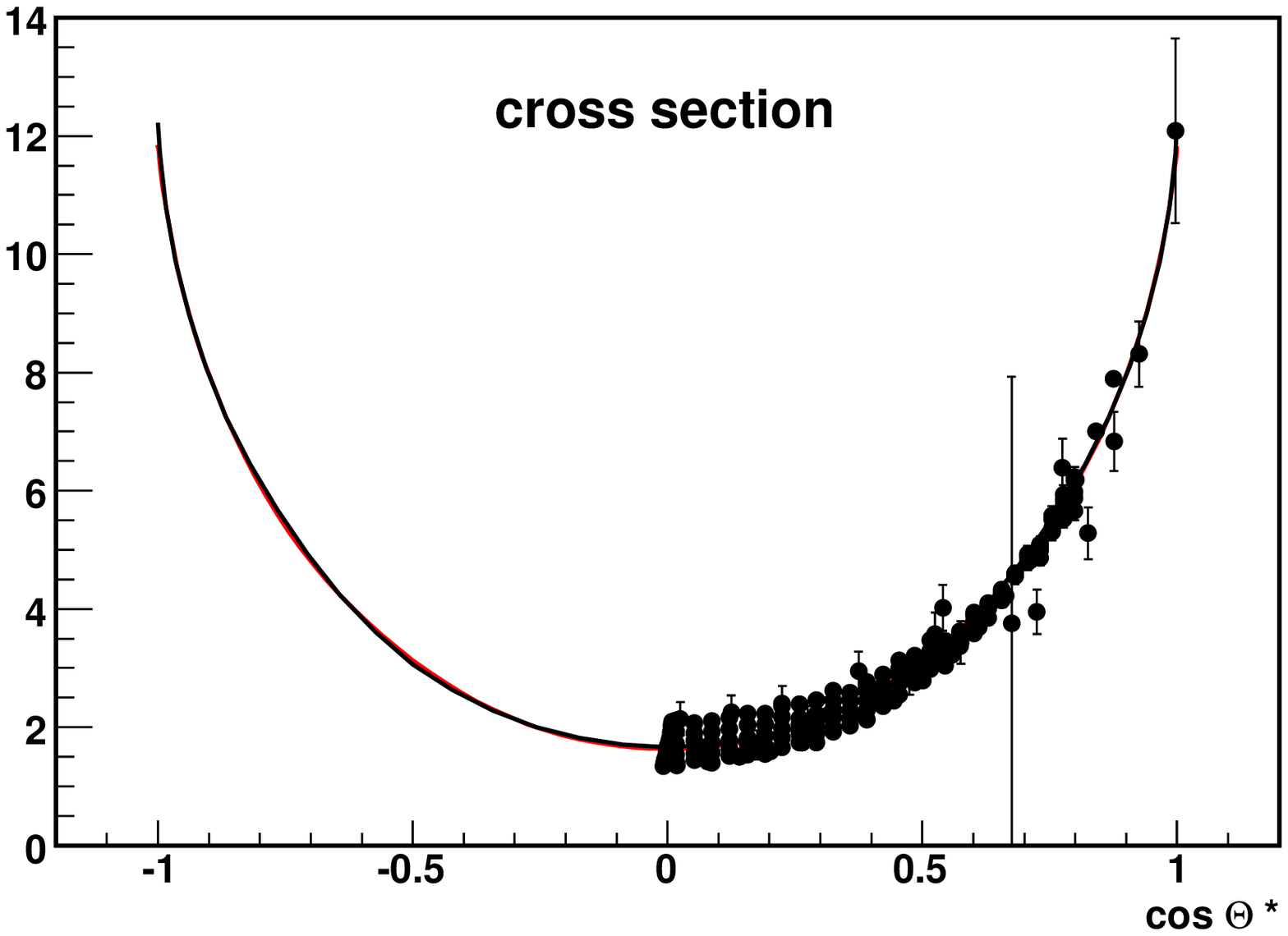}}
\end{figure}


\begin{figure}[hbtp]
 \centering
    \resizebox{9.8cm}{!}{\includegraphics{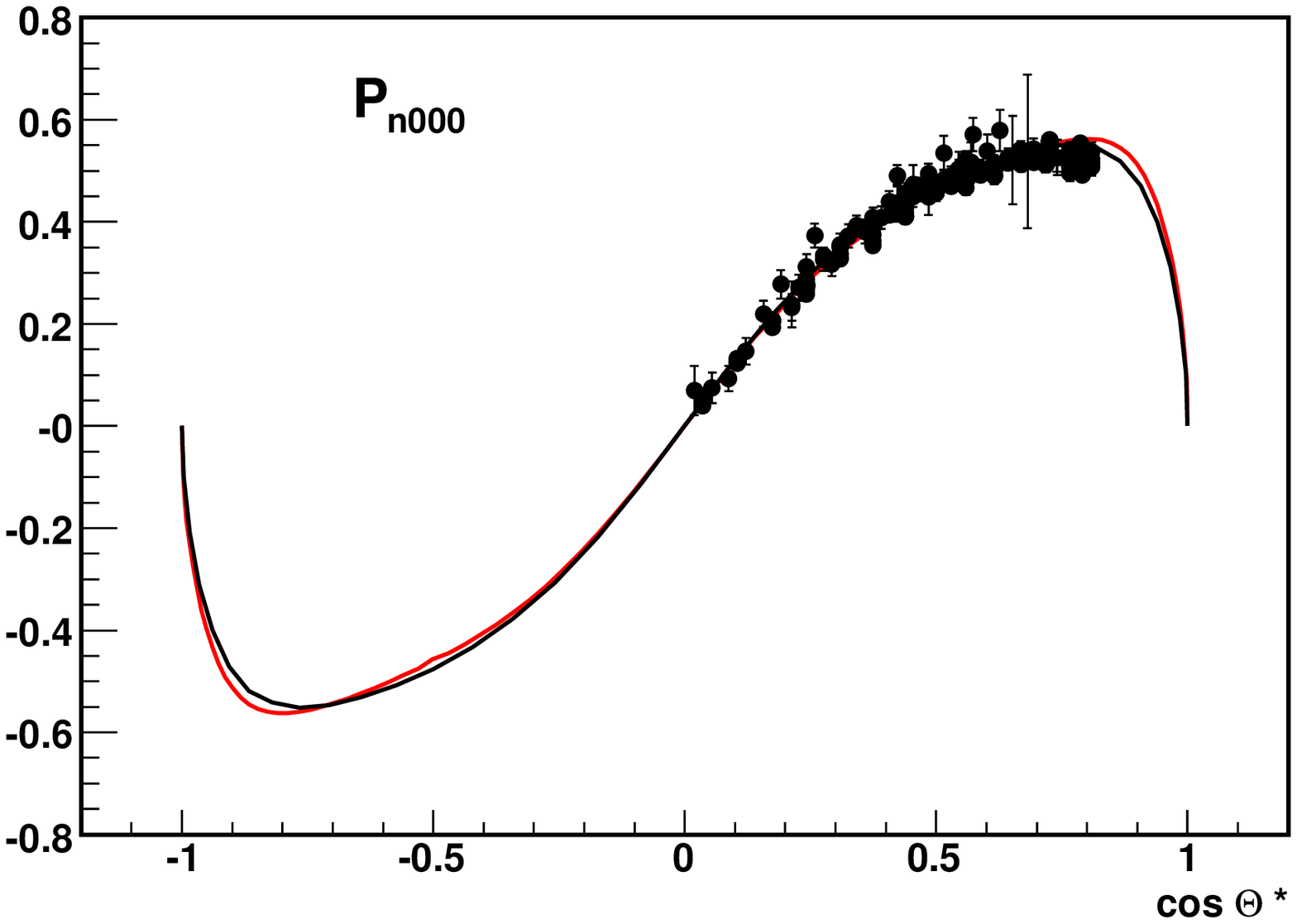}}
\end{figure}


\begin{figure}[hbtp]
 \centering
    \resizebox{9.8cm}{!}{\includegraphics{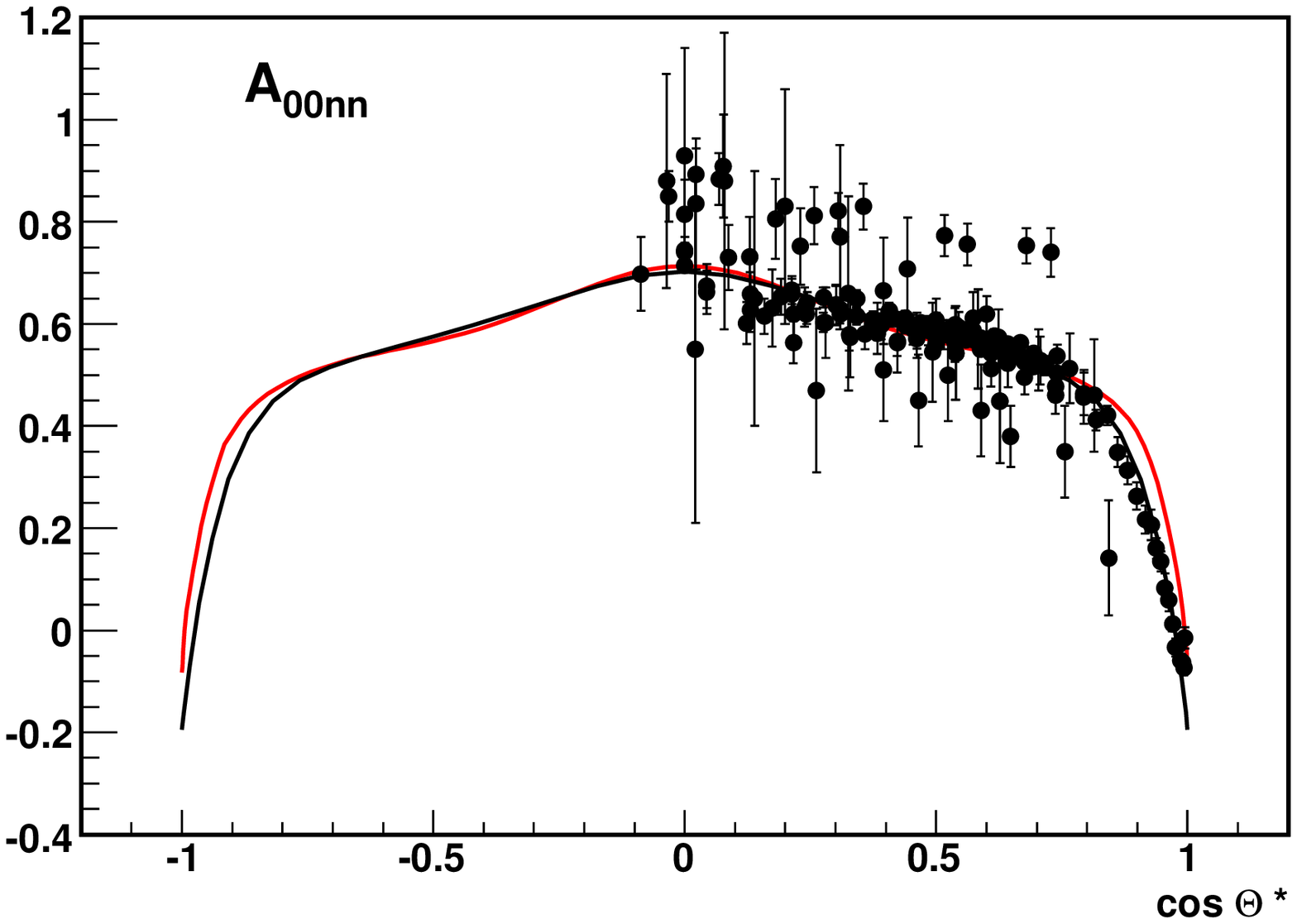}}
\end{figure}

\newpage
$T_{lab}=700 MeV$, np

\begin{figure}[hbtp]
 \centering
    \resizebox{9.8cm}{!}{\includegraphics{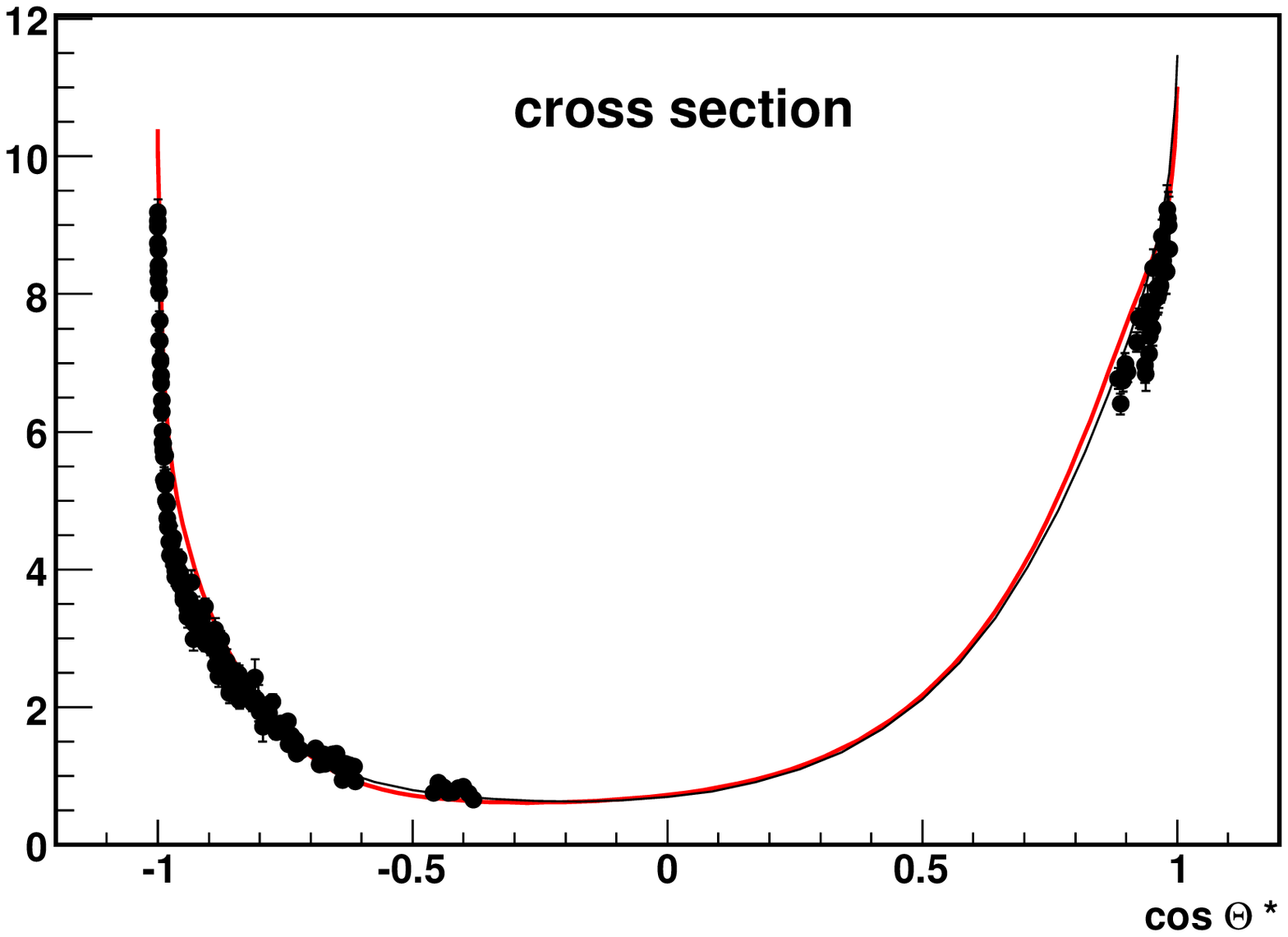}}
\end{figure}


\begin{figure}[hbtp]
 \centering
    \resizebox{9.8cm}{!}{\includegraphics{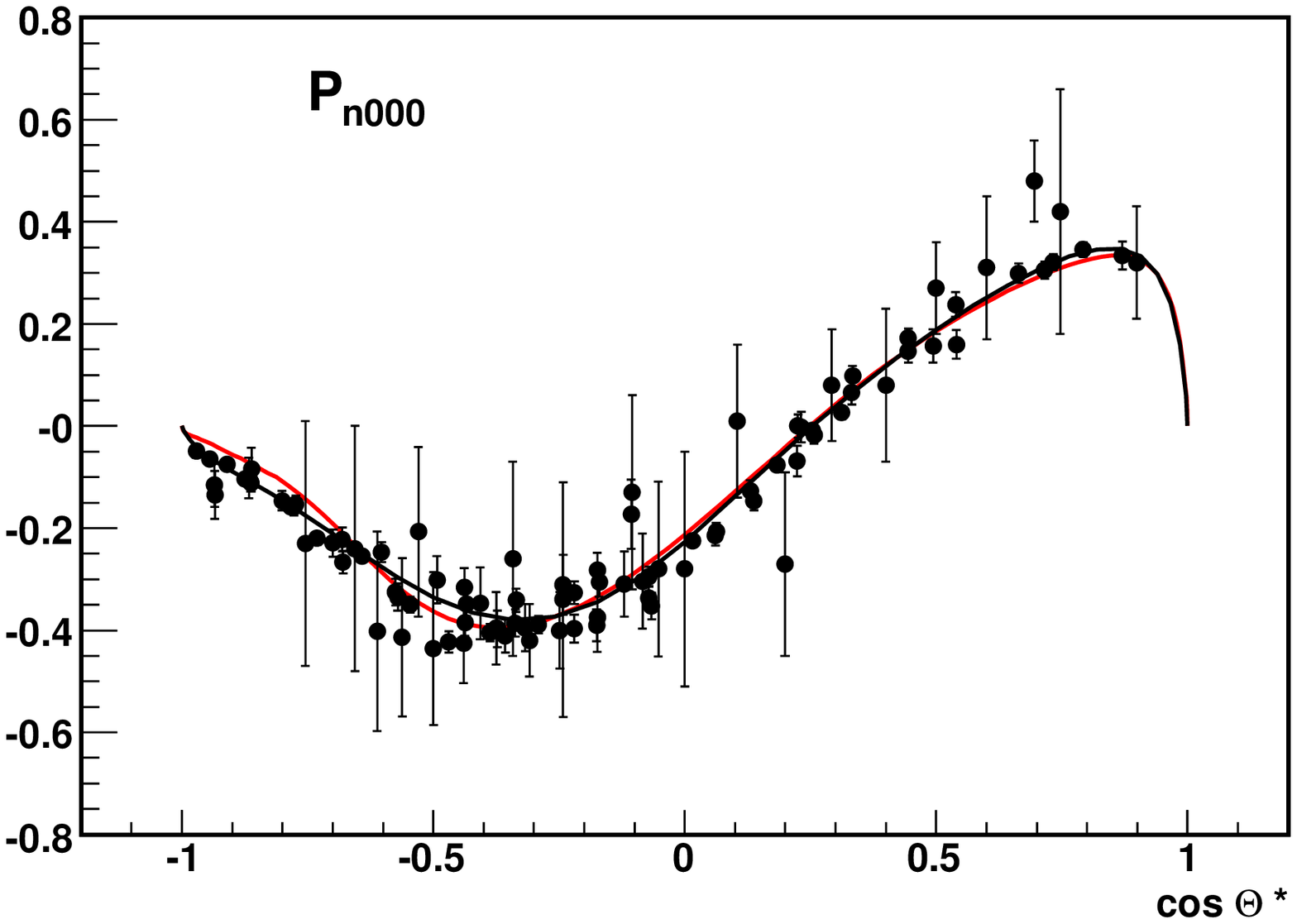}}
\end{figure}


\begin{figure}[hbtp]
 \centering
    \resizebox{9.8cm}{!}{\includegraphics{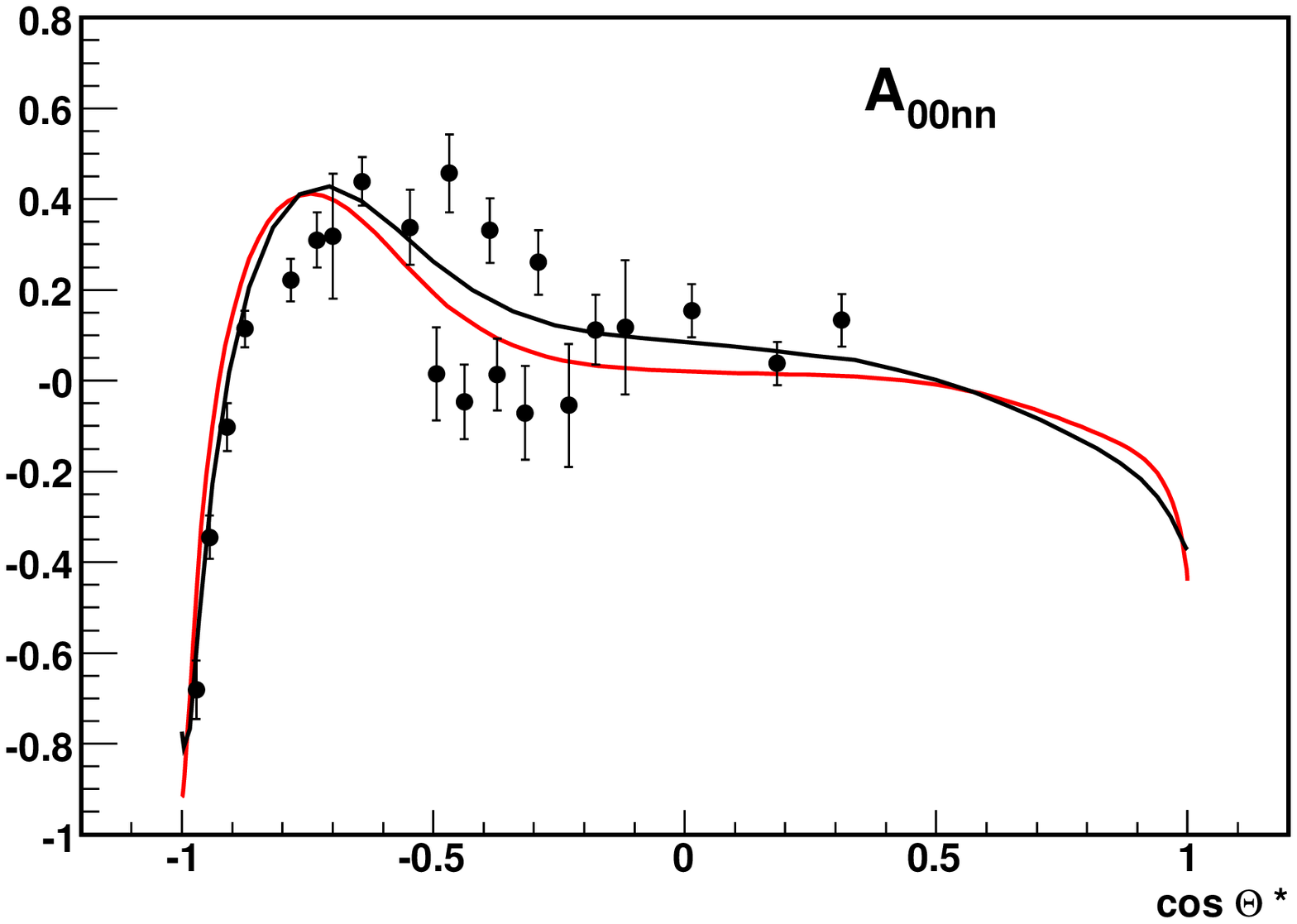}}
\end{figure}

\newpage

$T_{lab}=800 MeV$, Re part

\begin{tabular}{|l|l|l|l|l|}
\hline
R & VSE & VTO & VLSO & VTNO\\ \hline
0.11 & 45300.9 & 34887.9 & 0 & -800000\\
0.15 & -79869.3 & -917.097 & 0 & 320054\\
0.25 & 36206.3 & 6224.42 & 0 & -25343.3\\
0.4 & -10176.9 & -4412.2 & 0 & 2511.69\\
0.55 & 2127.49 & 1327.29 & 0 & -497.576\\
0.7 & -261.752 & -258.907 & 0 & 70.3886\\
1.4 & -2.43463 & 2.48021 & 0 & -0.128731\\
\hline
\end{tabular}

\vspace{2cm}
$T_{lab}=800 MeV$, Im part

\begin{tabular}{|l|l|l|l|l|}
\hline
R & VSE & VTO & VLSO & VTNO\\ \hline
0.11 & -56467.2 & -13305.2 & 25165.4 & -920000\\
0.15 & 99981 & 121165 & -18003.1 & 291900\\
0.25 & -36248.1 & -31106.7 & 3546.83 & -17421.9\\
0.4 & 6145.54 & 2806.01 & 119.925 & 1236.76\\
0.55 & -1034.03 & 146.599 & -25.1756 & -156.989\\
0.7 & 204.387 & -112.002 & 3.38124 & 2.05897\\
1.4 & 4.72574 & -1.29686 & -0.00598186 & -0.0544541\\
\hline
\end{tabular}

\vspace{2cm}
$T_{lab}=800 MeV$, Re part

\begin{tabular}{|l|l|l|l|l|}
\hline
R & VSO & VTE & VLSE & VTNE\\ \hline
0.11 & 39000 & -34171.7 & 0 & 808495\\
0.15 & -99999.9 & 29618 & 0 & -259064\\
0.25 & 27341.1 & -7497.5 & 0 & 19949.5\\
0.4 & -7133.84 & 3443.67 & 0 & -2646.22\\
0.55 & 1700.12 & -1865.03 & 0 & 516.977\\
0.7 & -103.451 & 400 & 0 & -82.5902\\
1.4 & 4.47402 & -0.481941 & 0 & 0.0493507\\
\hline
\end{tabular}

\vspace{2cm}
$T_{lab}=800 MeV$, Im part

\begin{tabular}{|l|l|l|l|l|}
\hline
R & VSO & VTE & VLSE & VTNE\\ \hline
0.11 & -82029.8 & 85585.8 & -97896.2 & -380000\\
0.15 & 54576.5 & -46437.8 & 40754.2 & 175805\\
0.25 & -8000.48 & 4000.48 & -4054.97 & -14975.3\\
0.4 & 3146.73 & -2561.78 & 1336.19 & 2303.39\\
0.55 & -2717.57 & 1632.37 & -380 & -398.557\\
0.7 & 799.87 & -498.918 & 67.7489 & 86.4692\\
1.4 & -35.4606 & 11.2218 & -0.464504 & 0.0050183\\
\hline
\end{tabular}

\newpage
$T_{lab}=800 MeV$, pp

\begin{figure}[hbtp]
 \centering
    \resizebox{9.8cm}{!}{\includegraphics{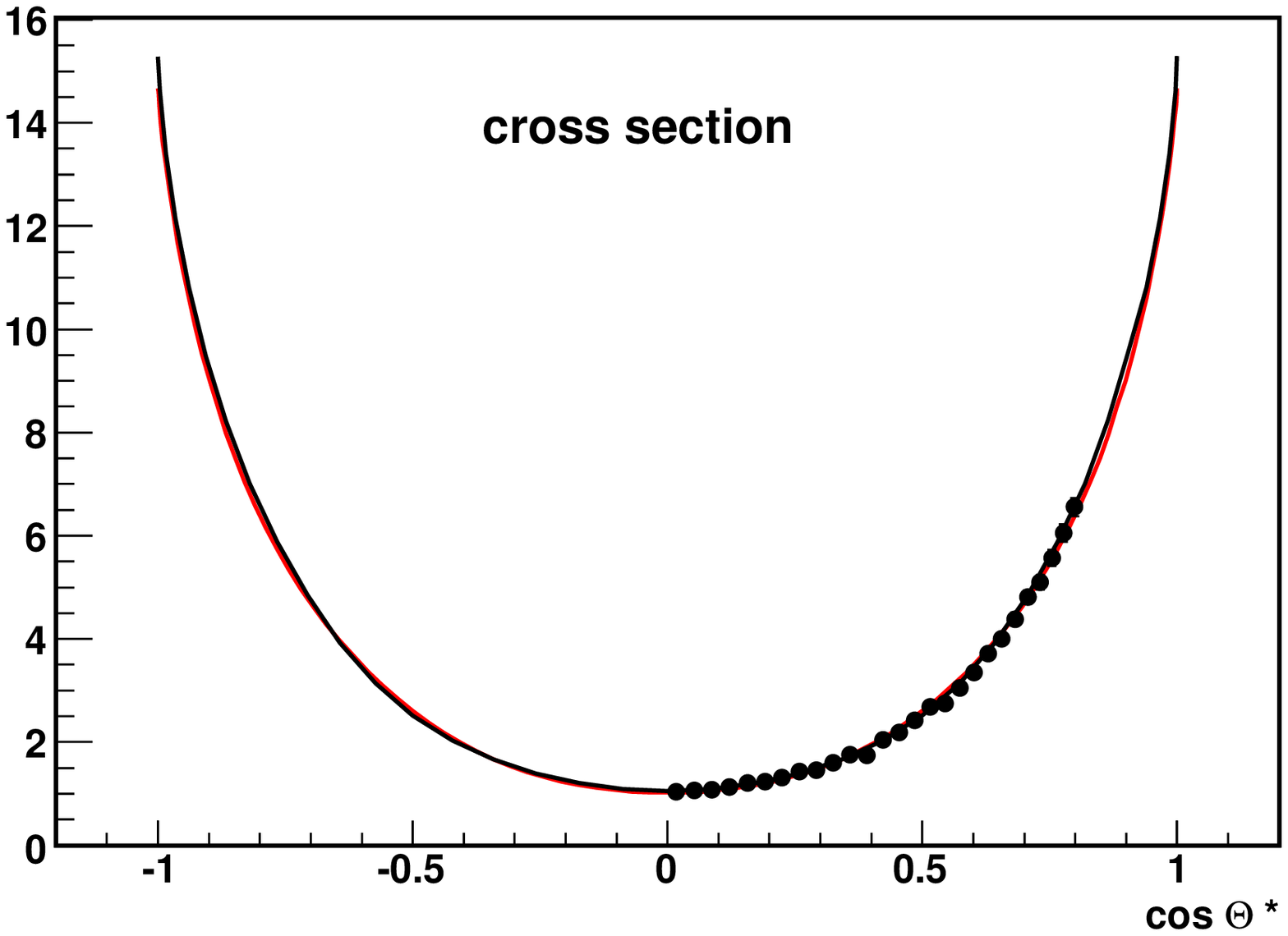}}
\end{figure}


\begin{figure}[hbtp]
 \centering
    \resizebox{9.8cm}{!}{\includegraphics{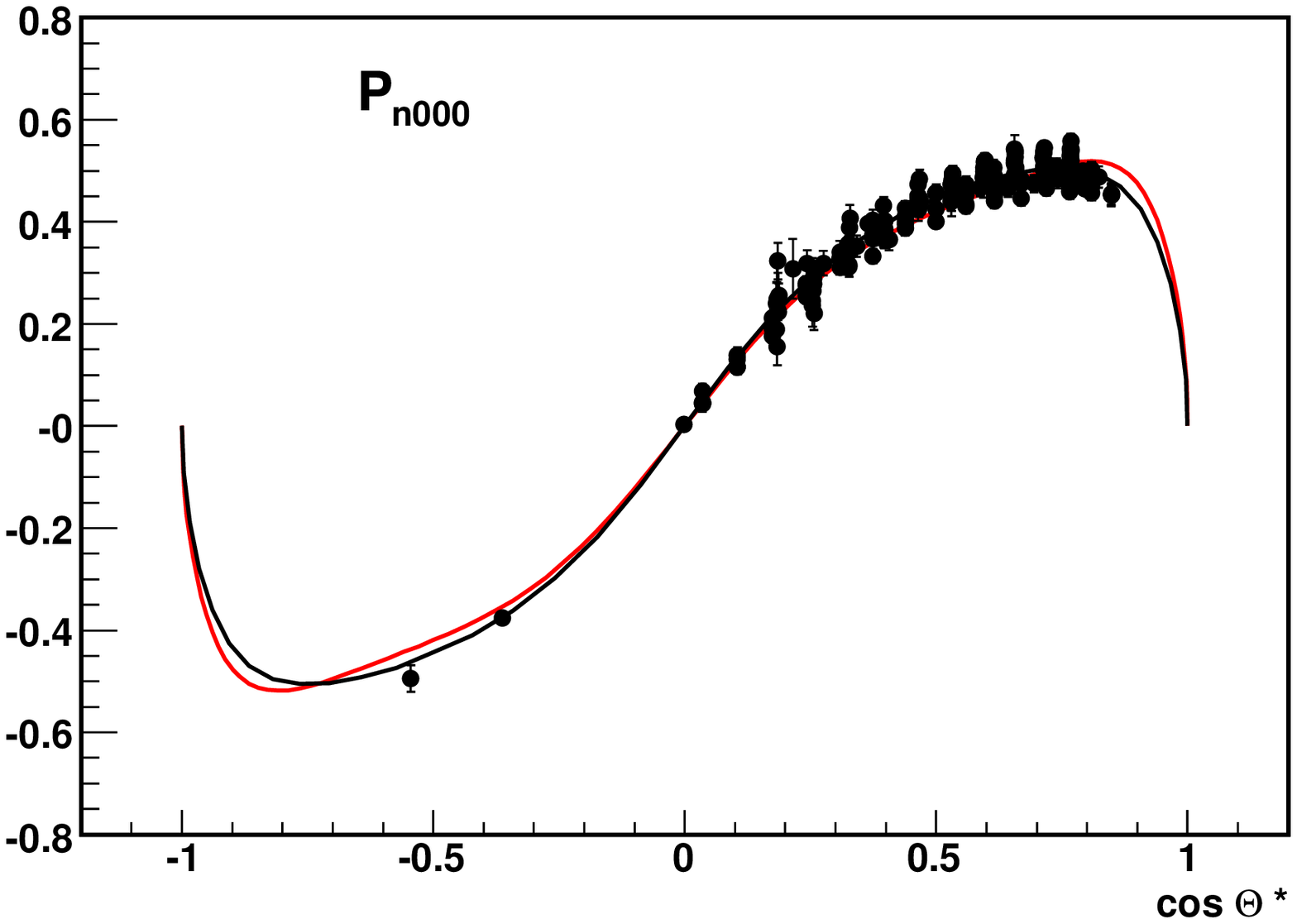}}
\end{figure}


\begin{figure}[hbtp]
 \centering
    \resizebox{9.8cm}{!}{\includegraphics{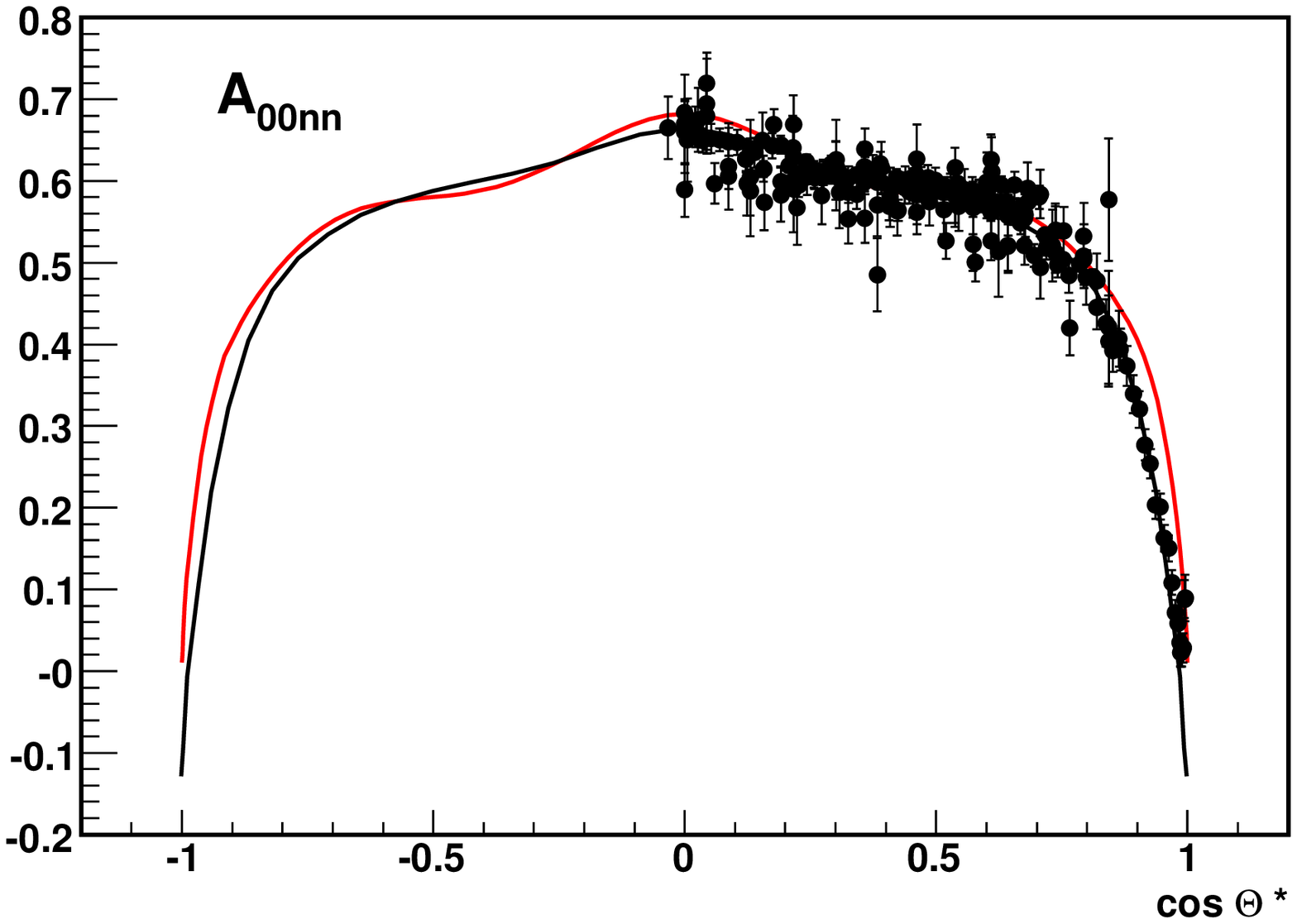}}
\end{figure}

\newpage
$T_{lab}=800 MeV$, np

\begin{figure}[hbtp]
 \centering
    \resizebox{9.8cm}{!}{\includegraphics{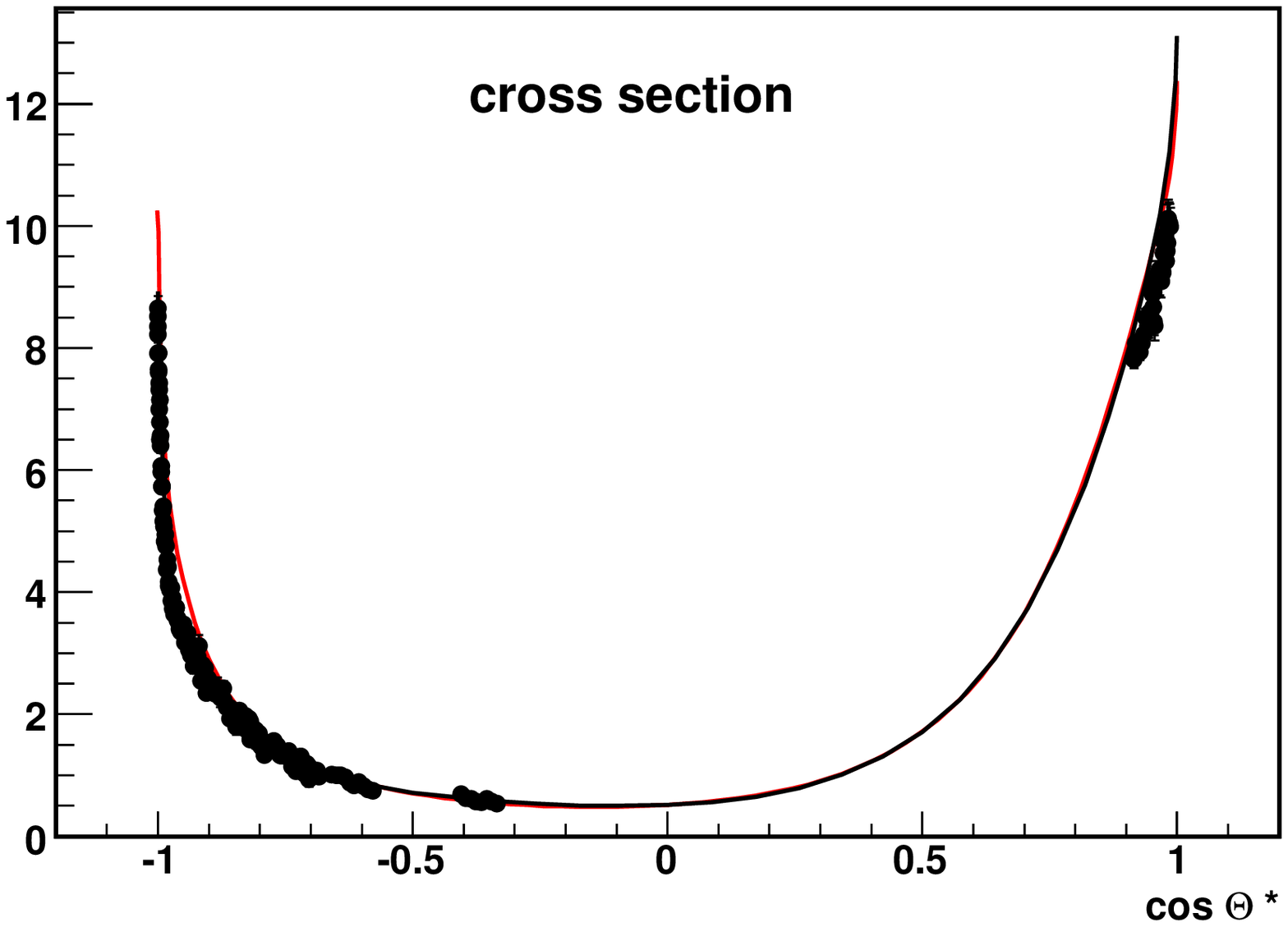}}
\end{figure}


\begin{figure}[hbtp]
 \centering
    \resizebox{9.8cm}{!}{\includegraphics{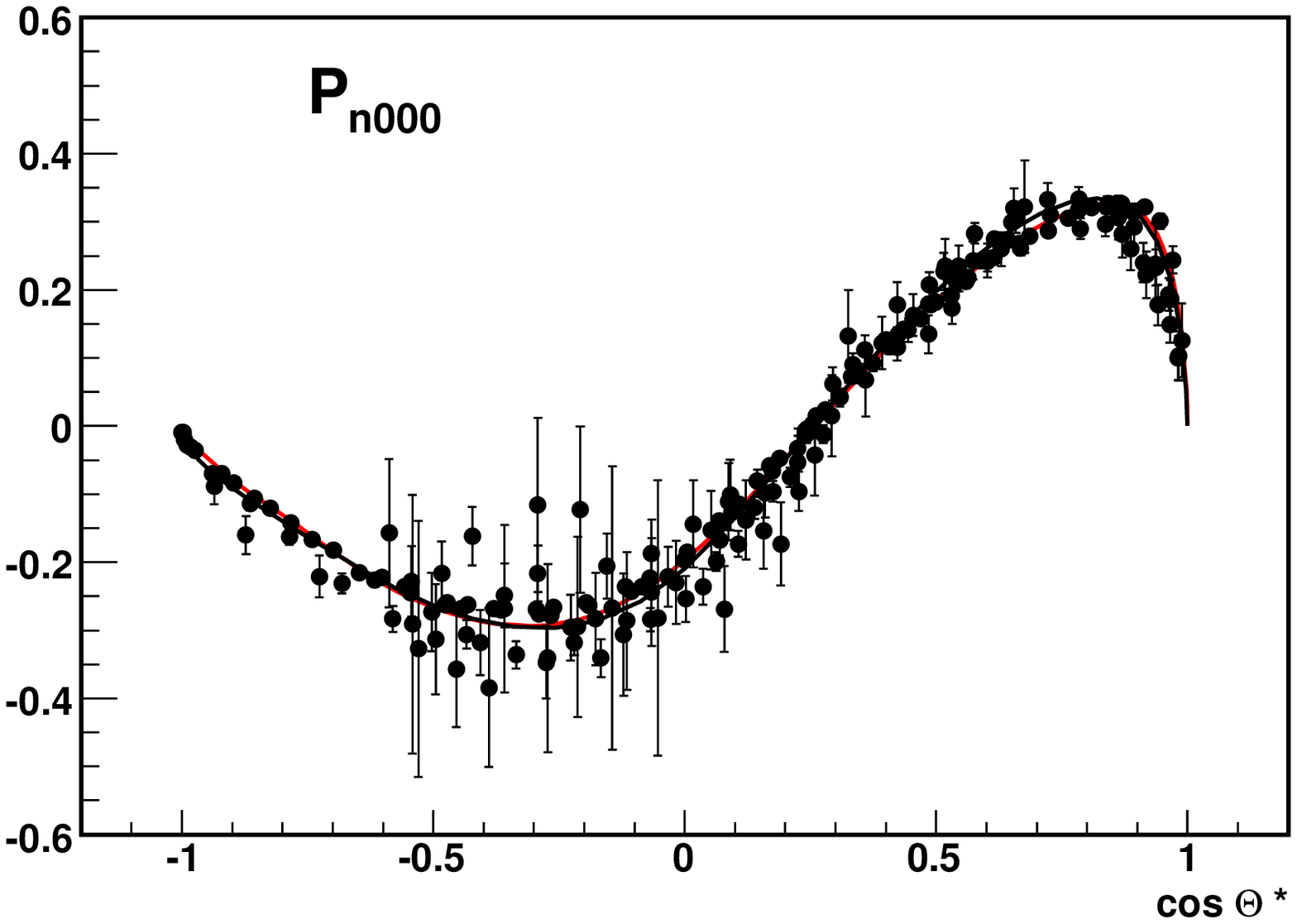}}
\end{figure}


\begin{figure}[hbtp]
 \centering
    \resizebox{9.8cm}{!}{\includegraphics{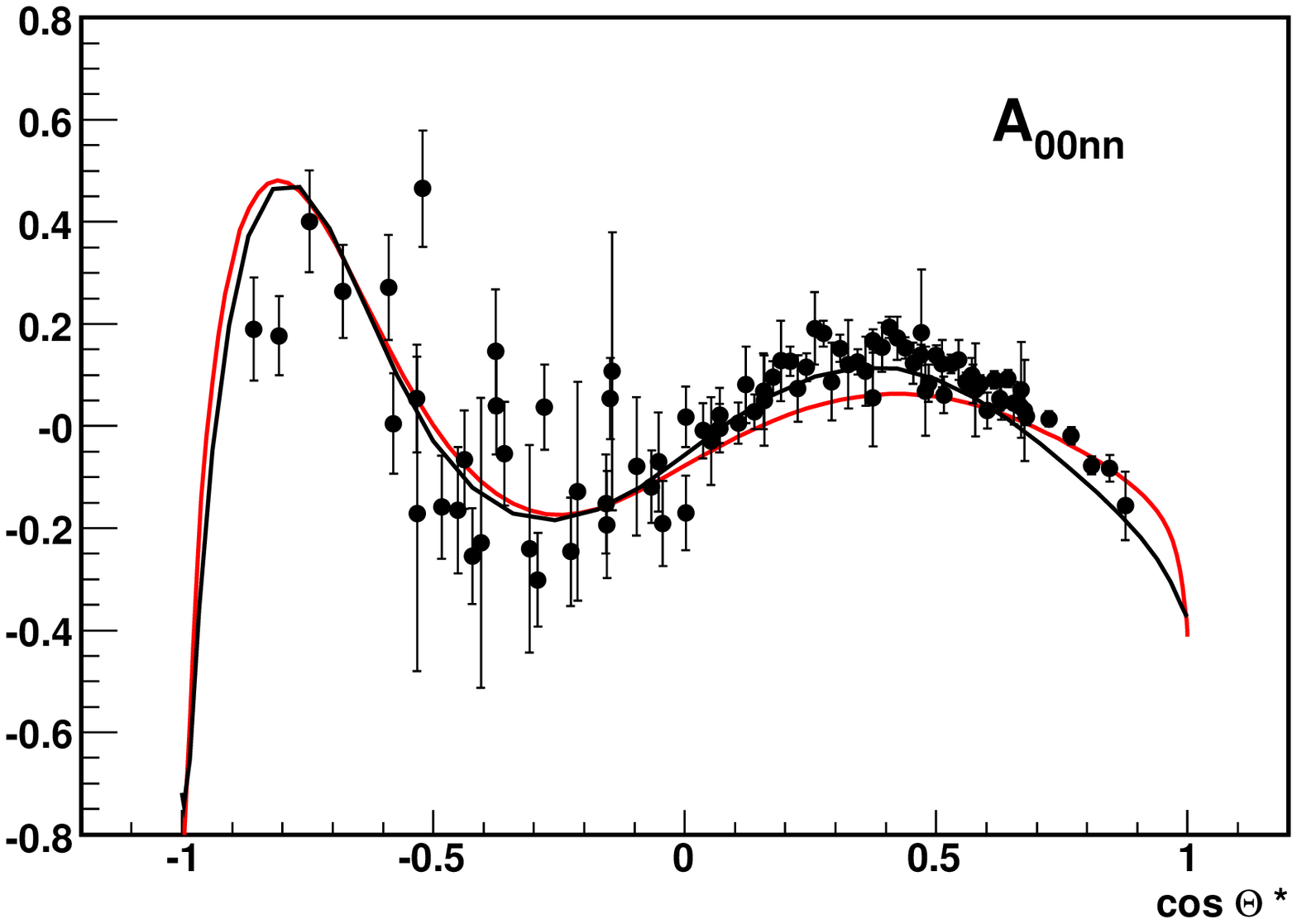}}
\end{figure}

\newpage

$T_{lab}=900 MeV$, Re part

\begin{tabular}{|l|l|l|l|l|}
\hline
R & VSE & VTO & VLSO & VTNO\\ \hline
0.11 & 45401.6 & 14223.1 & 0 & -280001\\
0.15 & -74533.3 & -909.33 & 0 & 139758\\
0.25 & 33478.5 & 4815.19 & 0 & -13235\\
0.4 & -9375.58 & -2655 & 0 & 1360.05\\
0.55 & 1809.97 & 429.457 & 0 & -305.773\\
0.7 & -189.67 & -47.8691 & 0 & 48.2382\\
1.4 & -2.9259 & 0.8333 & 0 & -0.0968045\\
\hline
\end{tabular}

\vspace{2cm}
$T_{lab}=900 MeV$, Im part

\begin{tabular}{|l|l|l|l|l|}
\hline
R & VSE & VTO & VLSO & VTNO\\ \hline
0.11 & -84664.2 & -9301.41 & 53744.1 & -600000\\
0.15 & 99433 & 116767 & -27889.2 & 225933\\
0.25 & -28765 & -30906 & 3551.32 & -17130.4\\
0.4 & 4099.44 & 2829.68 & 159.308 & 1316.68\\
0.55 & -600 & 114.117 & -41.1986 & -165.379\\
0.7 & 163.771 & -116.036 & 6.12693 & 2.03006\\
1.4 & 3.70942 & -0.897898 & -0.0157595 & -0.042234\\
\hline
\end{tabular}

\vspace{2cm}
$T_{lab}=900 MeV$, Re part

\begin{tabular}{|l|l|l|l|l|}
\hline
R & VSO & VTE & VLSE & VTNE\\ \hline
0.11 & 40555.1 & -28315.4 & 0 & 683621\\
0.15 & -108846 & 28844.1 & 0 & -244909\\
0.25 & 28152.1 & -6934.63 & 0 & 21046.9\\
0.4 & -6500.3 & 1971.89 & 0 & -2813.95\\
0.55 & 1500.04 & -1021.26 & 0 & 561.755\\
0.7 & -135.43 & 215.321 & 0 & -88.4074\\
1.4 & 8.8632 & -0.214113 & 0 & 0.0688726\\
\hline
\end{tabular}

\vspace{2cm}
$T_{lab}=900 MeV$, Im part

\begin{tabular}{|l|l|l|l|l|}
\hline
R & VSO & VTE & VLSE & VTNE\\ \hline
0.11 & -93840.9 & 65153.1 & -96000 & -259059\\
0.15 & 49515.6 & -36529.1 & 39991.4 & 134984\\
0.25 & -5111.04 & 3922.52 & -4000.1 & -11979\\
0.4 & 2153.8 & -2439.75 & 1272.95 & 1782.36\\
0.55 & -2070.87 & 1212.83 & -374.312 & -273.571\\
0.7 & 583.589 & -270.009 & 67.6282 & 66.716\\
1.4 & -31.7335 & 2.15297 & -0.456572 & 0.0514867\\
\hline
\end{tabular}

\newpage
$T_{lab}=900 MeV$, pp

\begin{figure}[hbtp]
 \centering
    \resizebox{9.8cm}{!}{\includegraphics{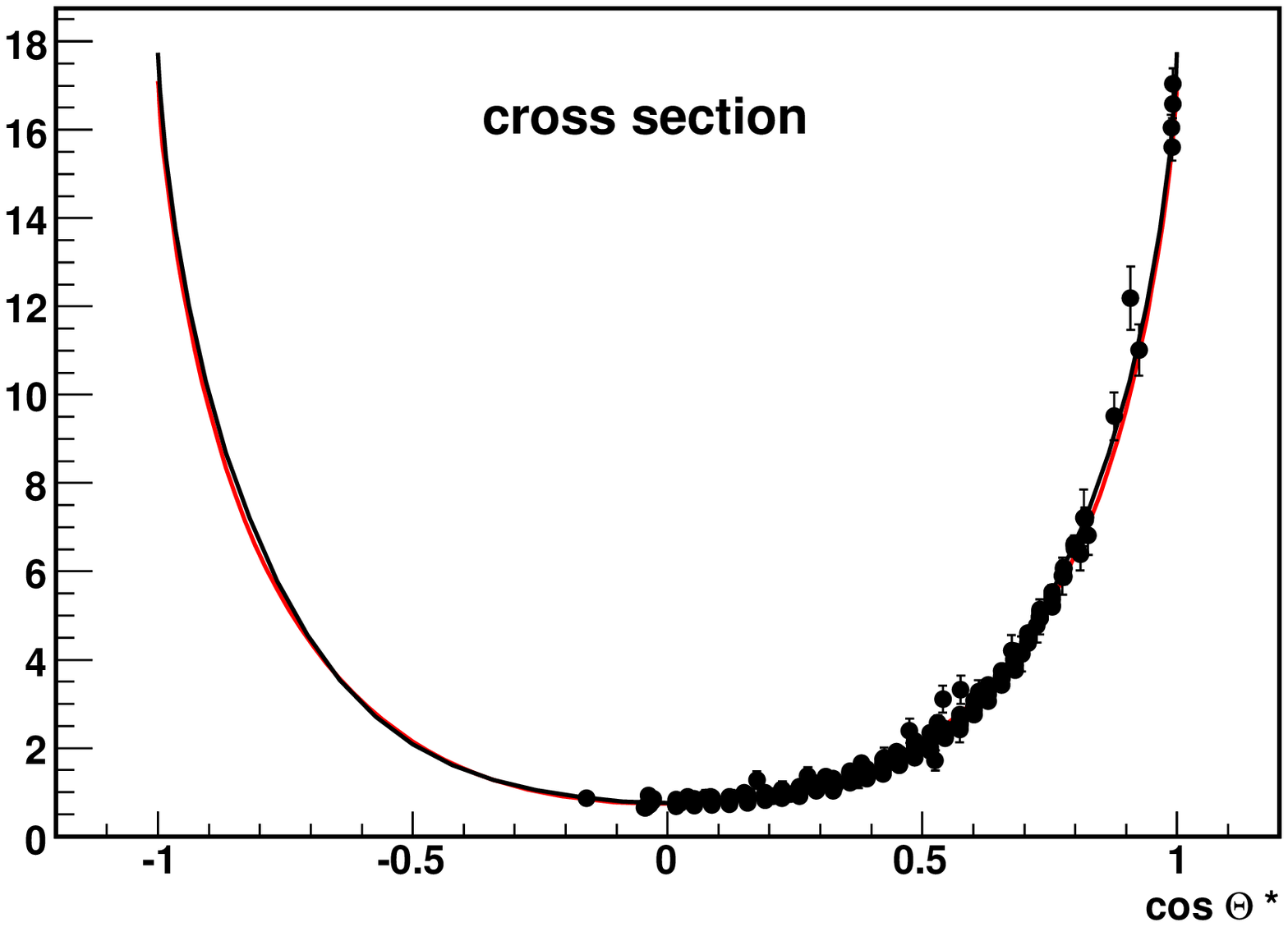}}
\end{figure}


\begin{figure}[hbtp]
 \centering
    \resizebox{9.8cm}{!}{\includegraphics{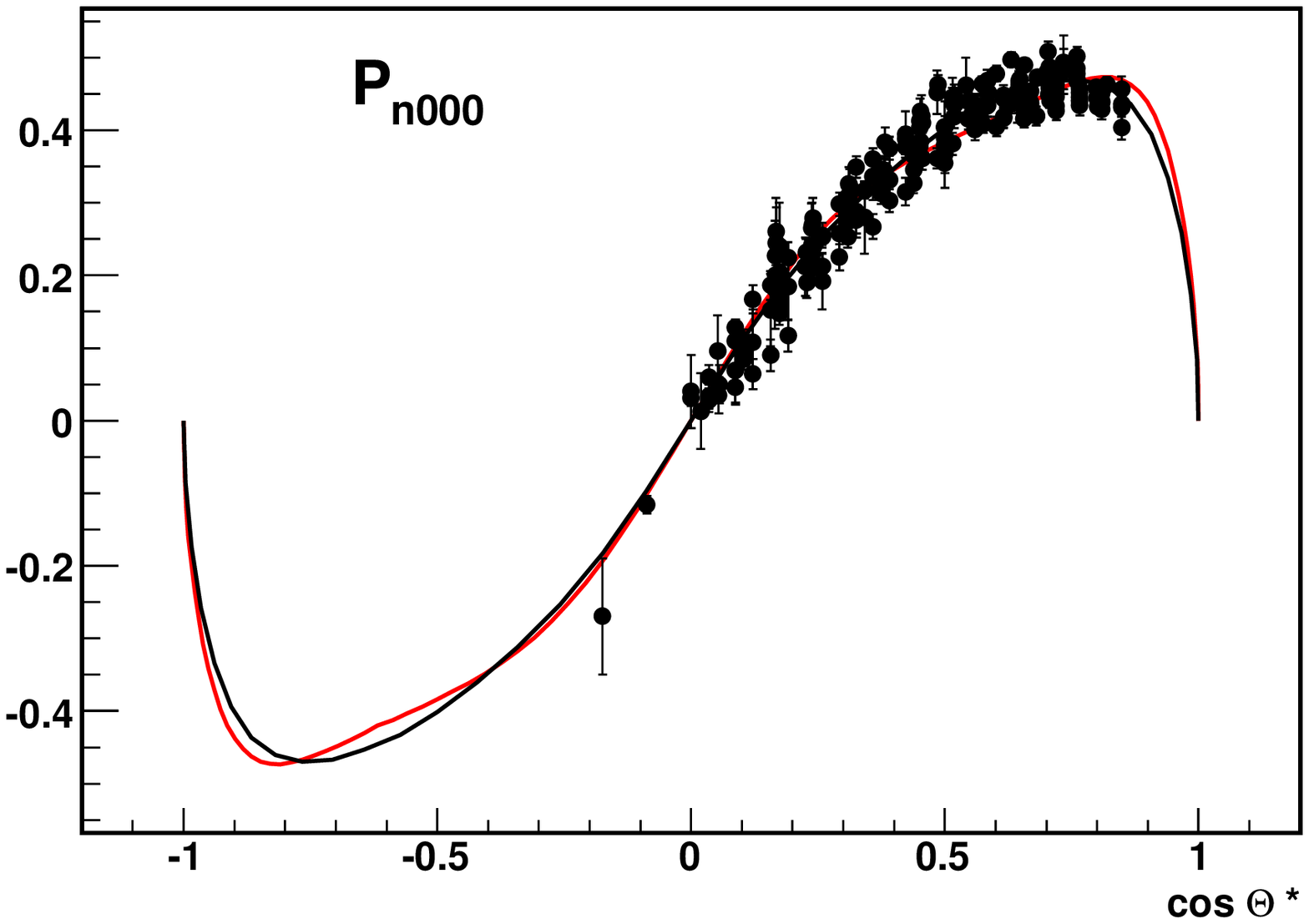}}
\end{figure}


\begin{figure}[hbtp]
 \centering
    \resizebox{9.8cm}{!}{\includegraphics{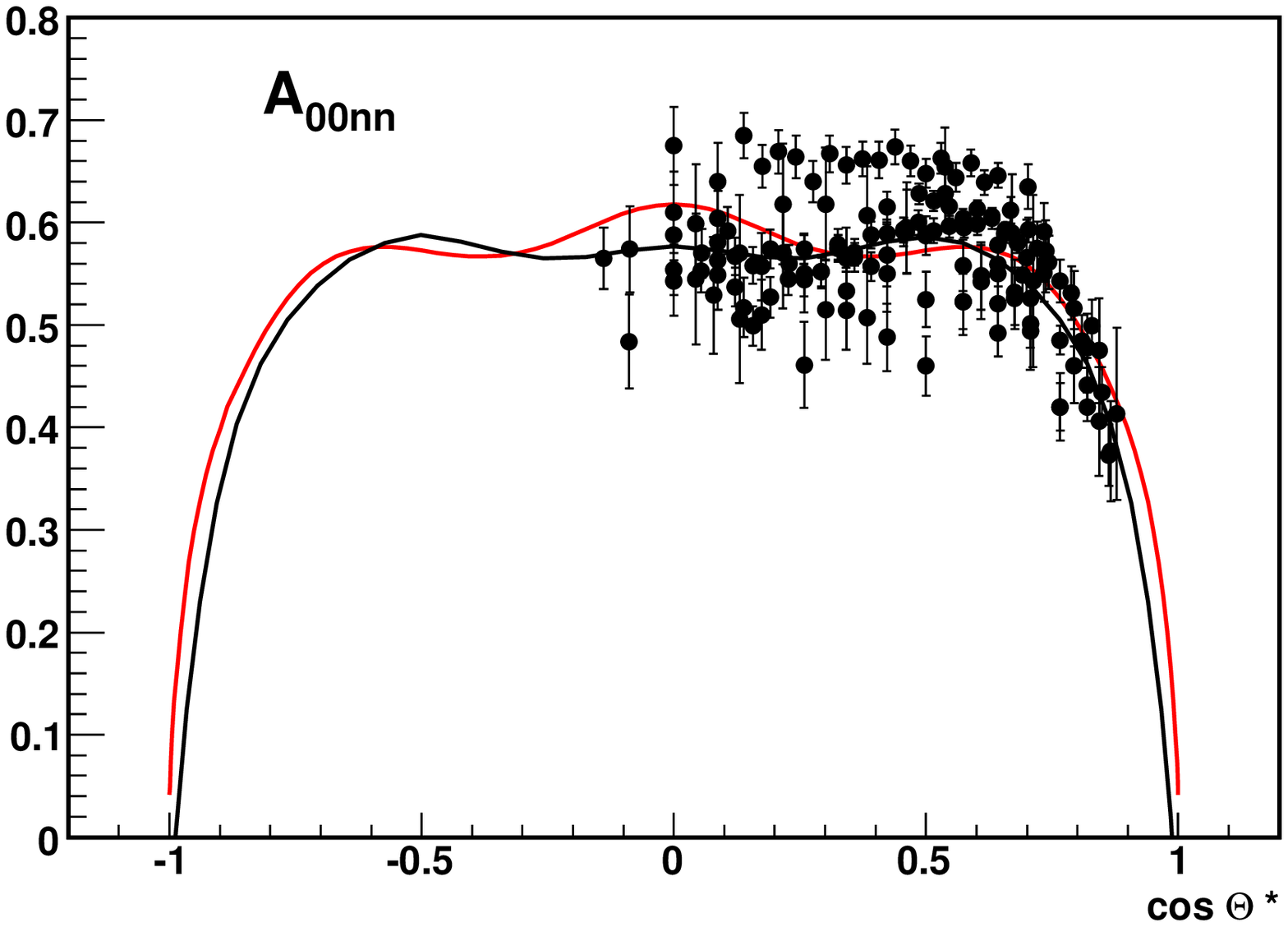}}
\end{figure}

\newpage
$T_{lab}=900 MeV$, np

\begin{figure}[hbtp]
 \centering
    \resizebox{9.8cm}{!}{\includegraphics{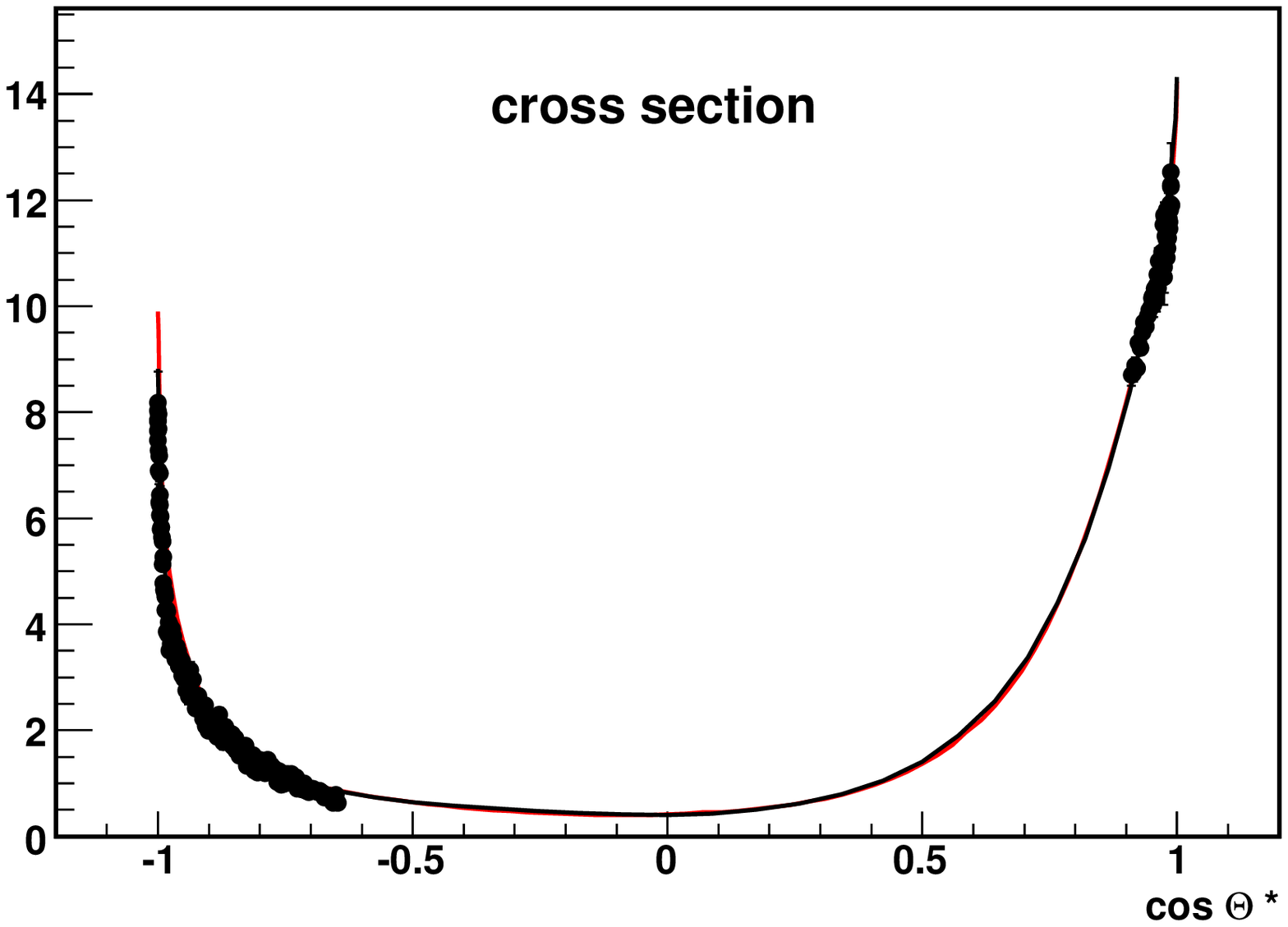}}
\end{figure}


\begin{figure}[hbtp]
 \centering
    \resizebox{9.8cm}{!}{\includegraphics{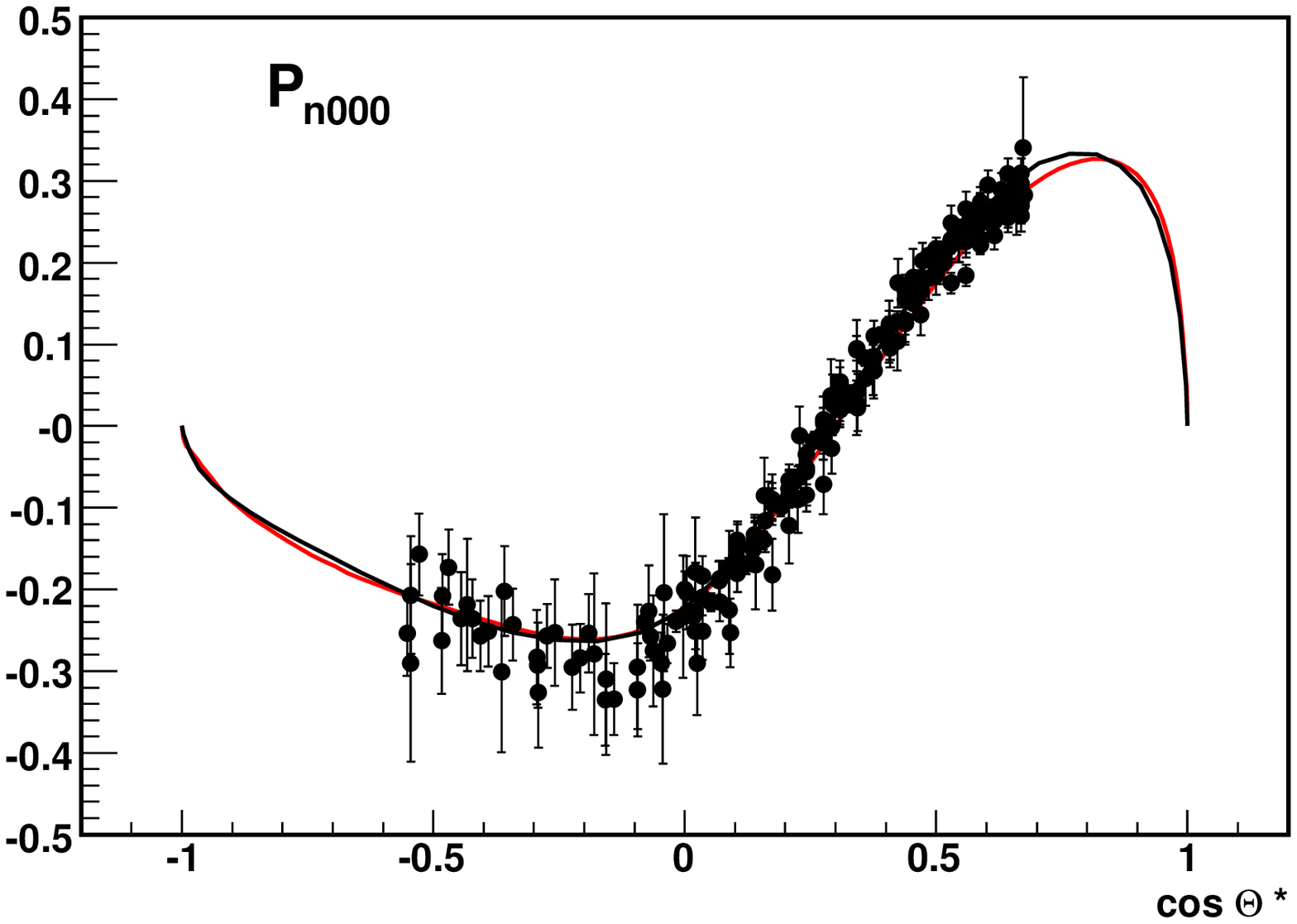}}
\end{figure}


\begin{figure}[hbtp]
 \centering
    \resizebox{9.8cm}{!}{\includegraphics{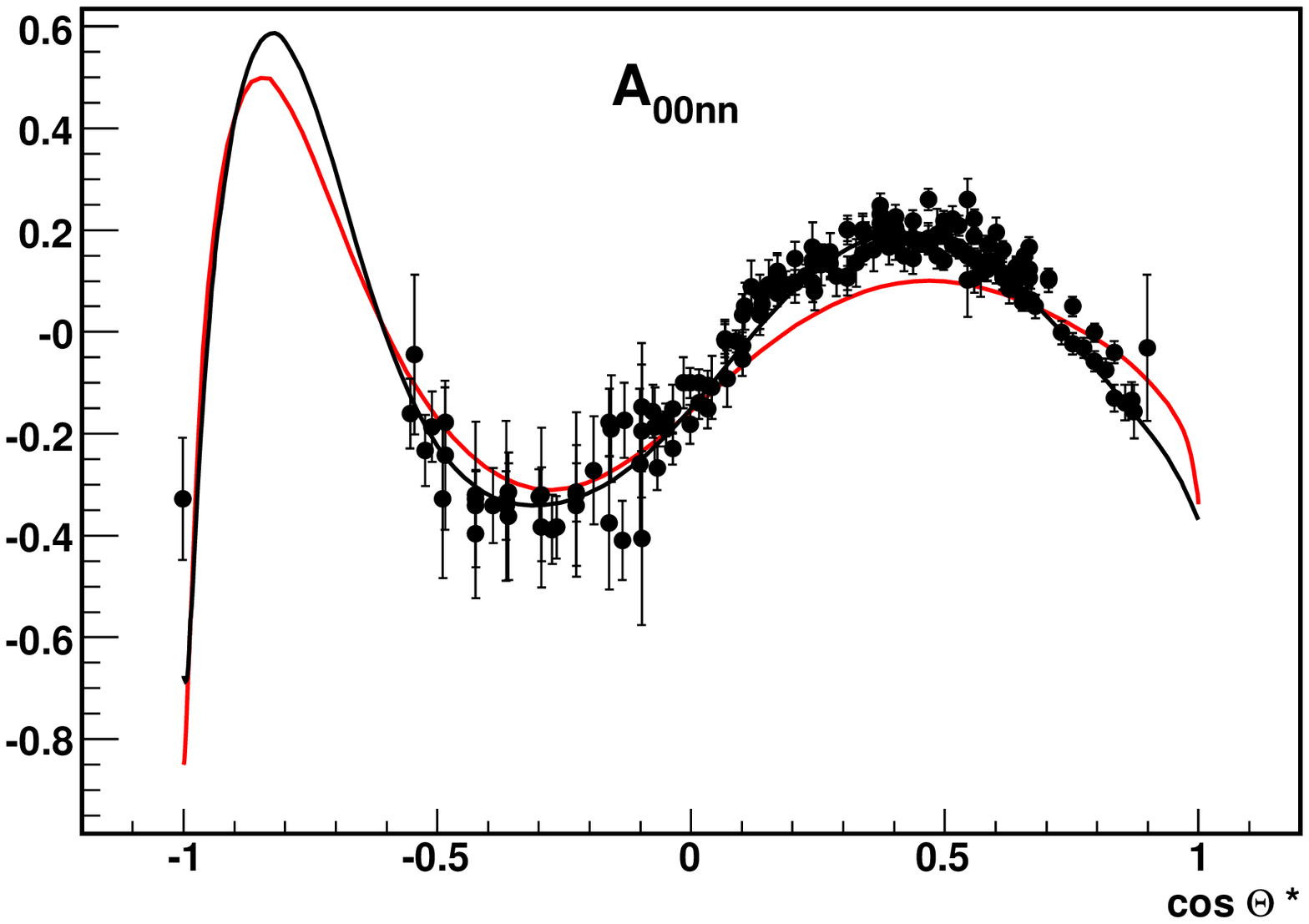}}
\end{figure}

\newpage

$T_{lab}=1000 MeV$, Re part

\begin{tabular}{|l|l|l|l|l|}
\hline
R & VSE & VTO & VLSO & VTNO\\ \hline
0.11 & 60666.9 & 2888.67 & 0 & -230463\\
0.15 & -72709.5 & -808.002 & 0 & 98816.7\\
0.25 & 28017.3 & 4547.87 & 0 & -7827.76\\
0.4 & -7447.1 & -2165.55 & 0 & 644.909\\
0.55 & 1117.53 & 230.777 & 0 & -164.754\\
0.7 & -50.0001 & -15.2833 & 0 & 29.8561\\
1.4 & -3.56802 & 0.979312 & 0 & -0.0626672\\
\hline
\end{tabular}

\vspace{2cm}
$T_{lab}=1000 MeV$, Im part

\begin{tabular}{|l|l|l|l|l|}
\hline
R & VSE & VTO & VLSO & VTNO\\ \hline
0.11 & -98711.2 & -9057.76 & 129026 & -300001\\
0.15 & 97688.4 & 96145.3 & -54998.6 & 143127\\
0.25 & -24155 & -26263.4 & 4350.96 & -14093.3\\
0.4 & 2897.23 & 2389.94 & 167.686 & 1192.58\\
0.55 & -390.274 & 113.728 & -60.9247 & -156.562\\
0.7 & 163.013 & -116.033 & 10.5639 & 2.00019\\
1.4 & 2.2639 & -0.415819 & -0.0387636 & -0.0420831\\
\hline
\end{tabular}

\vspace{2cm}
$T_{lab}=1000 MeV$, Re part

\begin{tabular}{|l|l|l|l|l|}
\hline
R & VSO & VTE & VLSE & VTNE\\ \hline
0.11 & 44901.6 & -15179 & 0 & 605715\\
0.15 & -70514.2 & 23029.8 & 0 & -240000\\
0.25 & 14013.4 & -6293.02 & 0 & 22538.7\\
0.4 & -3749.13 & 1127 & 0 & -2967.2\\
0.55 & 1270.4 & -600.001 & 0 & 592.673\\
0.7 & -199.734 & 137.29 & 0 & -91.3122\\
1.4 & 4.82993 & -0.142917 & 0 & 0.0768991\\
\hline
\end{tabular}

\vspace{2cm}
$T_{lab}=1000 MeV$, Im part

\begin{tabular}{|l|l|l|l|l|}
\hline
R & VSO & VTE & VLSE & VTNE\\ \hline
0.11 & -119999 & 64686.9 & -90000 & -120000\\
0.15 & 59750.4 & -34026.4 & 23438.1 & 93524\\
0.25 & -6000 & 2150.45 & -1500 & -10164.8\\
0.4 & 2797.14 & -1557.28 & 906.58 & 1576.55\\
0.55 & -2312.91 & 769.397 & -309.155 & -235.409\\
0.7 & 600 & -150.109 & 60.0015 & 60.937\\
1.4 & -30.1323 & 0.190335 & -0.439257 & 0.0573121\\
\hline
\end{tabular}

\newpage
$T_{lab}=1000 MeV$, pp

\begin{figure}[hbtp]
 \centering
    \resizebox{9.8cm}{!}{\includegraphics{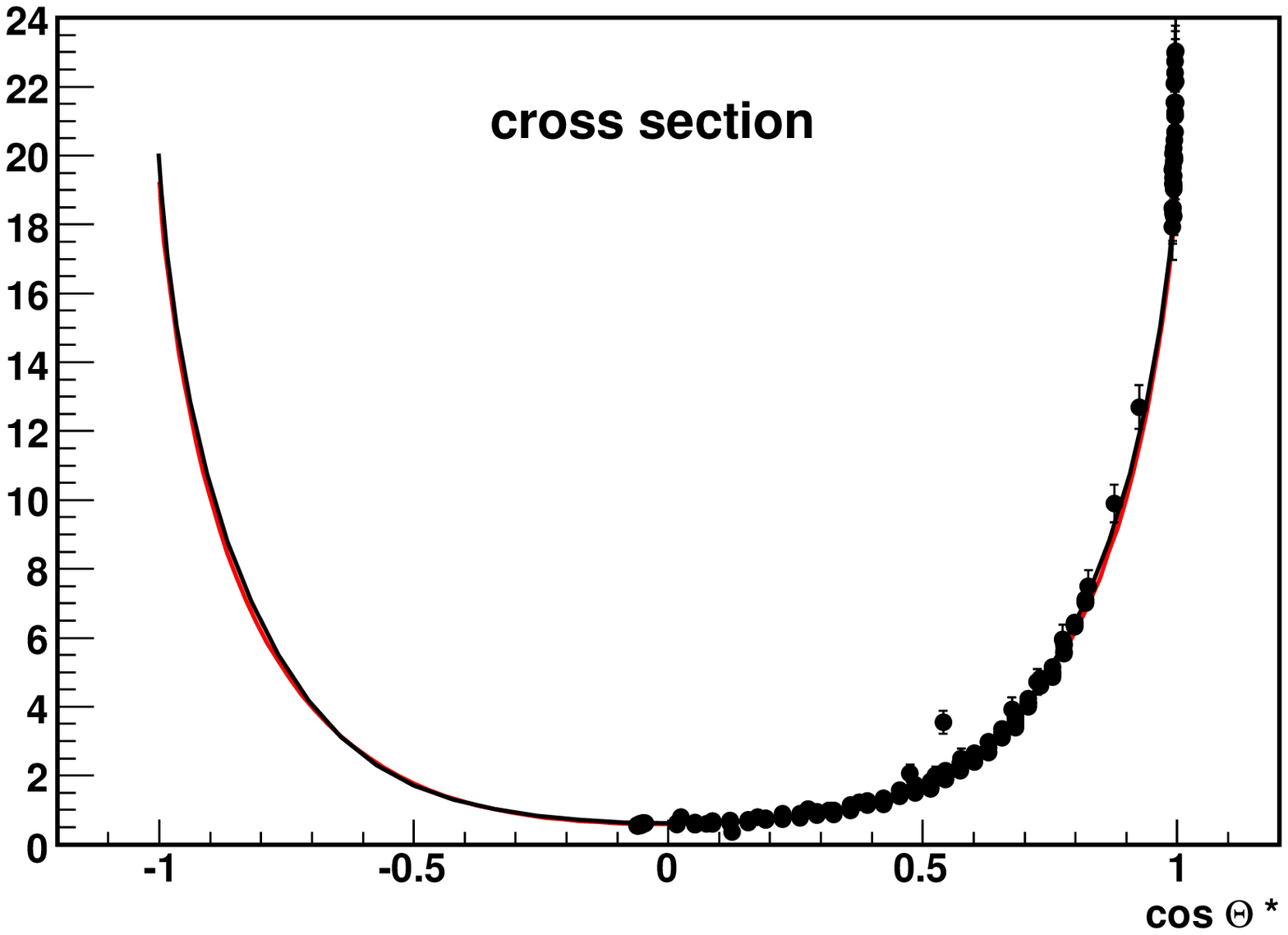}}
\end{figure}


\begin{figure}[hbtp]
 \centering
    \resizebox{9.8cm}{!}{\includegraphics{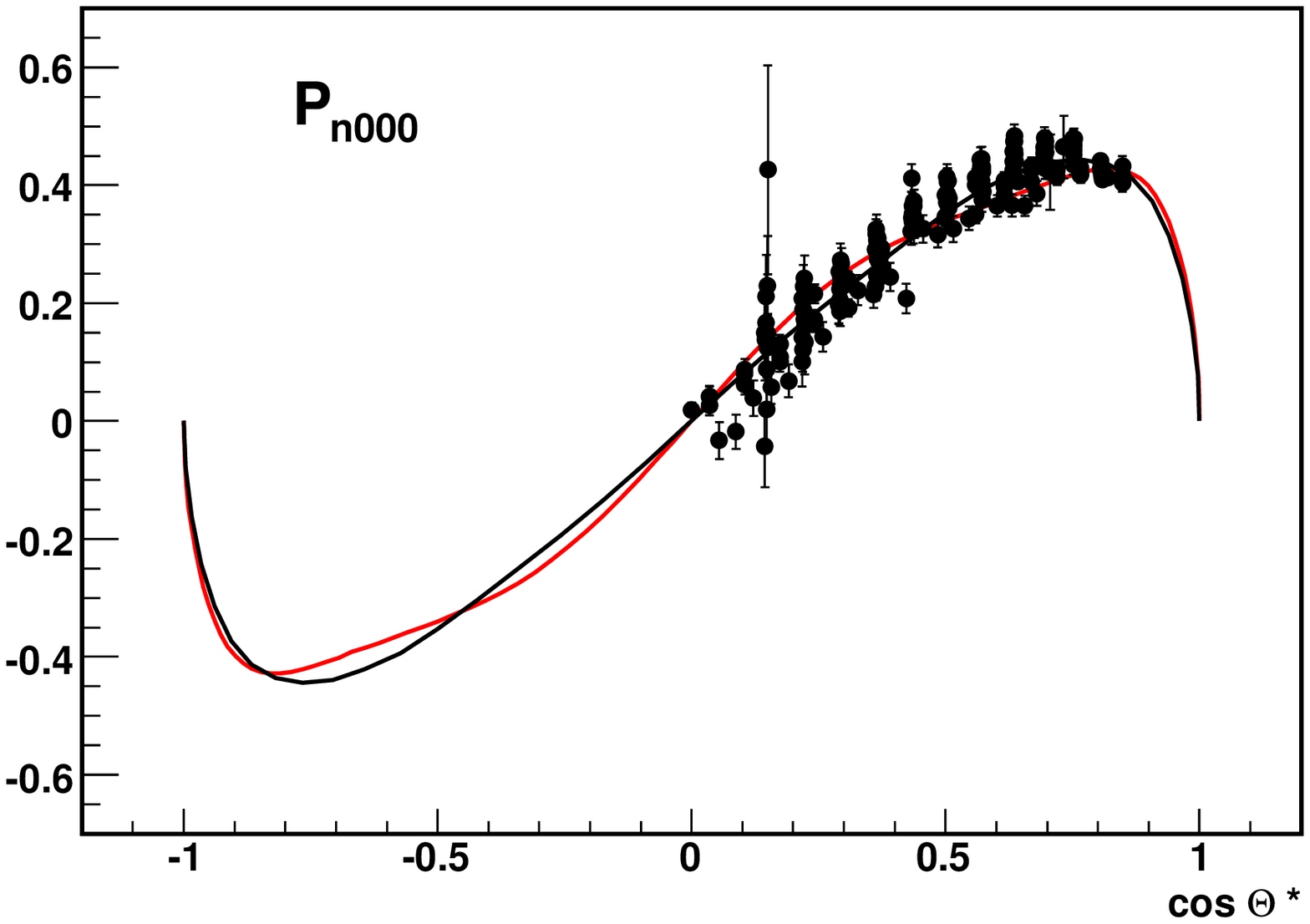}}
\end{figure}


\begin{figure}[hbtp]
 \centering
    \resizebox{9.8cm}{!}{\includegraphics{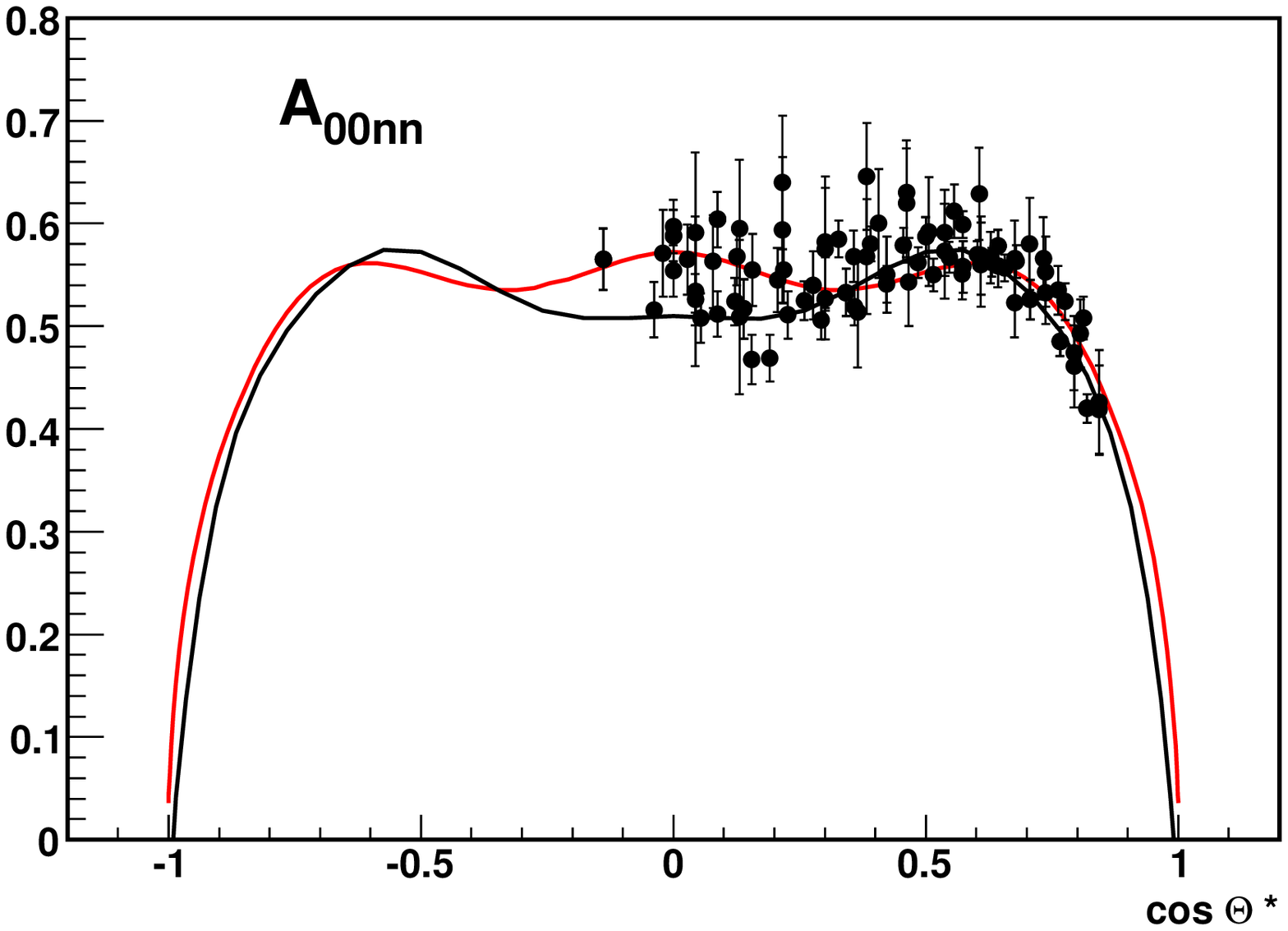}}
\end{figure}

\newpage
$T_{lab}=1000 MeV$, np

\begin{figure}[hbtp]
 \centering
    \resizebox{9.8cm}{!}{\includegraphics{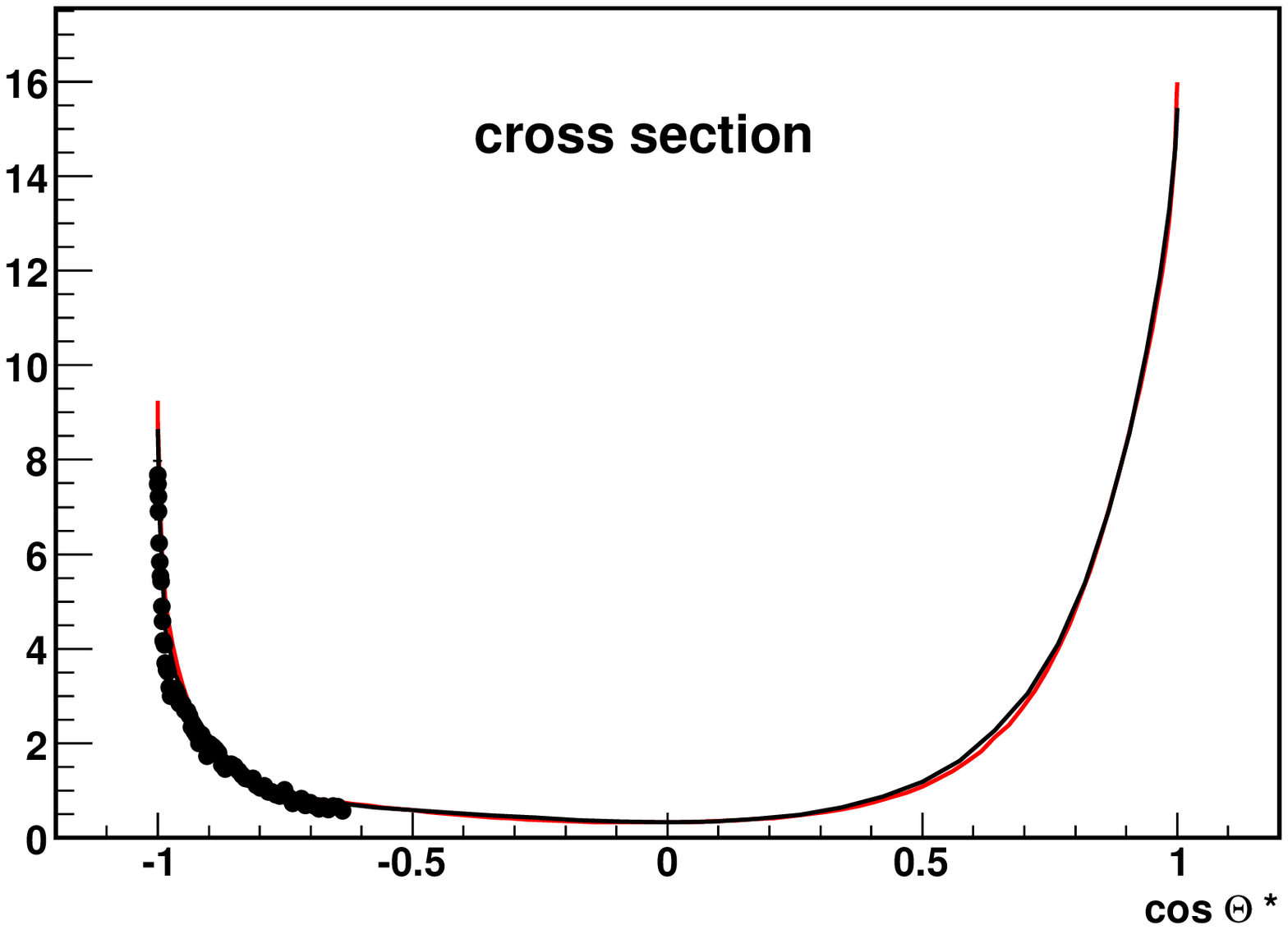}}
\end{figure}


\begin{figure}[hbtp]
 \centering
    \resizebox{9.8cm}{!}{\includegraphics{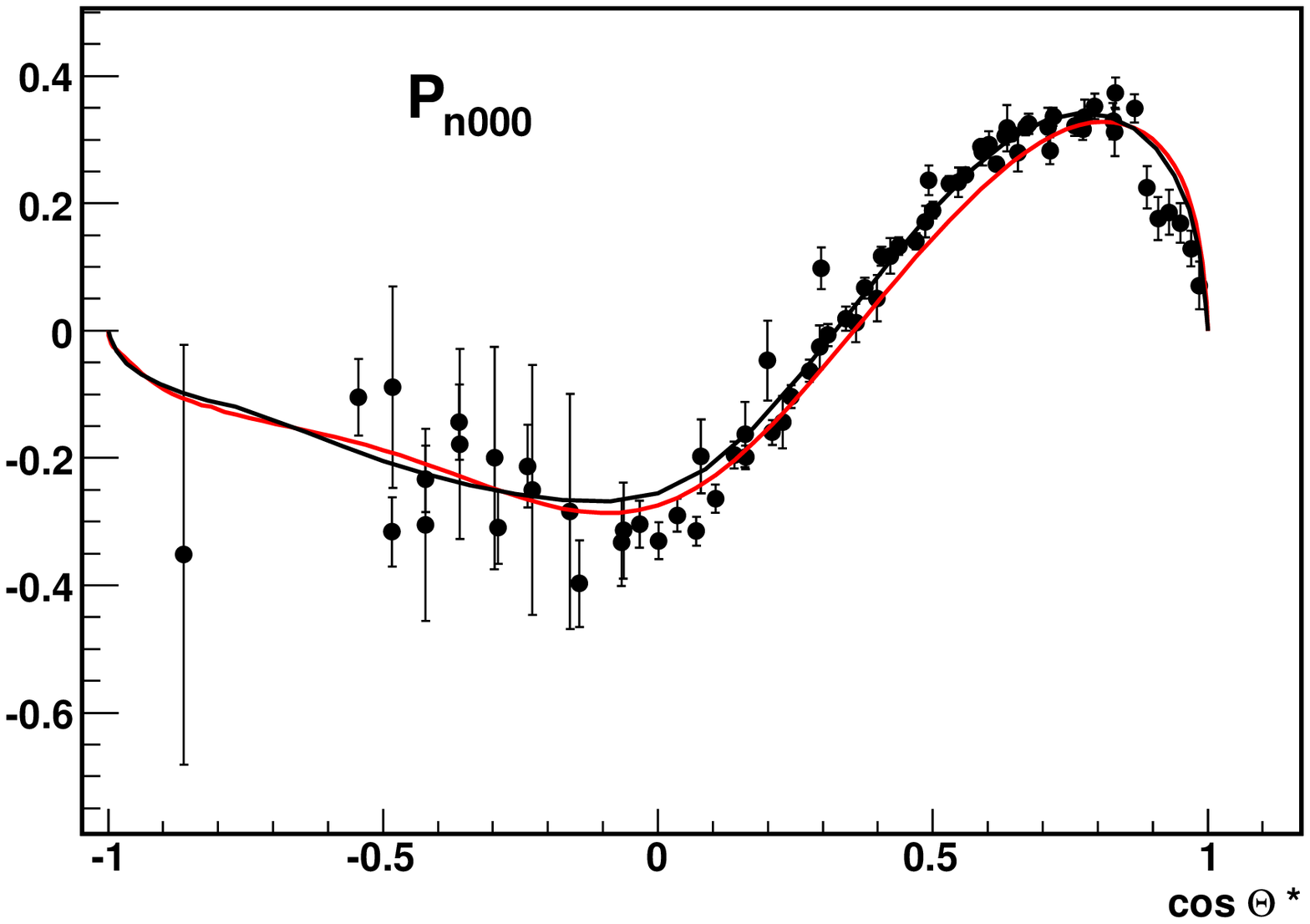}}
\end{figure}


\begin{figure}[hbtp]
 \centering
    \resizebox{9.8cm}{!}{\includegraphics{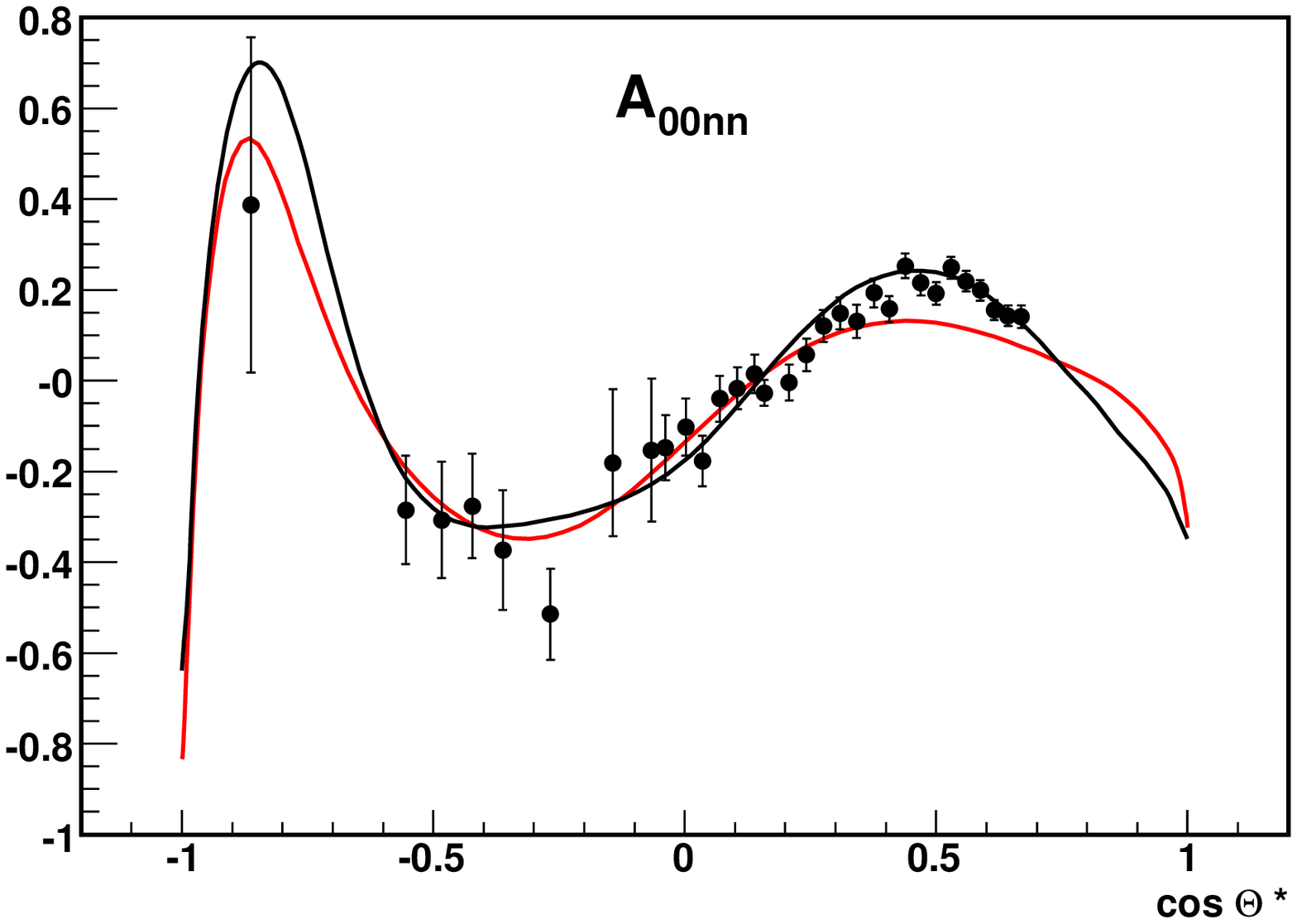}}
\end{figure}

\newpage

$T_{lab}=1100 MeV$, Re part

\begin{tabular}{|l|l|l|l|l|}
\hline
R & VSE & VTO & VLSO & VTNO\\ \hline
0.11 & 69410.4 & 550 & 0 & -185846\\
0.15 & -71738.4 & -808 & 0 & 72199.7\\
0.25 & 25344 & 4289.49 & 0 & -4635.81\\
0.4 & -6844.69 & -1986.87 & 0 & 250\\
0.55 & 993.501 & 175 & 0 & -98.0508\\
0.7 & -36.194 & -9.61922 & 0 & 22.7208\\
1.4 & -3.27555 & 1.13202 & 0 & -0.0626293\\
\hline
\end{tabular}

\vspace{2cm}
$T_{lab}=1100 MeV$, Im part

\begin{tabular}{|l|l|l|l|l|}
\hline
R & VSE & VTO & VLSO & VTNO\\ \hline
0.11 & -102796 & -9017.06 & 181874 & -120000\\
0.15 & 95875.2 & 81222.3 & -74712.7 & 85623.9\\
0.25 & -22050.9 & -23137.9 & 4967.06 & -11257.2\\
0.4 & 2184.63 & 2204.78 & 168 & 1058.8\\
0.55 & -100.491 & 87.9625 & -76.9202 & -149.126\\
0.7 & 94.7553 & -117.096 & 14.3827 & 2.58782\\
1.4 & 2.25999 & -0.100002 & -0.0596146 & -0.0456317\\
\hline
\end{tabular}

\vspace{2cm}
$T_{lab}=1100 MeV$, Re part

\begin{tabular}{|l|l|l|l|l|}
\hline
R & VSO & VTE & VLSE & VTNE\\ \hline
0.11 & 48439.2 & -13708.4 & 0 & 551305\\
0.15 & -59038.5 & 22873.1 & 0 & -235000\\
0.25 & 7698.42 & -6077.29 & 0 & 23953.4\\
0.4 & -1908.28 & 650.012 & 0 & -3143.67\\
0.55 & 1121.09 & -398.936 & 0 & 631.914\\
0.7 & -319.695 & 109.256 & 0 & -95.9452\\
1.4 & 4.6774 & -0.110743 & 0 & 0.0900328\\
\hline
\end{tabular}

\vspace{2cm}
$T_{lab}=1100 MeV$, Im part

\begin{tabular}{|l|l|l|l|l|}
\hline
R & VSO & VTE & VLSE & VTNE\\ \hline
0.11 & -218411 & 41234.9 & -86000 & -60000\\
0.15 & 86984 & -21169.5 & 15583.9 & 69533.6\\
0.25 & -6075.13 & 1340.98 & -600 & -8687.87\\
0.4 & 5346.6 & -1196.53 & 818.464 & 1358.4\\
0.55 & -4490.38 & 500.001 & -300 & -186.977\\
0.7 & 1269.38 & -85.6839 & 58.3339 & 53.0555\\
1.4 & -38.4122 & 0.104405 & -0.402291 & 0.0701362\\
\hline
\end{tabular}

\newpage
$T_{lab}=1100 MeV$, pp

\begin{figure}[hbtp]
 \centering
    \resizebox{9.8cm}{!}{\includegraphics{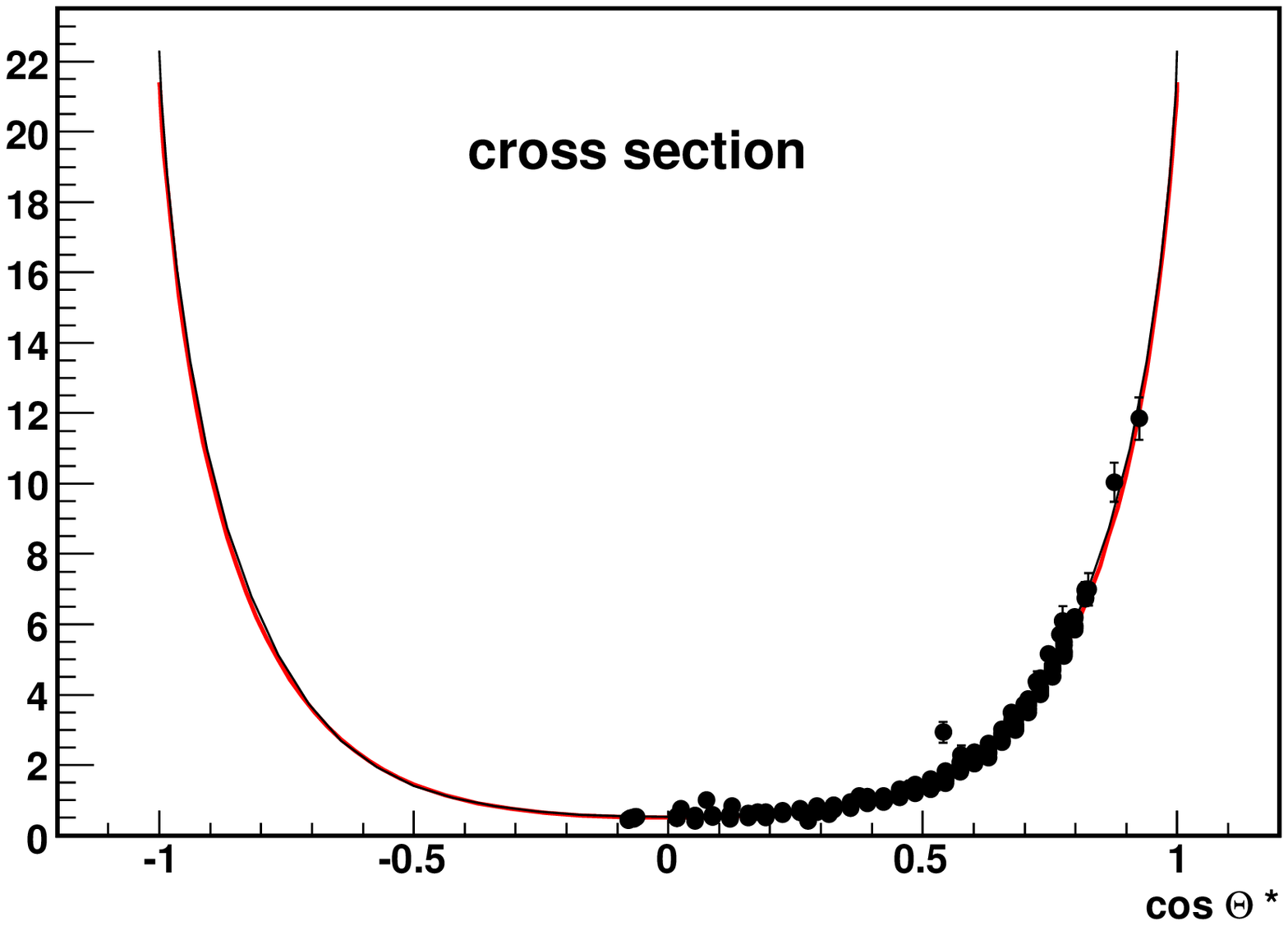}}
\end{figure}


\begin{figure}[hbtp]
 \centering
    \resizebox{9.8cm}{!}{\includegraphics{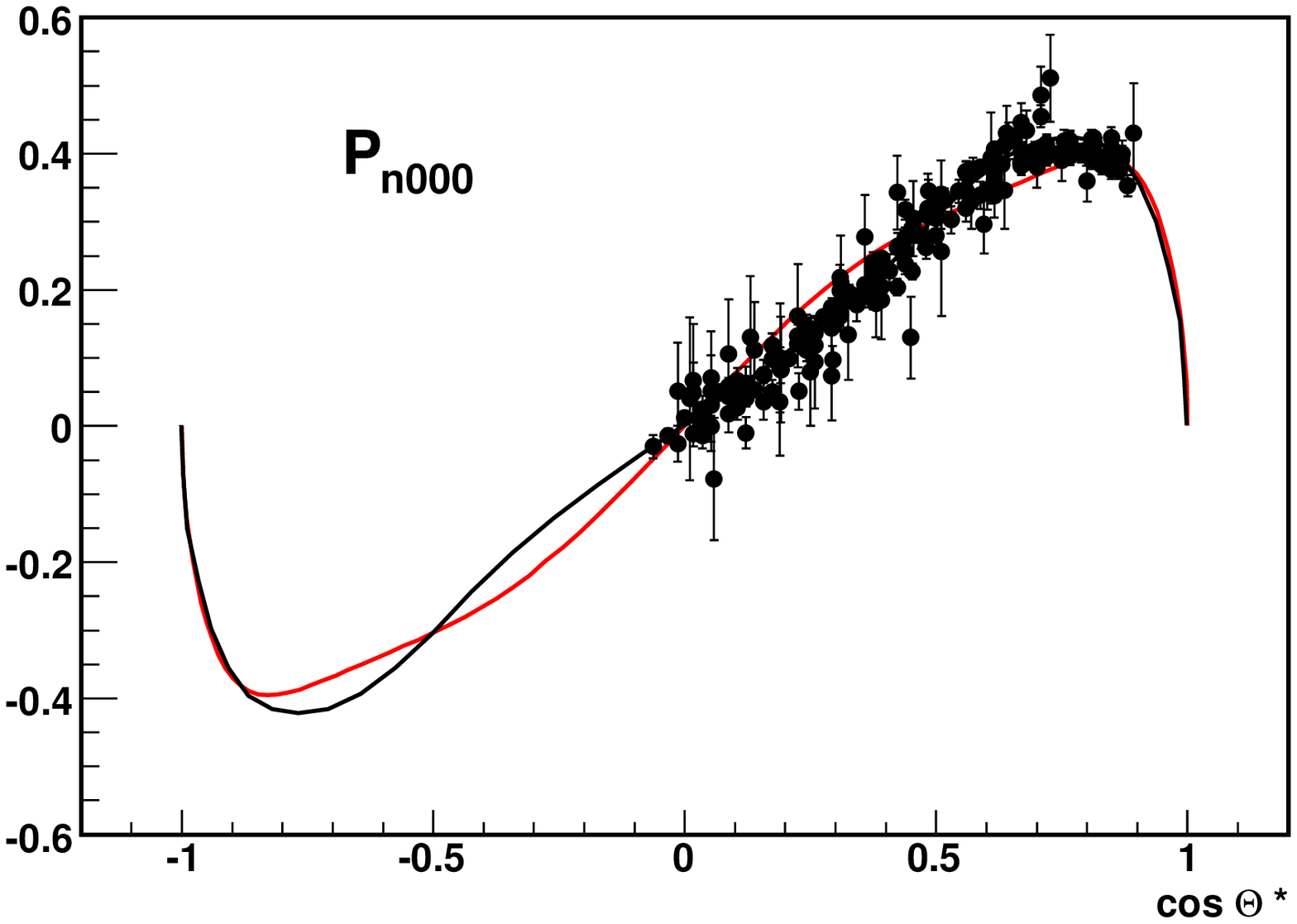}}
\end{figure}


\begin{figure}[hbtp]
 \centering
    \resizebox{9.8cm}{!}{\includegraphics{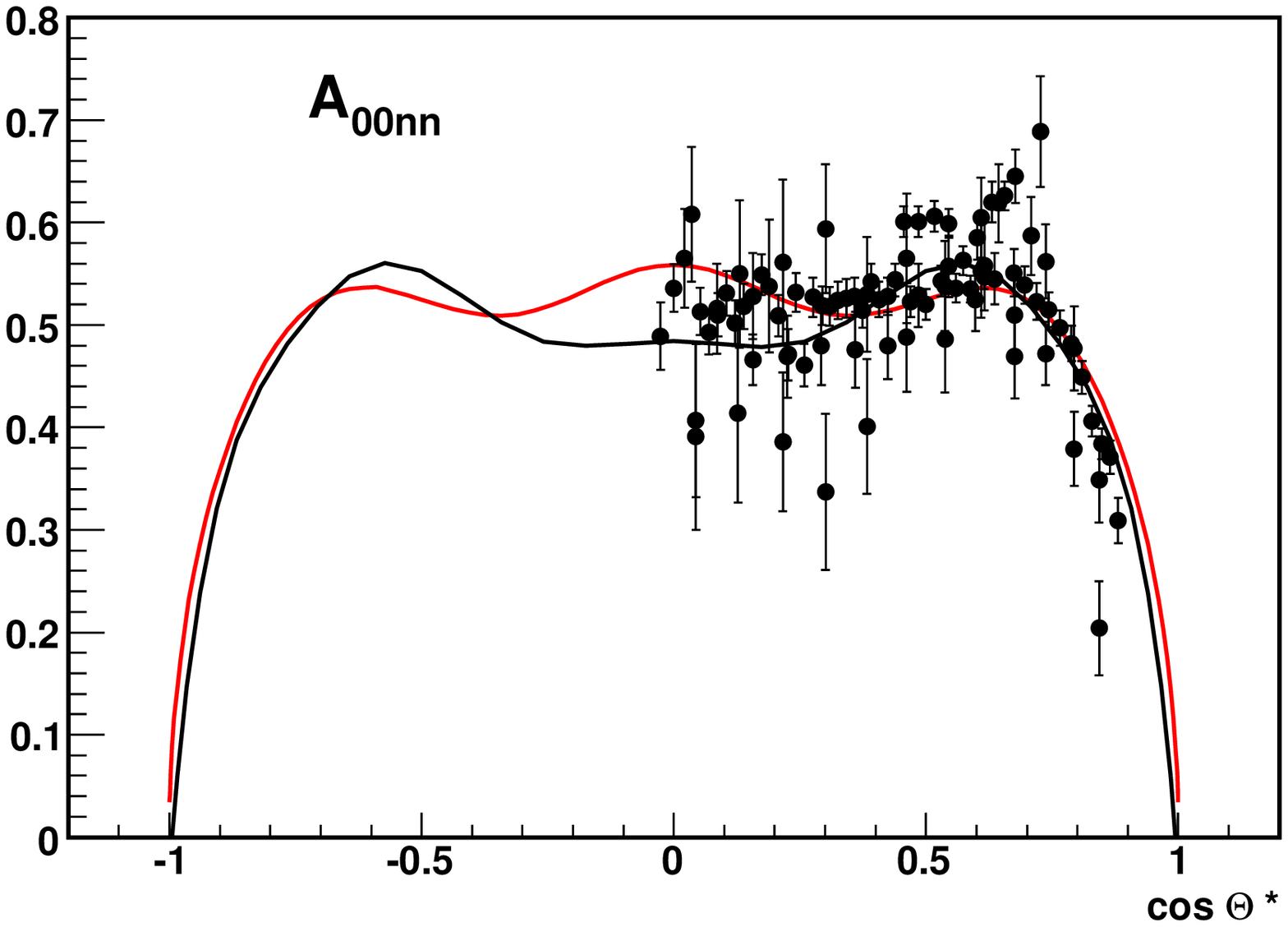}}
\end{figure}

\newpage
$T_{lab}=1100 MeV$, np

\begin{figure}[hbtp]
 \centering
    \resizebox{9.8cm}{!}{\includegraphics{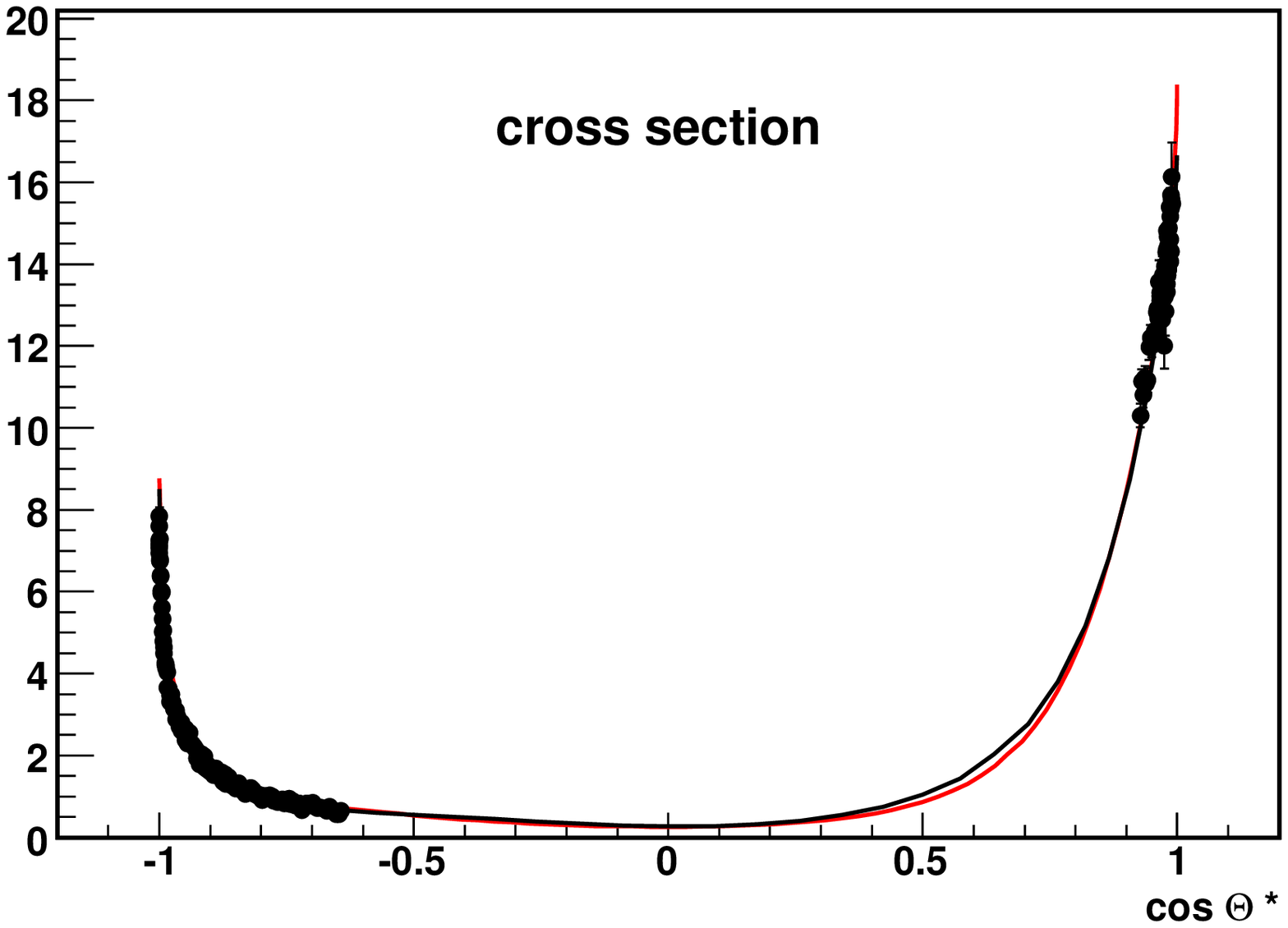}}
\end{figure}


\begin{figure}[hbtp]
 \centering
    \resizebox{9.8cm}{!}{\includegraphics{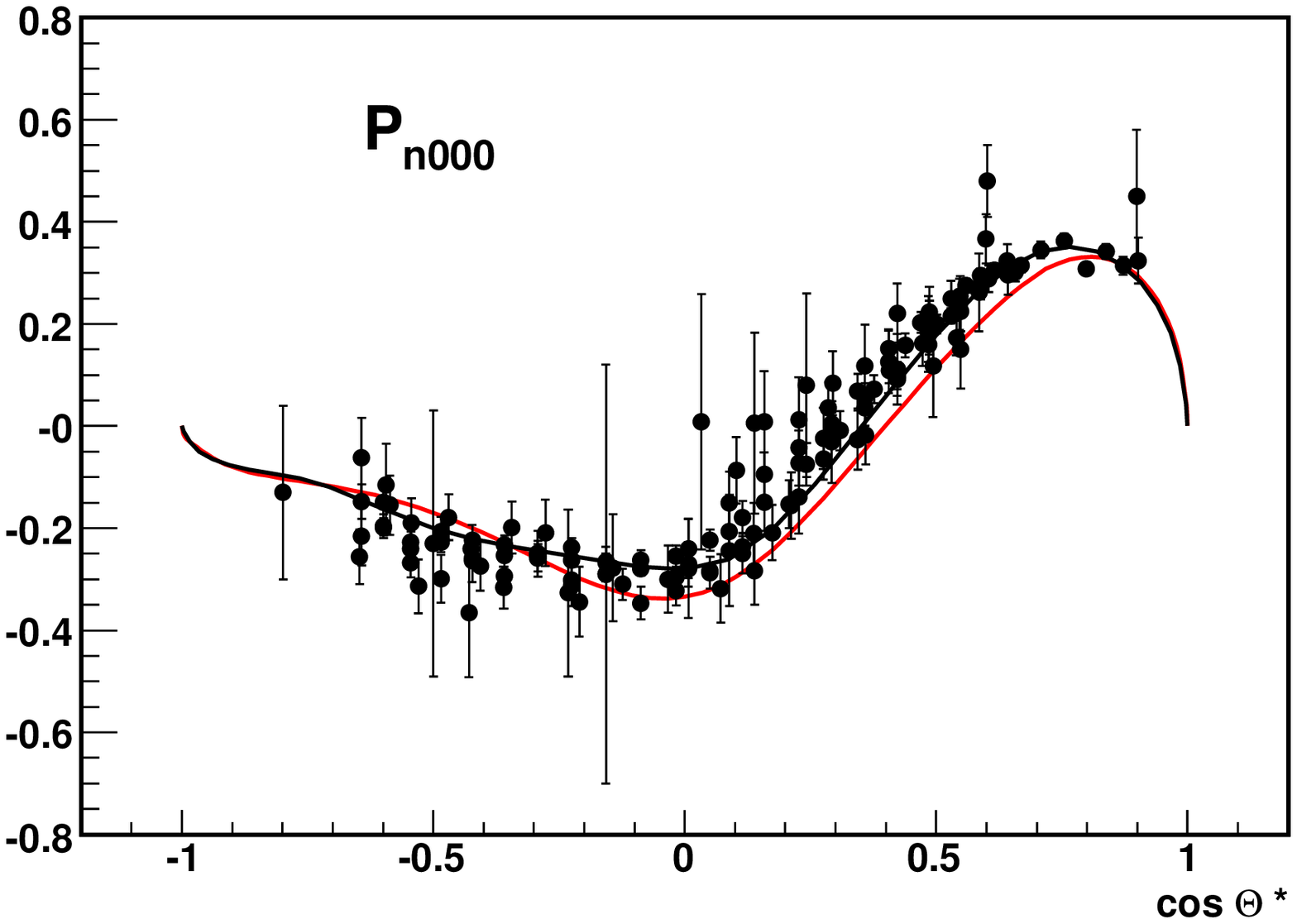}}
\end{figure}


\begin{figure}[hbtp]
 \centering
    \resizebox{9.8cm}{!}{\includegraphics{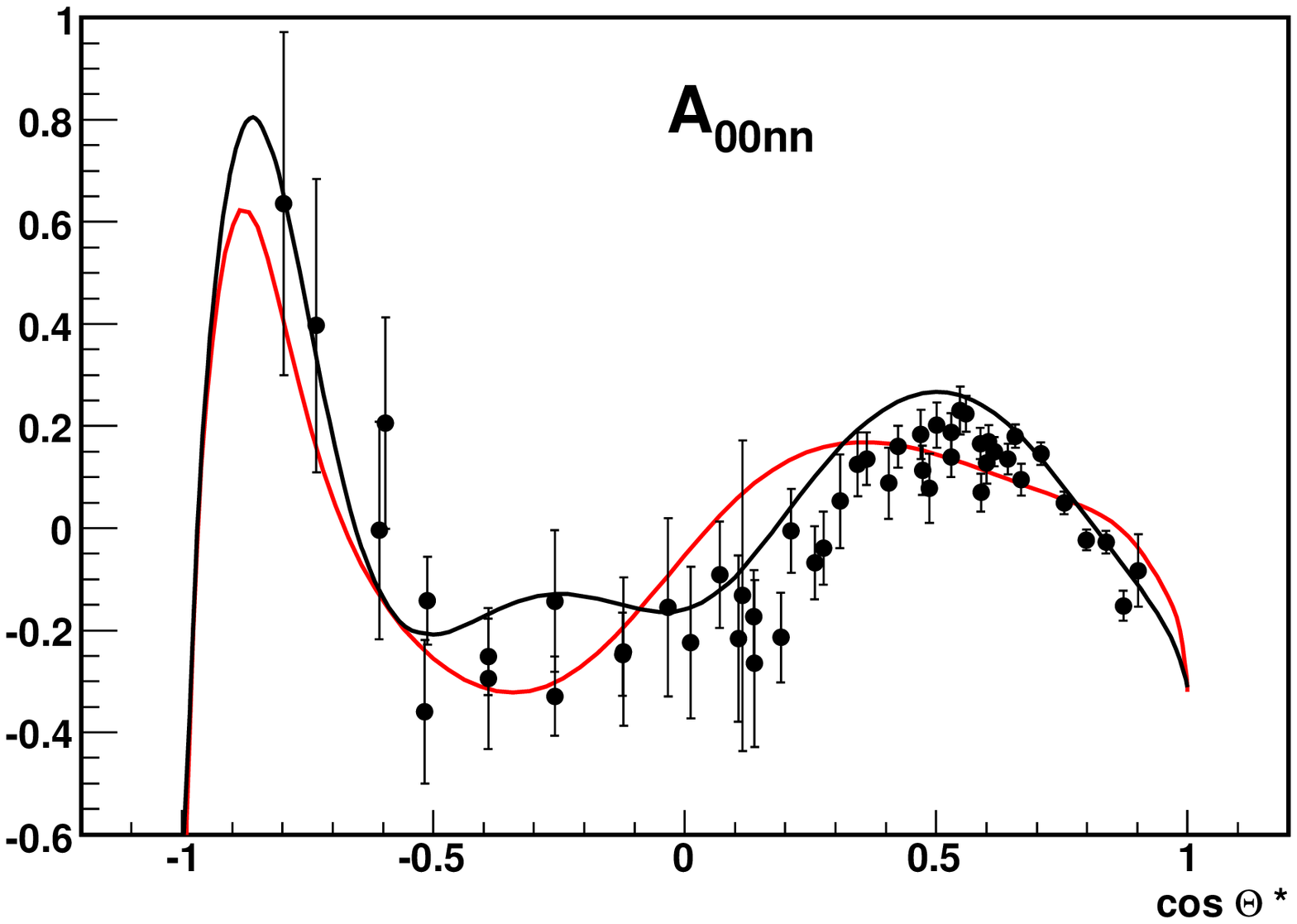}}
\end{figure}

\end{document}